\documentclass[fleqn,usenatbib]{mnras}

\usepackage{newtxtext,newtxmath}

\usepackage[T1]{fontenc}

\DeclareRobustCommand{\VAN}[3]{#2}
\let\VANthebibliography\thebibliography
\def\thebibliography{\DeclareRobustCommand{\VAN}[3]{##3}\VANthebibliography}

\usepackage{lineno}

\usepackage{graphicx}	
\usepackage{amsmath}	
\graphicspath{{./}{figures/}}
\usepackage{subcaption}
\title[z$\sim$2 Type-2 QSO Spectra]{Luminous Mid-IR Selected Obscured Quasars at Cosmic Noon in SDSS Stripe82 II: Spectroscopic Diversity and Broad H$\alpha$ Emissions }

\author[Ben Wang et al.]{{Ben Wang$^{1,2}$\thanks{E-mail: bwang@mail.strw.leidenuniv.nl},}
{Yuzo Ishikawa$^{3}$,}
{Joseph F. Hennawi$^{2,4}$,}
{Zheng Cai$^{1}$,}
{Gordon T. Richards$^{5}$,}
{Nadia L. Zakamska$^{6,7}$,}
\newauthor
{Daming Yang$^{2}$,}
{Jan-Torge Schindler$^{8}$}
\\
$^{1}$Department of Astronomy, Tsinghua University, Beijing 100084, China\\
$^{2}$Leiden Observatory, Leiden University, Leiden 2333 CA, Netherland\\
$^{3}$MIT Kavli Institute for Astrophysics and Space Research, Massachusetts Institute of Technology, Cambridge, MA 02139, USA\\
$^{4}$Department of Physics, Broida Hall, University of California, Santa Barbara Santa Barbara, CA 93106-9530, USA\\
$^{5}$Department of Physics, Drexel University, 32 S. 32nd Street, Philadelphia, PA 19104, USA\\
$^{6}$Department of Physics and Astronomy, Bloomberg Center, Johns Hopkins University, Baltimore, MD 21218, USA \\
$^{7}$Institute for Advanced Study, Princeton University, Princeton, NJ 08544, USA\\
$^{8}$Hamburg Observatory, Gojenbergsweg 112, 21029 Hamburg, Germany\\
\\
}

\date{Accepted XXX. Received YYY; in original form ZZZ}

\pubyear{2015}

\begin{document}
\label{firstpage}
\pagerange{\pageref{firstpage}--\pageref{lastpage}}
\maketitle

\begin{abstract}

We present a multiwavelength spectroscopic survey of 23 luminous mid-infrared–selected Type-2 quasars at the redshifts of $z=0.88$–$3.49$. The targets were 
selected in the SDSS Stripe 82 field based on their bright WISE 22~$\mu$m detections 
(flux $>5$ mJy) and extremely faint or red optical counterparts (e.g., $r>23$ or $r - W4 > 8.4$), 
designed to identify heavily obscured quasars. 
Deep near-infrared (Gemini/GNIRS) and optical (Keck/LRIS and KCWI) 
spectroscopy confirm 23 out of 24 candidates as Type-2 quasars in this redshift range, 
including 12 objects at $z>2$. The spectra exhibit strong rest-frame UV and optical emission 
lines (Ly$\alpha$, C IV, [O III], H$\alpha$) with a wide range of line widths, 
indicating significant spectral diversity. Approximately one-third of the sample 
(8 of 23) shows broad H$\alpha$ emission (FWHM $>2000$ km s$^{-1}$) despite their 
Type-2 classification, while the rest have only narrow lines (FWHM $<2000$ km s$^{-1}$) 
characteristic of classical obscured quasars. Notably, these broad-line Type-2 quasars 
share similar spectral energy distributions with the JWST-discovered “little red dot” (LRD) AGNs, 
suggesting that our sample could be lower-redshift analogues of the heavily obscured 
broad-line AGNs uncovered by JWST. We also find that the [O III] $\lambda$5007 emission 
is relatively weak for their high bolometric luminosities, deviating from trends 
seen in lower-$z$ Type-2 QSOs. 
A new composite spectrum for Type-2 QSOs is built using our sample. 
Overall, our results demonstrate that mid-IR selection 
efficiently uncovers a diverse population of obscured quasars and that spectroscopic 
follow-up is crucial for revealing their true nature. This study provides new insights 
into heavily obscured SMBH growth at cosmic noon and bridges the gap to the obscured 
AGN populations now being revealed by JWST.

\end{abstract}






\section{Introduction}
Quasars, or quasi-stellar objects (QSOs), are among the most luminous objects in the universe, powered by accretion onto supermassive black holes (SMBHs) at their centers \citep{Schmidt1963}. Their intense radiation originates from the accretion of matter onto the SMBH, producing high-energy emission across multiple wavelengths. 
According to the classical  “unified” model \citep{Antonucci1993,Urry1995} for active galactic nuclei (AGN), there are two different types of AGNs. In Type-1, or unobscured, AGNs, both continuum emission from the accretion disk region and broad emission lines from the broad line region can be detected. The other type are called Type-2, or obscured, AGN. In this case, the line of sight to the broad emission line and the central continuum regions are obscured by the dusty torus, so only the narrow emission lines can be detected. In the classical  “unified” model, the viewing angle is the main reason to cause the difference of these two types of AGNs. 
 Another scenario is that the difference between Type-1 and Type-2 QSOs is due to evolutionary effects. In this picture, Type-1 and Type-2 QSOs 
 reflect different evolutionary phases,
 especially after galaxy mergers \citep[e.g.][]{Sanders88, Hopkins06}. Some researchers even find that the environment around Type-2 AGNs could be denser than Type-1 AGNs \citep{Cai2017,Zhang2023}. Investigating the relative abundances of 
 of Type-1 and Type-2 QSOs across cosmic time can help us to better understand the physical processes producing these two different kinds of QSOs, and further understand the SMBH accretion and SMBH/galaxy co-evolution. 
 
In the past decades, large-scale surveys have discovered substantial numbers of Type-1 QSOs, with more than 750,000 identified through dedicated programs such as the Sloan Digital Sky Survey (SDSS; e.g., \citealt{Schneider2007,Lyke2020}) and the ongoing Dark Energy Spectroscopic Instrument (DESI) survey (e.g., \citealt{Chaussidon2023,DESI2024}). Dedicated efforts have pushed the Type-1 QSO frontier to  $z>7$ \citep[e.g.][]{Ba2018,Wangf2021}. For Type-2 QSOs, dust obscuration makes them faint in the UV/optical,; thus, multi-wavelength data are needed to isolate these objects. Sizable samples of Type-2 QSOs have been identified from mid-IR, X-ray, and the SDSS prior to the launch of JWST \citep[e.g.][]{Zakamska2003,Haas2004,Mart2006,Alonso-Herrero2006,Brand2007,Alexandroff2013,Lacy2013,Yuan2016}. 
However, the majority of these Type-2 quasars are low-luminosity objects at z $<$ 1. 
The number of spectroscopically confirmed luminous Type-2 QSO at $z > 2$  is still very limited \citep{Lacy2013}, leading to a poorly understood QSO luminosity function (QLF) for obscured QSO populations \citep{Glikman2018}.

The identification of obscured quasars relies on multi-wavelength selection techniques combined with spectroscopic confirmation. Optical Type-2 QSO samples from SDSS spectroscopy represent the most extensive collections at low redshifts $z<1$
\citep{Zakamska2003,Zakamska2006,Reyes2008,Yuan2016}. X-ray surveys offer an alternative method to uncover Type-2 AGN \citep{Polletta+2006,Brandt15,Peca2023,Peca2024}; however, these surveys often depend on uncertain or poorly calibrated photometric redshifts and X-ray surveys will miss Compton-thick objects with high photoelectric obscuring columns $N_{\rm H} > 10^{24}\,{\rm cm^{-2}}$.

Nevertheless, previous multi-wavelength surveys of Type-2 quasars have assembled substantial samples that allow measurements of the obscured-to-unobscured quasar ratio (Type-2:Type-1) at $z\lesssim 2$. 
According to the AGN unification model, the number density of Type-1 and Type-2 QSOs should be comparable through cosmic time,  both peaking at $z \sim 2.5$ \citep{Richards2006,Ross2013}. Decades of AGN/QSO censuses across the electromagnetic spectrum \citep[e.g.][]{Zakamska2003,Assef13,Lacy2013,Yan2013,Brandt15} 
have led to the consensus that the Type-2: Type-1 ratio is 
comparable at low redshift $z<1$ \citep[e.g.][]{Reyes2008,Lawrence2010,Merloni2014}, consistent with the unification models. 

However, recent JWST observations suggest the presence of a previously undetected, substantial population of potentially obscured AGN at high redshift (${4 \lesssim z \lesssim 9}$), often referred to as "little red dots" (LRDs; e.g., \citep{Matthee+23,Greene+23,Kocevski2023}). 
Some of these sources show broad emission lines in their spectra, suggestive of active accretion onto a supermassive black hole\citep[e.g.][]{Kokorev2023,Harikane2023}. Intriguingly, the inferred obscured-to-unobscured ratio from these JWST detections is possibly as extreme as $\sim 10^{3-4}:1$ \citep{Pizzati2024}, far exceeding expectations from previous models.  Following the So\l tan argument \citep{Soltan1982}, this would suggest as much SMBH growth at $z \gtrsim 4$ as at later times \citep{Inayoshi2025}. But the redshift distribution of these LRDs is mainly at $z = 4- 9$\citep[e.g.][]{Akins2024}. It remains uncertain whether a similarly large obscured population exists at intermediate redshifts at $2 < z < 4$. This gap introduces substantial uncertainty in our understanding of the cosmic history of SMBH growth and the full evolution of the quasar population. 

Identifying the obscured QSO population at intermediate redshift $2<z<4$ is the key to bridging this gap. 
Mid-infrared (mid-IR) selection provides a more complete and less biased approach to identifying obscured quasars. Color-selection criteria, such as mid-IR color wedges defined using {\it Spitzer} \citep{Lacy2004,Stern2005,Polleta+2008,Donley2012} and later
WISE \citep{Assef13,Glikman2018} have been used to select AGN candidates that are subsequently confirmed via optical and IR spectroscopy \citep{Lacy2013,Lacy13b,Glikman2018}. 
However, the redshift distribution of mid-IR-selected Type-2 quasars shows a sharp decline beyond $z>1$, with only 18 spectroscopically confirmed luminous Type-2 QSOs known at $z>2$ in \citet{Lacy2013,Glikman2018}. 

A very simple selection requiring a red optical to mid-IR
$r-[24\,\mu\rm{m}] \gtrsim 8$ color (AB), has been shown to
efficiently identify $z\gtrsim2$ dust obscured galaxies (DOGs)
\citep{Brand2007,Dey2008,Yan2013}.  The effectiveness of DOG selection is intuitive: adopting the reddest possible mid-IR band probes the peak of the AGN IR SED ($\lambda_{\rm rest} \sim 10\,\mu\rm{m}$), while demanding a
red $r-[24\,\mu\rm{m}]$ color excludes objects with AGN emission at
bluer wavelengths. The DOG selection uncovers a large population of the obscured populations with all diversity, and the number density $\rm \sim 0.9 deg^{-2}$ is comparable to the Type-1 QSO population with the same W4 flux level\citep{Yan2013}. 

A notable subset of WISE-selected AGNs are the so-called "W1W2 dropouts" or Hot Dust Obscured Galaxies (Hot DOGs), characterized by extremely red mid-IR spectral energy distributions. Published spectra of $\sim 60$ Hot DOGs indicate that $\sim 40$ are type-2s at $z\gtrsim 2$ \citep{Wu12,Tsai2015}. However, these objects are nearly an order of magnitude less abundant in the sky \citep[252 Hot-DOGs in 32,000 $\rm deg^2$][]{Eisenhardt2012,Assef2015}  than DOGs, 
implying that they likely constitute the (mid-IR)
reddest subset of the Type-2 population. 
Most WISE-selected obscured quasars are still located at $z<2$, and their number density is well below Type-1 QSOs. 
Some extremely red quasars (ERQs) and heavily reddened quasars (HRQs) are also selected using the IR selection \citep[e.g.][]{Ross2015,Banerji2015,Zakamska2016,Hamann2017,Alexandroff2018,Temple2019,Zakamska2023}. However, these samples often include optical magnitude cuts like $\rm i <20.5$, introducing significant luminosity biases and making them a subset of the whole obscured population. 

In this project, we present spectroscopic observations of a sample of obscured QSOs at $z \sim 2$ \citep{Ishikawa2023,wb2025}, using the DOG selection in the SDSS Stripe82 region. 
The spectra of these targets are obtained with state-of-the-art ground-based optical and near-infrared telescopes: Keck and Gemini. These 24 candidates are selected using a red color selection ($r - W4 >8.38$). Gemini/GNIRS is used for identifying the 24 targets, 20 of which are further observed with Keck to better constrain the redshift and reveal the rest-UV spectra. 
The SED fitting of this sample reveals the IR emission is dominated by the hot dust torus, and the rest-UV/optical is the combination of the AGN and galaxy emission \citep{wb2025}.
This paper is divided as follows. 
In Section \ref{sec:data} , we introduce the target selection, observation, and data reduction. 
In Section \ref{sec:Results}, we conduct the emission line fitting, discuss the line property and spectral diversity based on the line fitting. 
In Section \ref{sec:Discussion}, we discuss interesting, individual sources. 
In Section \ref{sec:composite}, we generate the composite spectra for all 23 identified targets. 

\section{Data Collection}\label{sec:data}
In this section, we present the target selection, spectroscopic observation using Keck and Gemini, and the data reduction. 

\subsection{Target Selection}
The selection of Type-2 QSO candidates in SDSS Stripe82 \citep{wb2025} 
is designed using the $r$-band photometry from SDSS/DESI Legacy Survey and $W4$ photometry from ALLWISE:
\begin{equation}
    \begin{aligned}
    &  12.62 < \text{W4} < 14.62 \, \text{AB mag} \\
    & \texttt{AND } \text{SNR}_{\text{W4}} > 5 \\
    & \texttt{AND } \left( r > 23 \, \text{AB mag} \texttt{ OR } r - \text{W4} > 8.38 \, \text{AB mag} \right)
    \end{aligned}
\end{equation}
The SED of the hot dust torus peaks at rest-frame $\rm 10 \mu$m. The WISE $W4$ bands probe the peak of the SED at $z \sim 2$. 
The upper magnitude limit of the $W4$ band is to select the most IR-luminous Type-2 QSOs. The lower limit of the magnitude range excludes bright stars or bright galaxies at low redshift. 
We further require candidates to be either non-detected in SDSS  ($r > 23$ in the SDSS Stripe 82 catalog) or with a red  
color cut: $r-W4 > 8.38$ to exclude the Type-1 QSOs. This selection leads to 24 candidates in 164 $\rm deg^2$ SDSS Stripe82 region. 

\subsection{Spectroscopic Observation}

\subsubsection{Gemini/GNIRS spectra}
To confirm the spectroscopic redshift of all 24 candidates, we use the Gemini Near-InfraRed Spectrograph \citep[GNIRS]{Elias2006} on the Gemini North 
telescope to conduct the spectroscopic observation. These candidates are expected to be at $z \sim 2$ so the H$\alpha$ and [O III] emission lines are expected to be covered by GNIRS. The candidates were observed under GN-2017B-Q-51 (PI: Gordon Richards) for several nights in 2017-09, 2017-10, and 2018-01. The grating was $ \rm 32/mmSB$ which covers $0.9-2.4 \rm \mu$m with spectral resolution of R $\sim$ 1100. The exposure time for each candidate was 2400s, and the observations were performed using an ABBA sequence. The spectroscopic classifications were published first in \citet{Ishikawa2023} and \citet{wb2025}.

\subsubsection{Keck/LRIS spectra}
We choose 18 targets (17 Type-2 QSO candidates and one inconclusive target) to conduct a follow-up observation using the Keck Low Resolution Imaging Spectrometer \citep[LRIS]{Oke1995}, to better confirm the redshifts and reveal the rest-UV spectra. 
The observations were taken on the 
night of 27 September 2022 and over the course of a 3-night run from 26-28 October 2022. 
We observed these targets using a 1.0$\farcs$  long-slit, 
2$\times$2 binning, 560 dichroic and $600/4000$ 
grating for the blue band with resolution $\sim$ 4\rm{\AA} 
covering 3040 to 5630\rm{\AA}. The grating for the red band is: 400/8500, 
1.0 $\farcs$ long slit, resolution around 7\rm{\AA}, 
$\Delta \lambda = 4762$\rm{\AA} with the central wavelength 7980\rm{\AA}. The average exposure time for each candidate is 1800s. The Ly$\alpha$ and CIV emission lines are detected in the LRIS spectra. The typical seeing was 1.2$\farcs$

\subsubsection{Keck/KCWI spectra}
We further observed two targets with the Keck Cosmic Web Imager Integral Field Spectrograph \citep[KCWI]{KCWI2018}, to better constrain their redshift and reveal extended Ly$\alpha$ emission around these QSOs. We used Large Slicer, BL grating which covers 3258 - 5600 \rm{\AA} on the blue. The large slicer covers a sky area with $30\farcs \times 16\farcs$. For the red side KCRM, we used RL grating covering 5550-9239 \rm{\AA}. The exposure time is $2\times 900s$ and $4\times 430s$ on the blue (KCWI) and red (KCRM), respectively. The observation was conducted on the night of 28 January 2025. The typical seeing was 1.2$\farcs$

\subsection{Data Reduction}
The Gemini and Keck observation data were reduced using the Pypeit pipeline \citep{Prochaska2020}. This pipeline can automatically produce calibrated 2-D spectra. Then we coadd the individual 2-D spectra and use manual extraction to get the 1-D spectra. The flux calibration and telluric correction are conducted using a standard star observed on the same night. 

The Gemini/GNIRS observation contains two ABBA sequences, eight exposures for every target. The Pypeit pipeline produces four 2-D spectra for each object. We check these 2-D spectra to find the object manually and then coadd the 2-D spectra using the object as a reference. This results in one coadded 2-D spectrum. The 1-D spectra are extracted manually or automatically from the coadded 2-D spectra. 
Then we apply the flux calibration and telluric correction on the 1-D spectra using the standard star. 

The Keck/LRIS observation contains two
exposures in the blue and four exposures in the red for each object. Pypeit can produce 2-D spectra for each individual exposure. Then we coadd all the 2-D spectra and then extract the 1-D spectra from that. 

The reduction for Keck/KCWI is more complex since this is an Integral Field Unit (IFU) instrument. Firstly, Pypeit can produce 2-D spectra for each individual exposure. Then we manually check the 2-D spectra and find any emission line signal. After that, we combine all the 2-D spectra to generate a data cube. The flux calibration is performed by extracting the spectrum of a standard star from the data cube. The wavelength of the emission line signal is used to generate the wavelength range of the white light. Finally, we manually extract the 1-D spectra from the data cube based on the position of the target in the white light image. 

\begin{figure*}
\begin{subfigure}[b]{0.65\linewidth}
         \includegraphics[width=\textwidth]{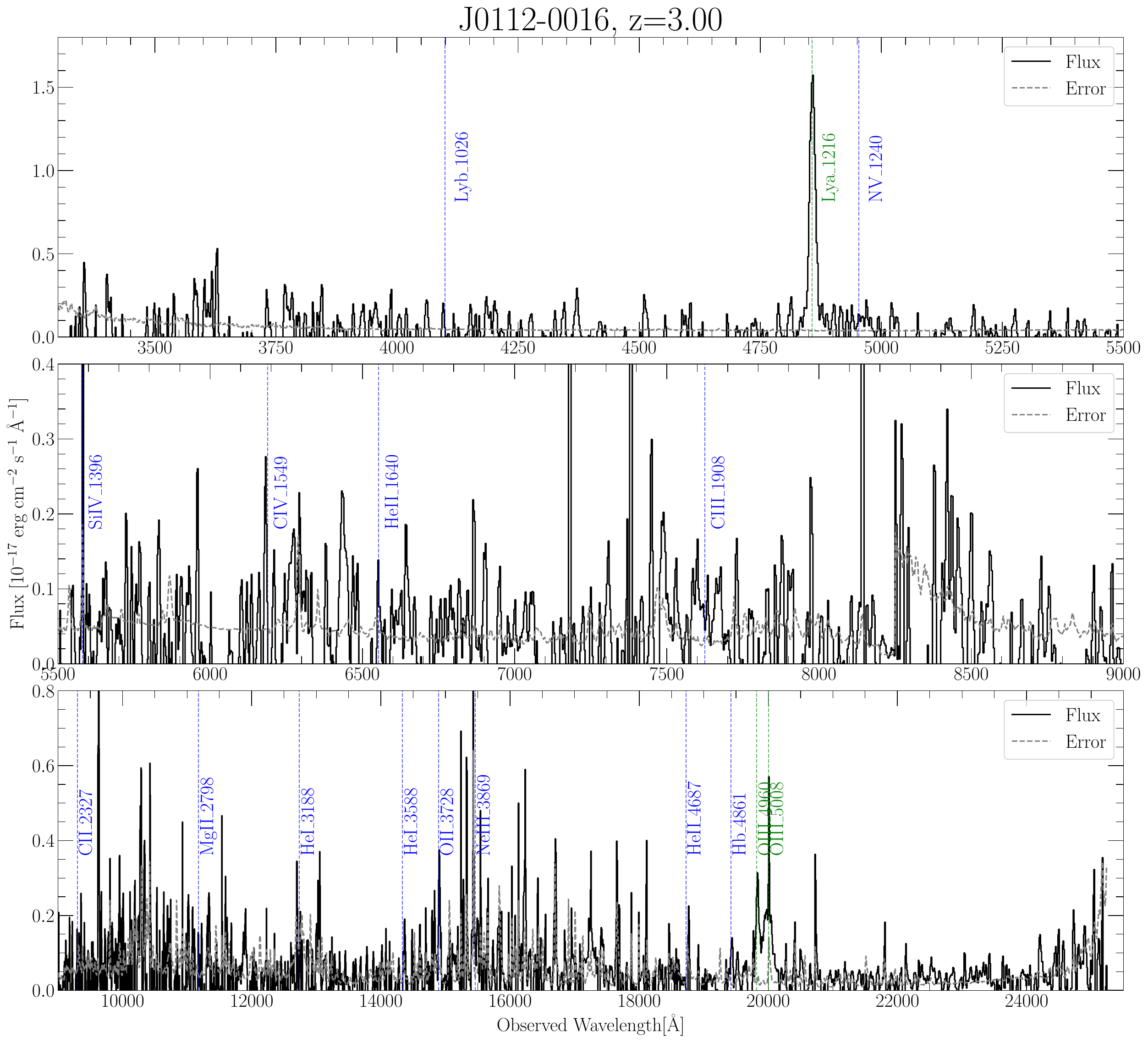}
    \end{subfigure}
     \begin{subfigure}[b]{0.34\linewidth}
         \includegraphics[width=\textwidth]{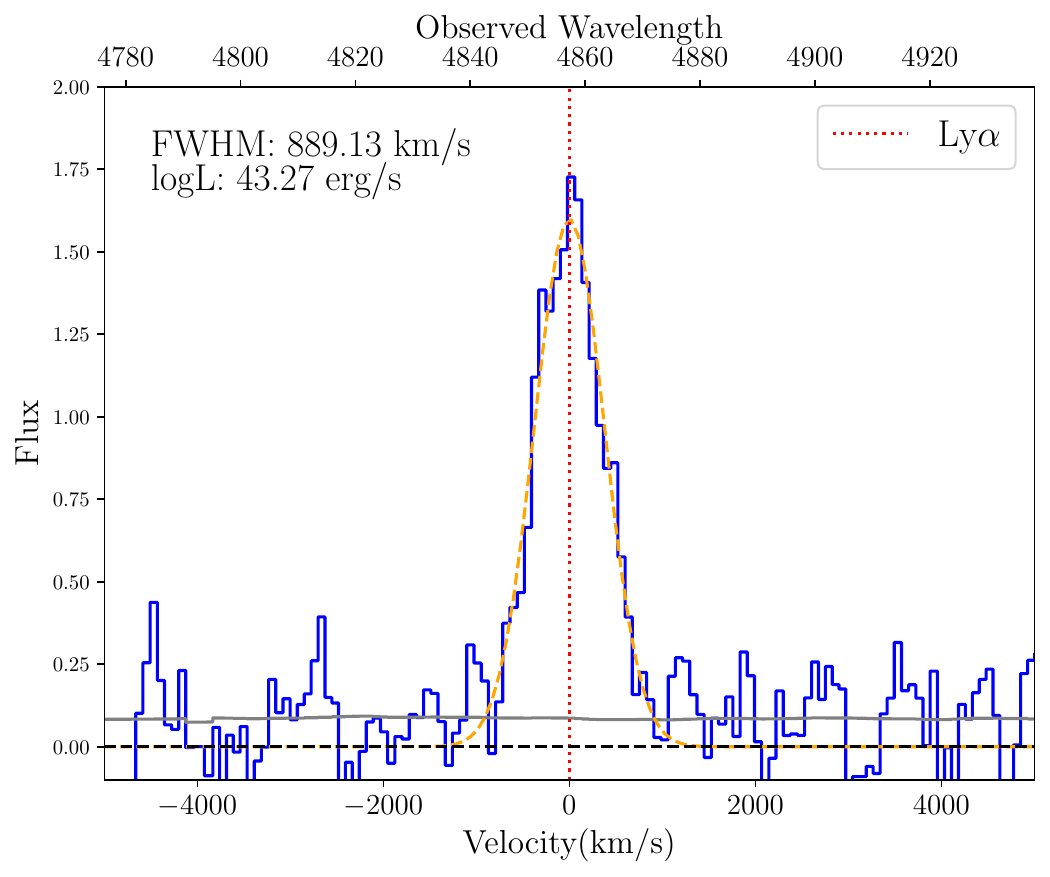}
         \hfill
         \includegraphics[width=\textwidth]{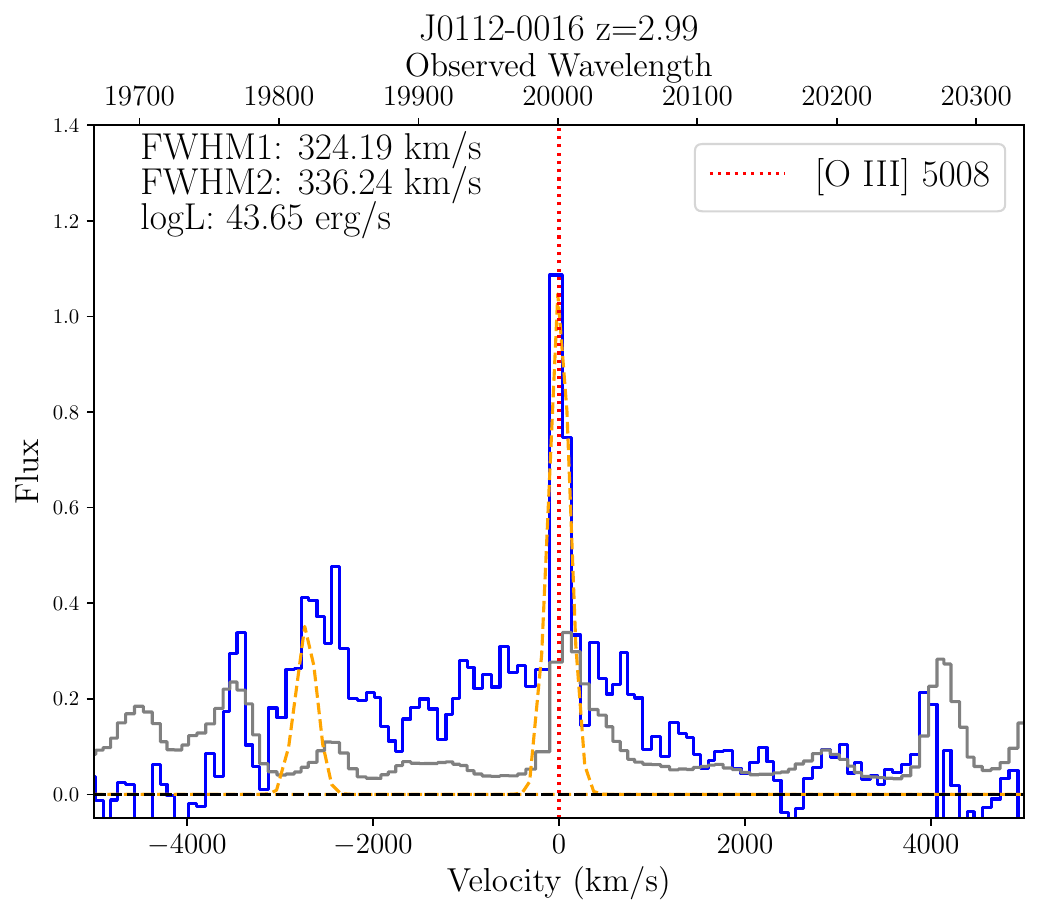}
    \end{subfigure} 
    \hfill
    \begin{subfigure}[b]{0.65\linewidth}
         \includegraphics[width=\textwidth]{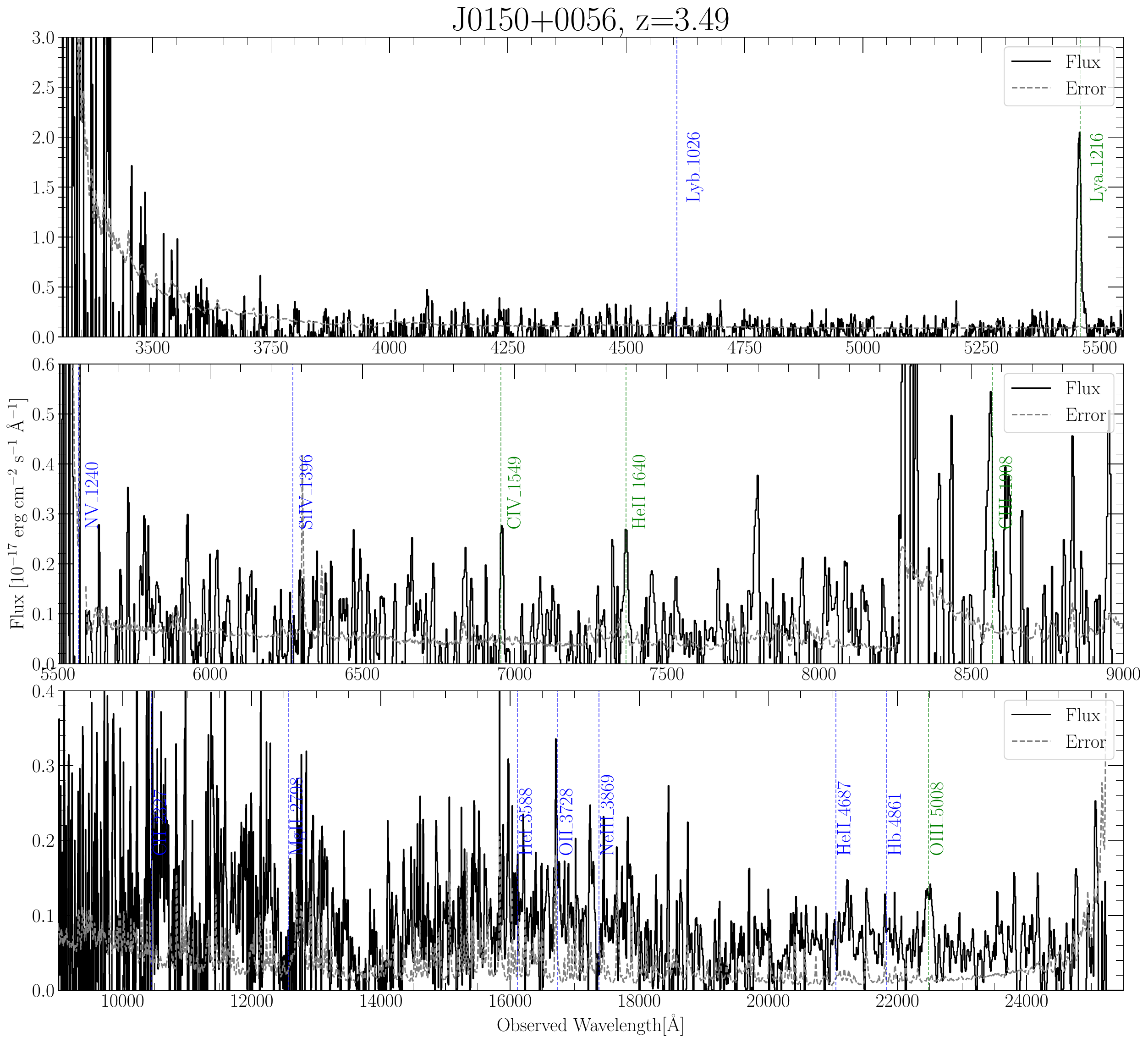}
    \end{subfigure}
     \begin{subfigure}[b]{0.34\linewidth}
         \includegraphics[width=\textwidth]{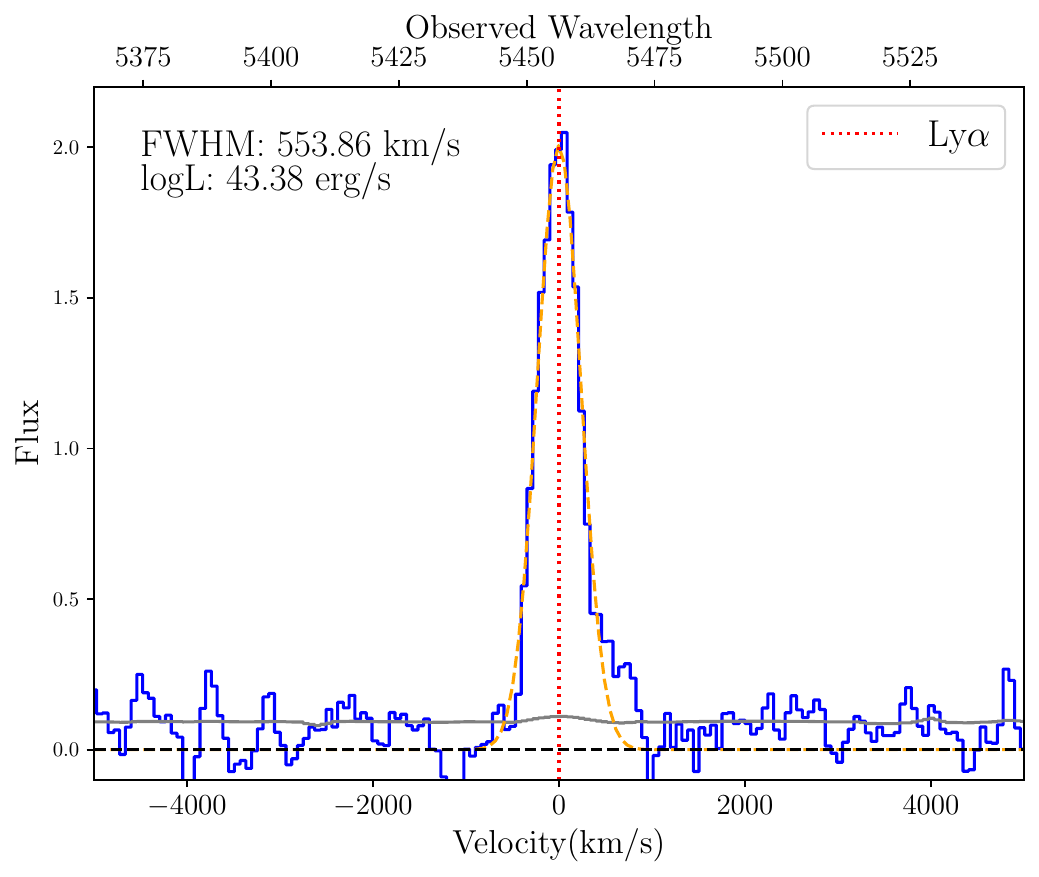}
         \hfill
         \includegraphics[width=\textwidth]{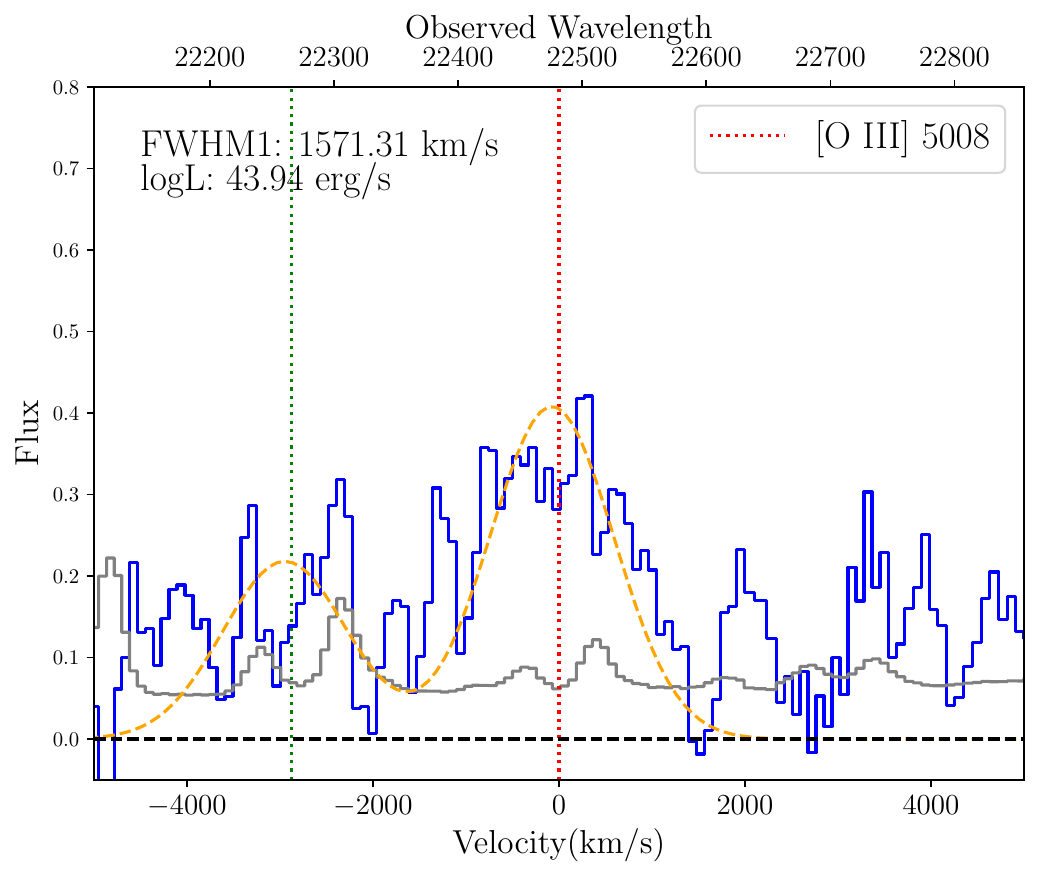}
    \end{subfigure}     
\caption{Example spectra from Keck and Gemini, and the fitted emission lines for 2 highest-$z$ ($z>3$) Type-2 QSO targets in our sample. 
The Left panel shows the spectra from Keck/LRIS(or Keck/KCWI) and Gemini/GNIRS, the green lines mark the detected emission line and the blue dashed lines are the undetected emission lines. The target name and redshift are shown in the title of the subplots. The right panel shows the Gaussian fitted emission lines for Ly$\alpha$ and  [O III]. The flux is shown in blue, and the fitted Gaussian profile is shown in orange. The red lines mark the lince center of Ly$\alpha$ or [O III]$\rm \lambda 5007$, and the green line marks the line center of [O III]$\rm \lambda 4960$. The FWHM and the line luminosity are labeled in the plots. 
}.\label{fig:gnirs+lris_v1}
     

\end{figure*}

\begin{figure*}
\begin{subfigure}[b]{0.65\linewidth}
         \includegraphics[width=\textwidth]{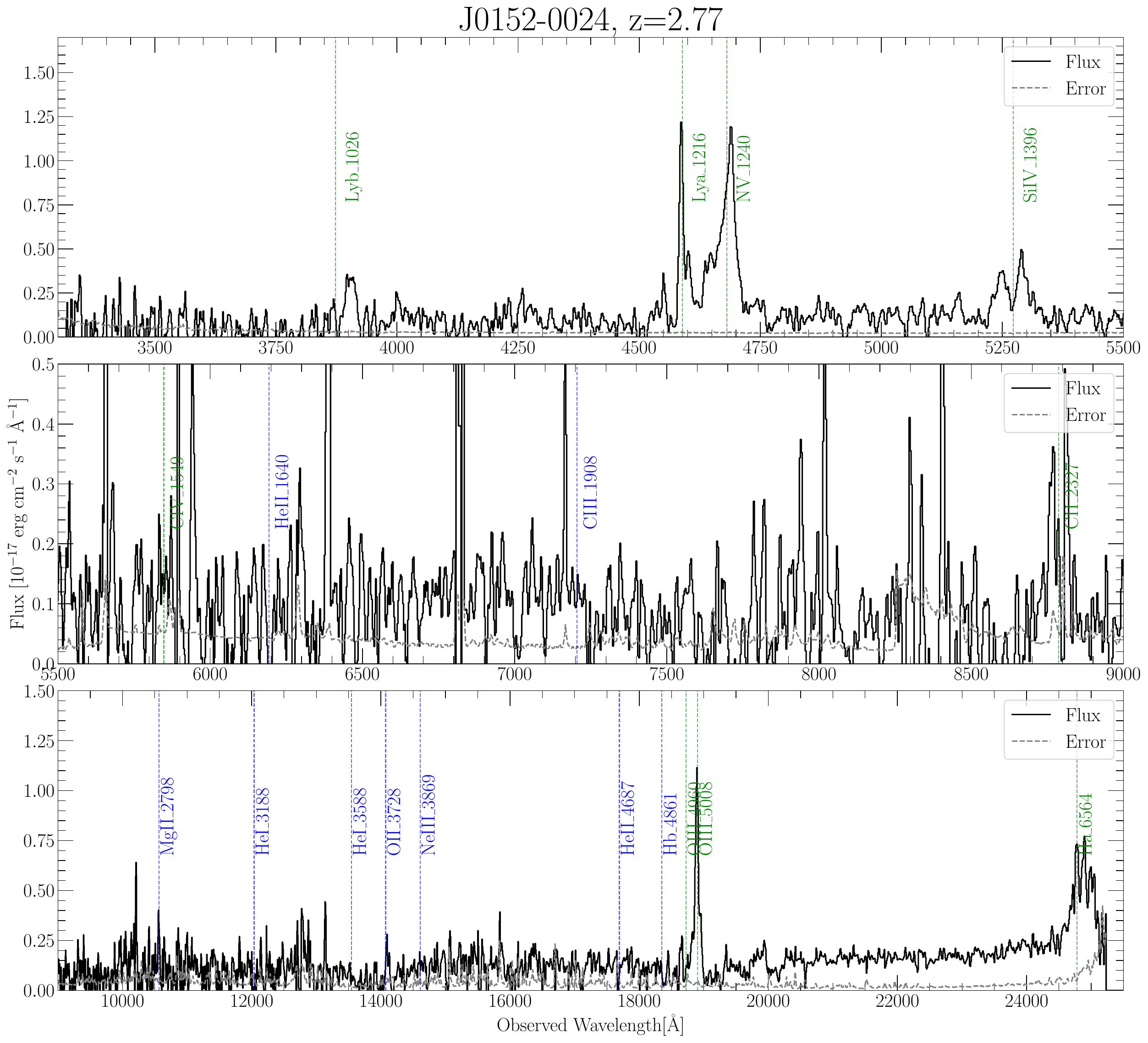}
    \end{subfigure}
     \begin{subfigure}[b]{0.34\linewidth}
         \includegraphics[width=\textwidth]{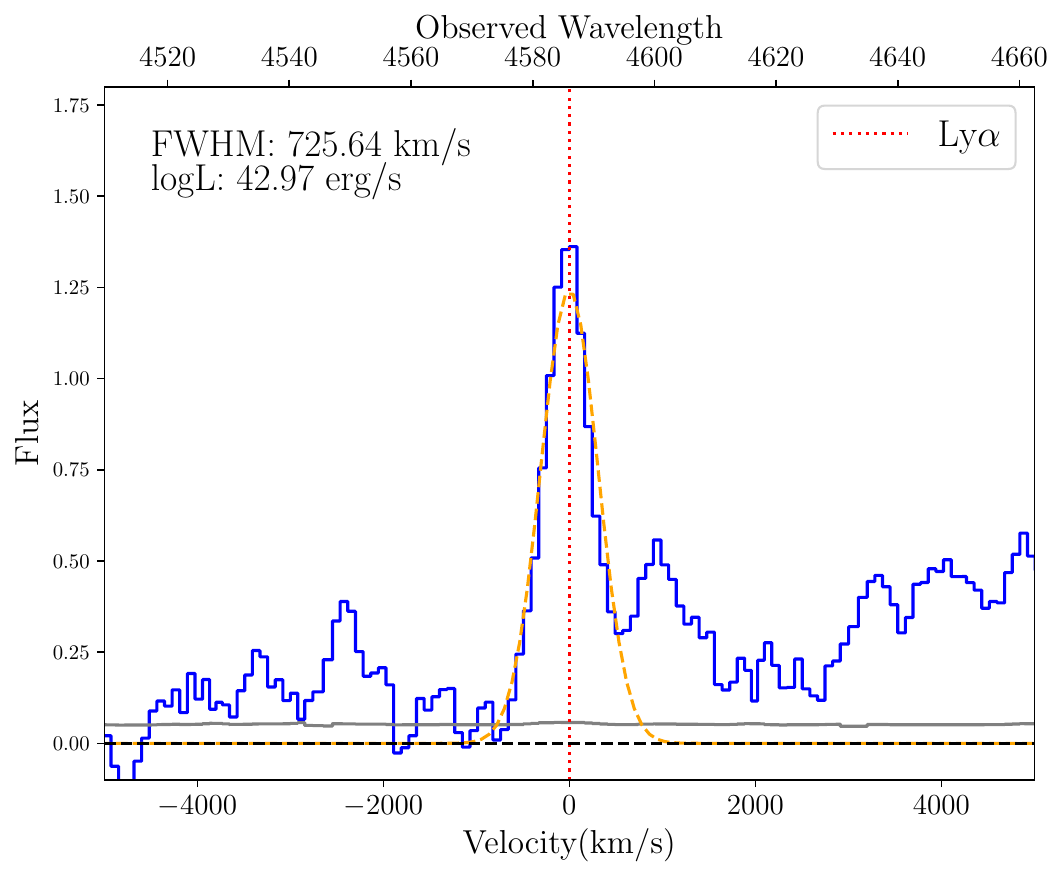}
         \hfill
         \includegraphics[width=\textwidth]{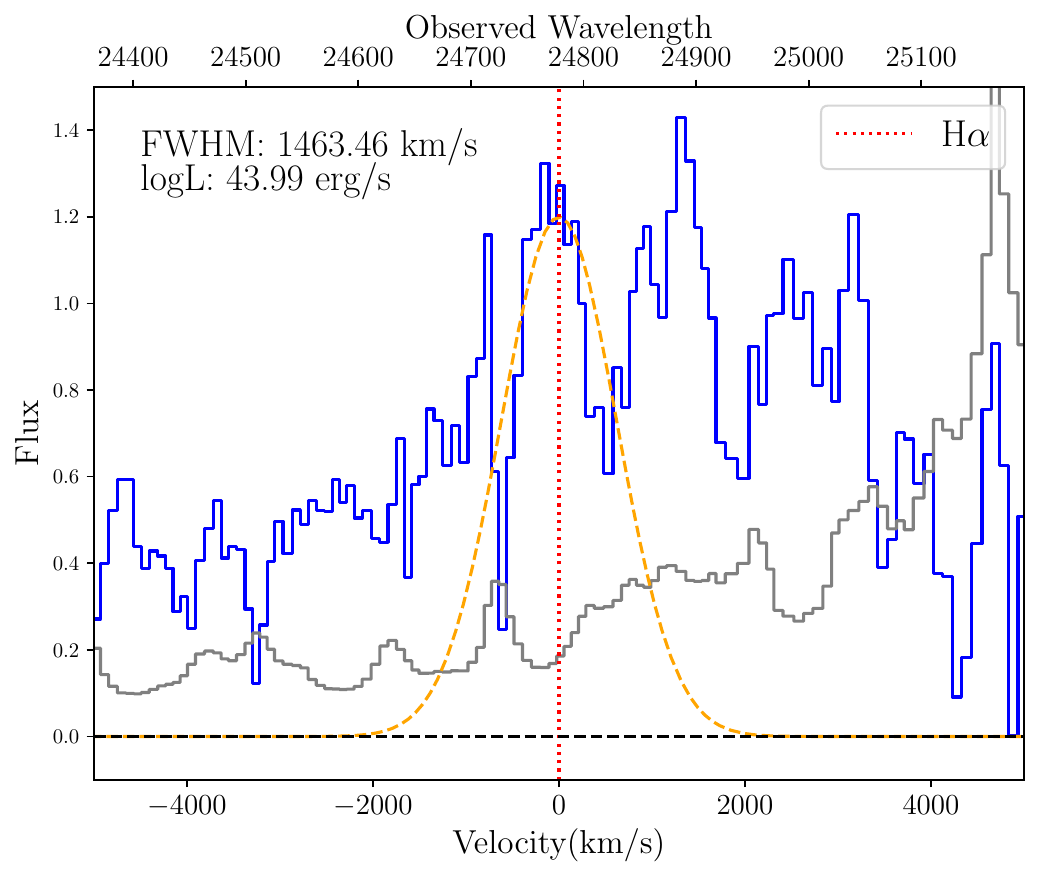}
    \end{subfigure} 
    \hfill
    \begin{subfigure}[b]{0.65\linewidth}
         \includegraphics[width=\textwidth]{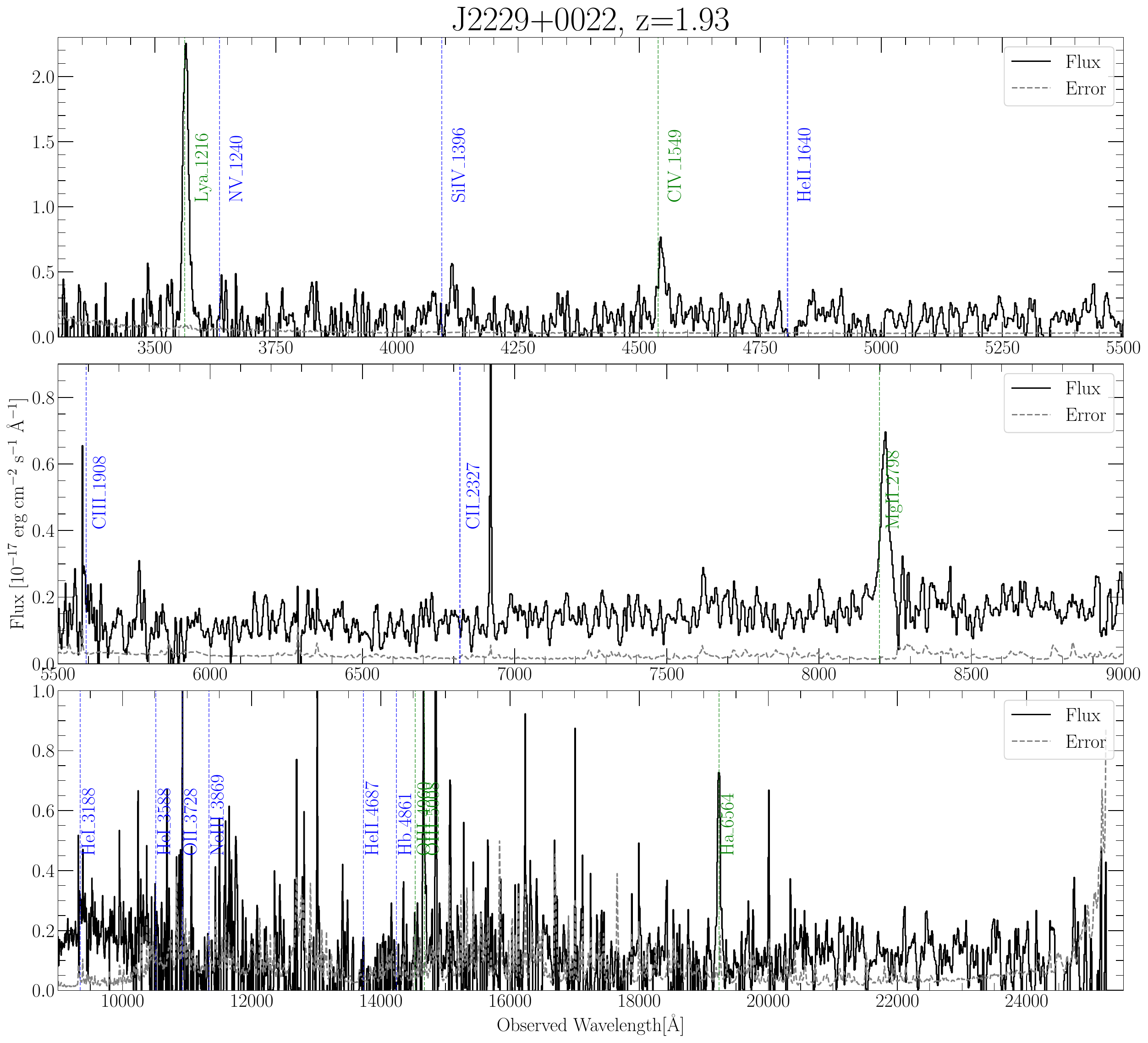}
    \end{subfigure}
     \begin{subfigure}[b]{0.34\linewidth}
         \includegraphics[width=\textwidth]{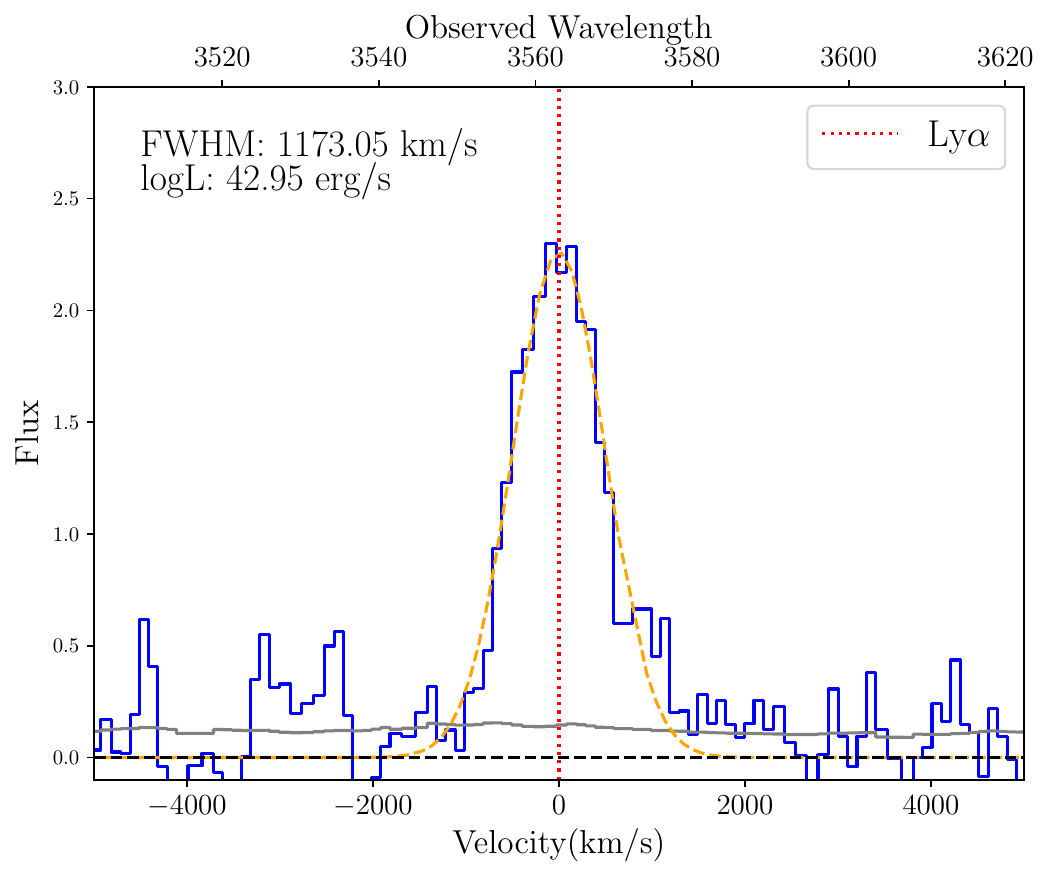}
         \hfill
         \includegraphics[width=\textwidth]{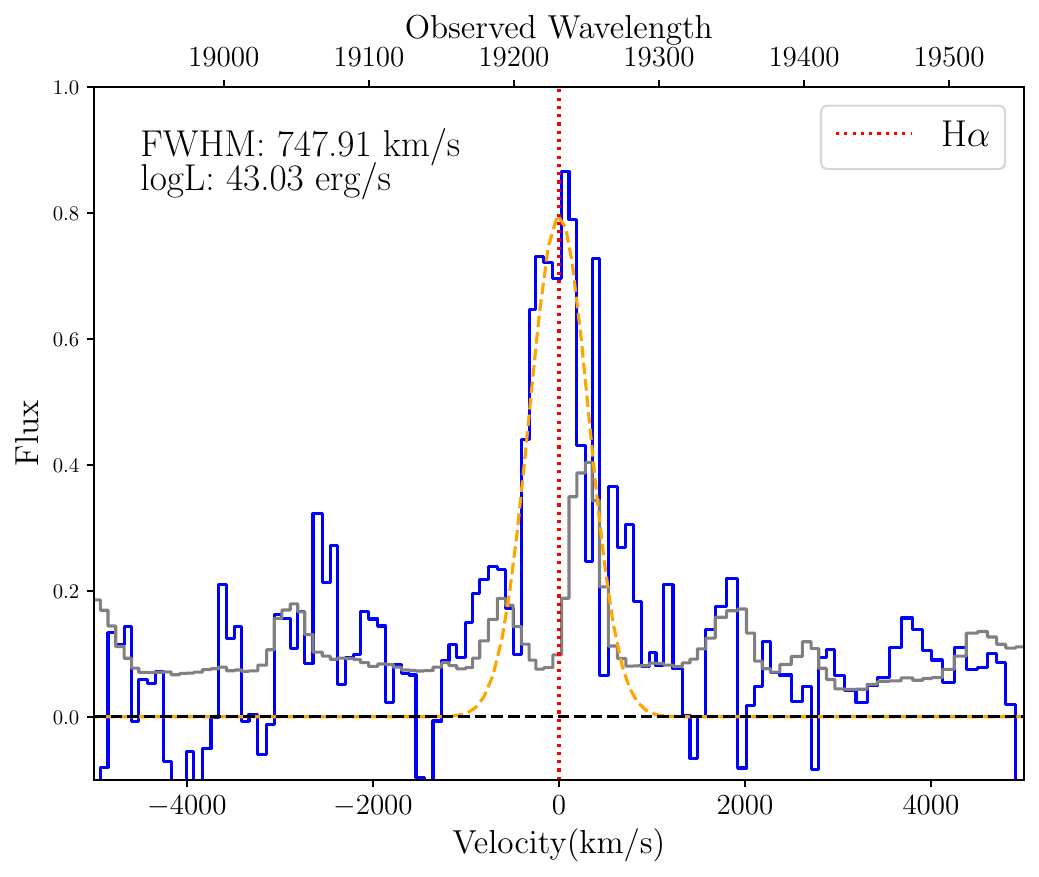}
    \end{subfigure} 
     
\caption{Example spectra and line fittings for 2 targets with narrow($\rm < 2000 km/s$) H$\alpha$ emission line. The top target has a strong absorption feature at the wavelength of Ly$\alpha$ and NV emission. The H$\alpha$ is located at the edge of the GNIRS coverage, making it difficult to measure the line width. The bottom panel shows a target with both narrow Ly$\alpha$ and narrow H$\alpha$ emission lines. 
}.\label{fig:gnirs+lris_v2}

\end{figure*}

\begin{figure*}
\begin{subfigure}[b]{0.65\linewidth}
         \includegraphics[width=\textwidth]{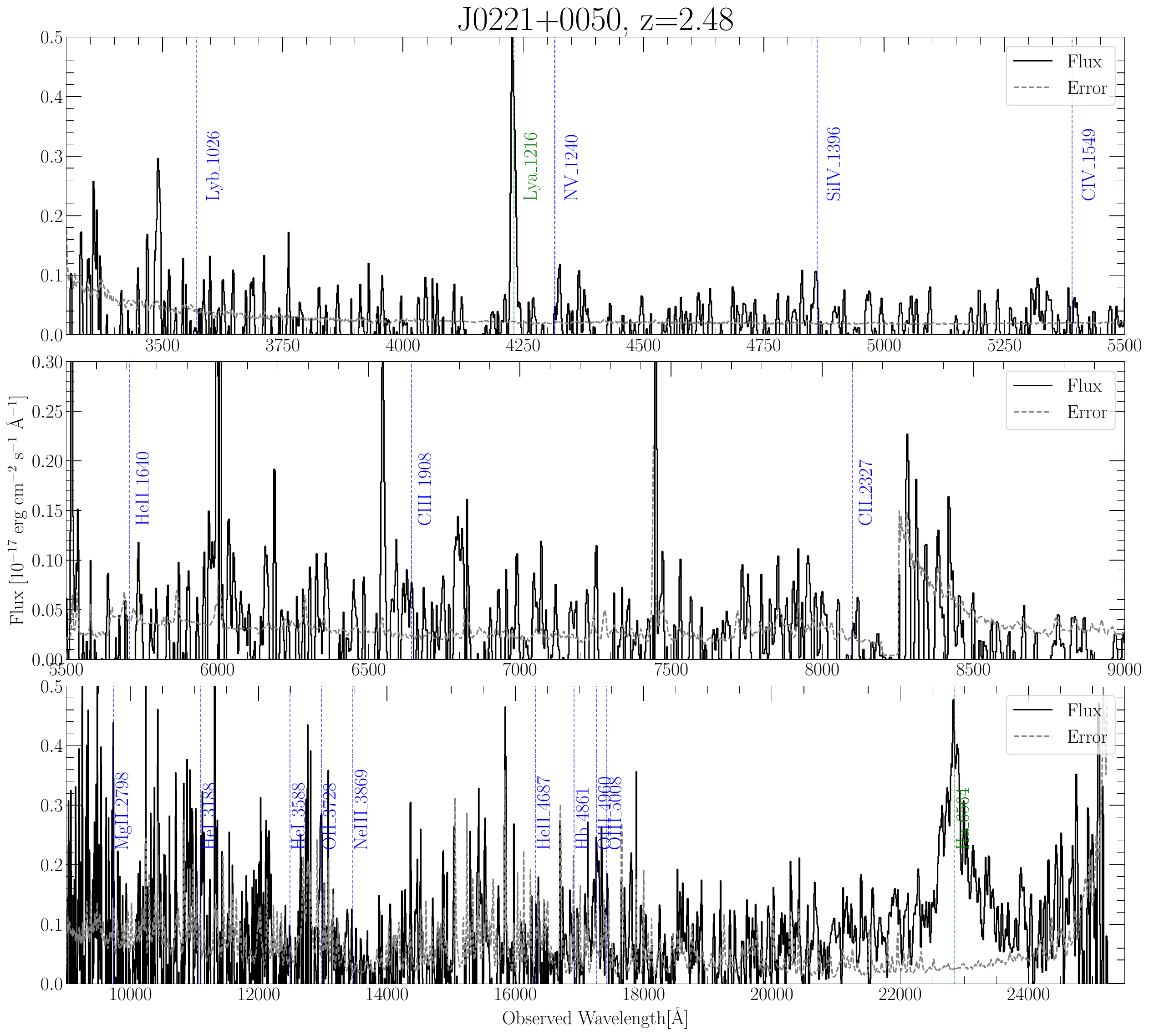}
    \end{subfigure}
     \begin{subfigure}[b]{0.34\linewidth}
         \includegraphics[width=\textwidth]{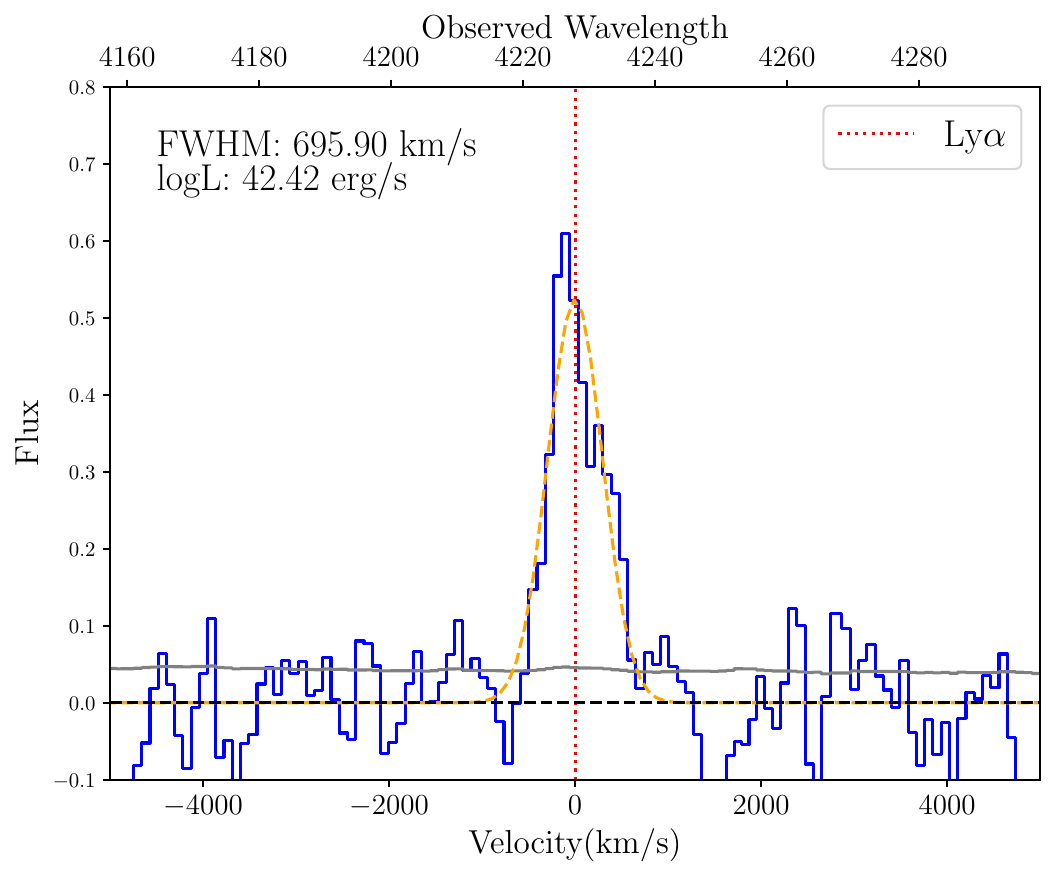}
         \hfill
         \includegraphics[width=\textwidth]{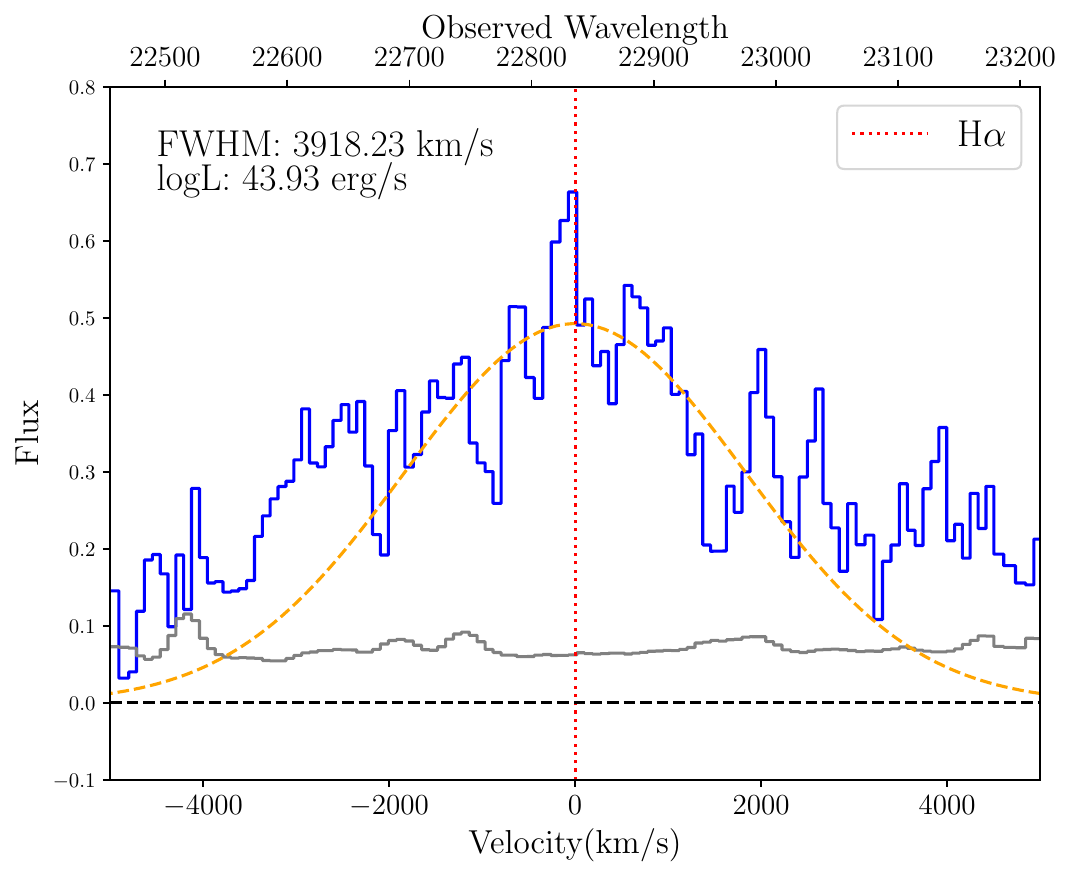}
    \end{subfigure}    
    \hfill
    \begin{subfigure}[b]{0.65\linewidth}
         \includegraphics[width=\textwidth]{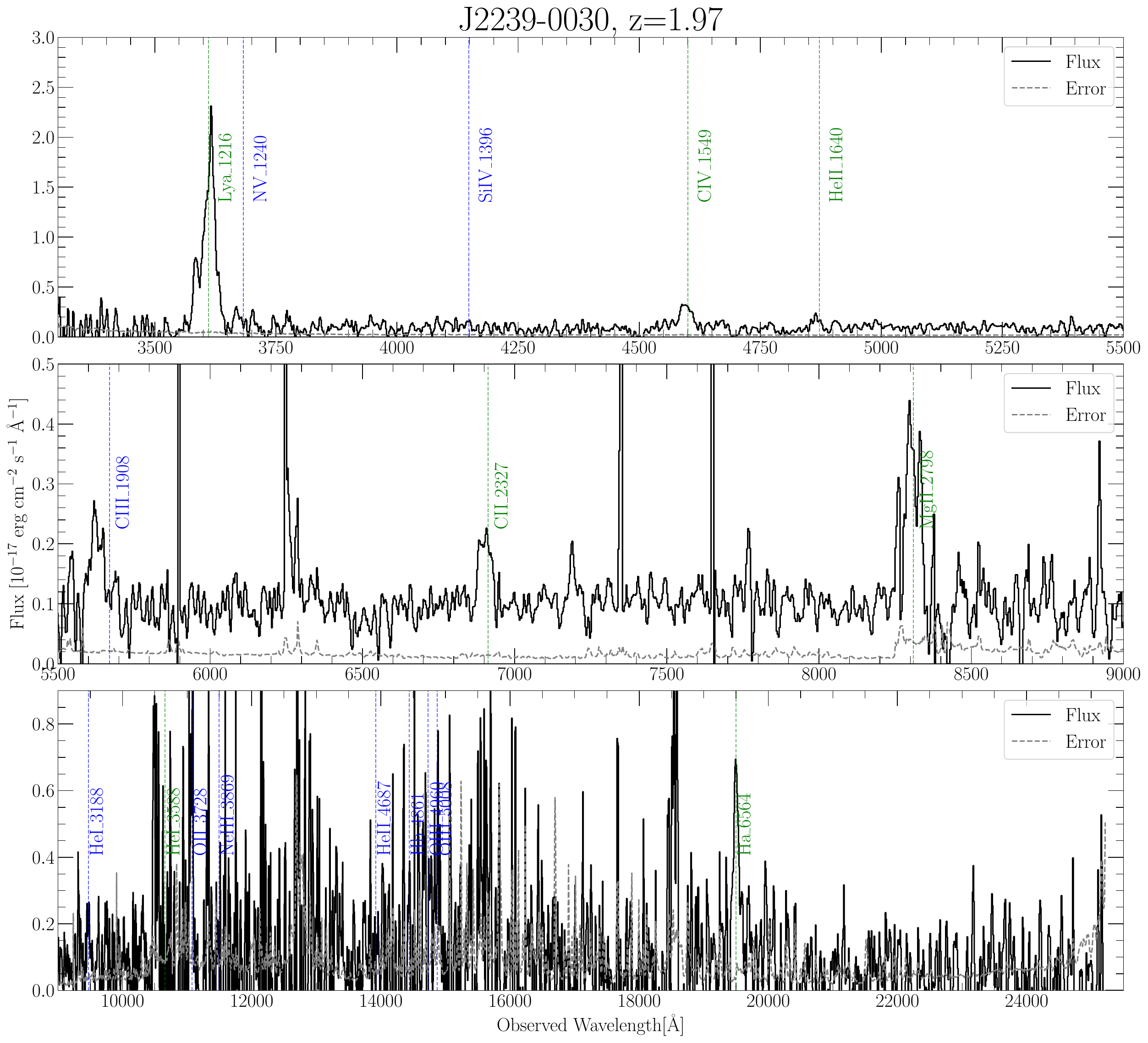}
    \end{subfigure}
     \begin{subfigure}[b]{0.34\linewidth}
         \includegraphics[width=\textwidth]{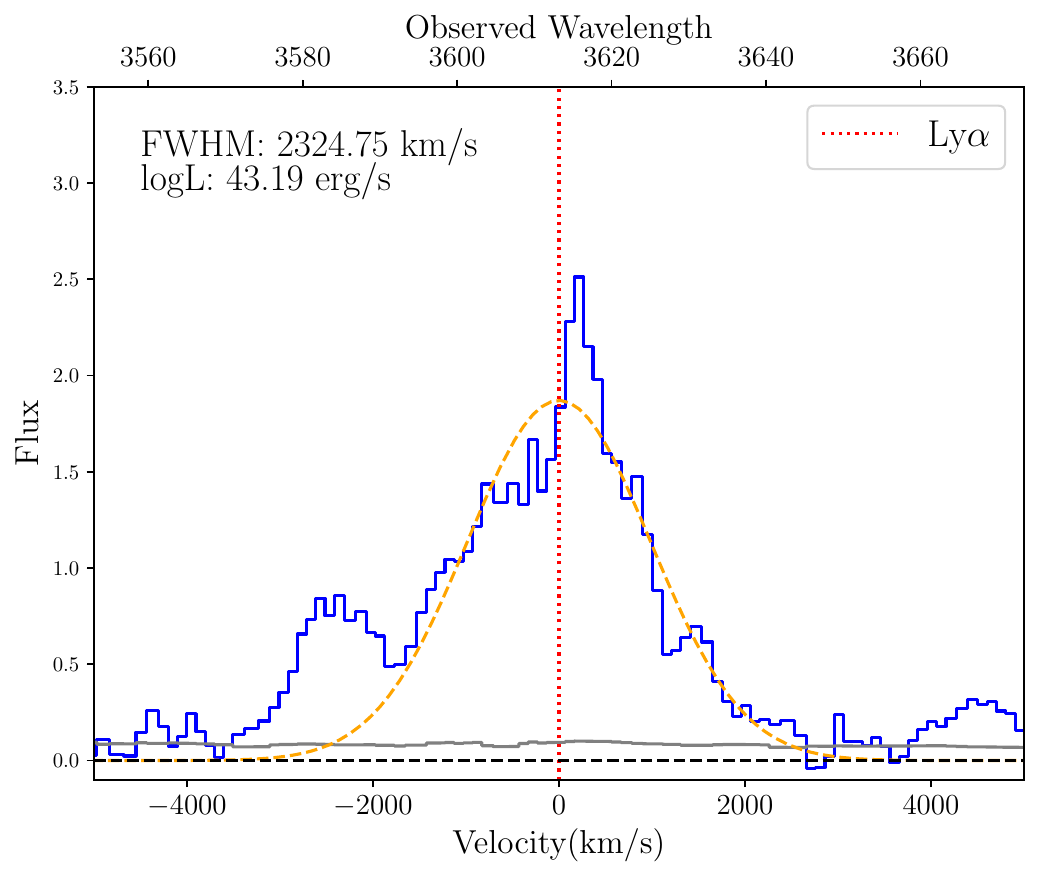}
         \hfill
         \includegraphics[width=\textwidth]{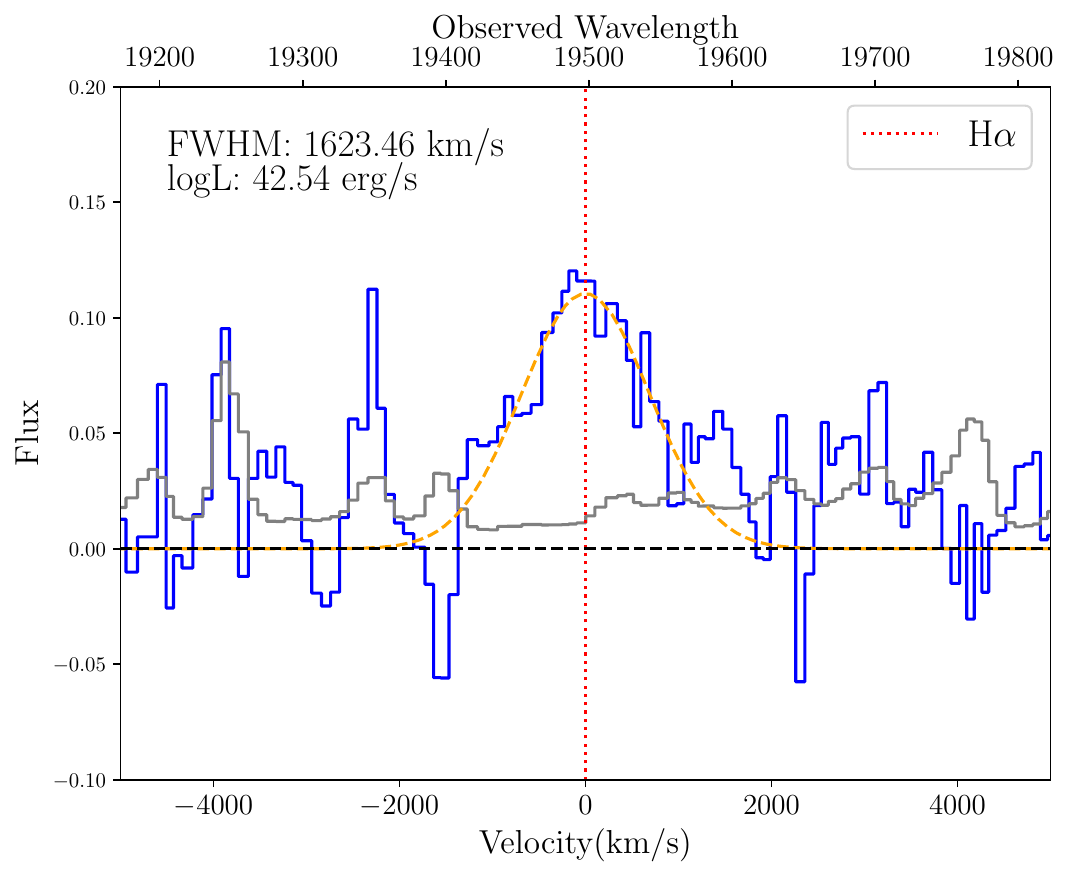}
    \end{subfigure}     

\caption{Example spectra and line fitting for 2 targets with broad H$\alpha$ or Ly$\alpha$ emissions. The upper panel shows a target J0221+0050 with broad H$\alpha$ and narrow Ly$\alpha$ emission. The broad emission lines are not expected for typical obscured QSOs, but it is a common feature in some ERQs and JWST AGNs. The lower panel shows a target with broad Ly$\alpha$ but narrow H$\alpha$ emission line. 
}.\label{fig:gnirs+lris_v3}     

\end{figure*}

\begin{figure*}
\begin{subfigure}[b]{0.65\linewidth}
         \includegraphics[width=\textwidth]{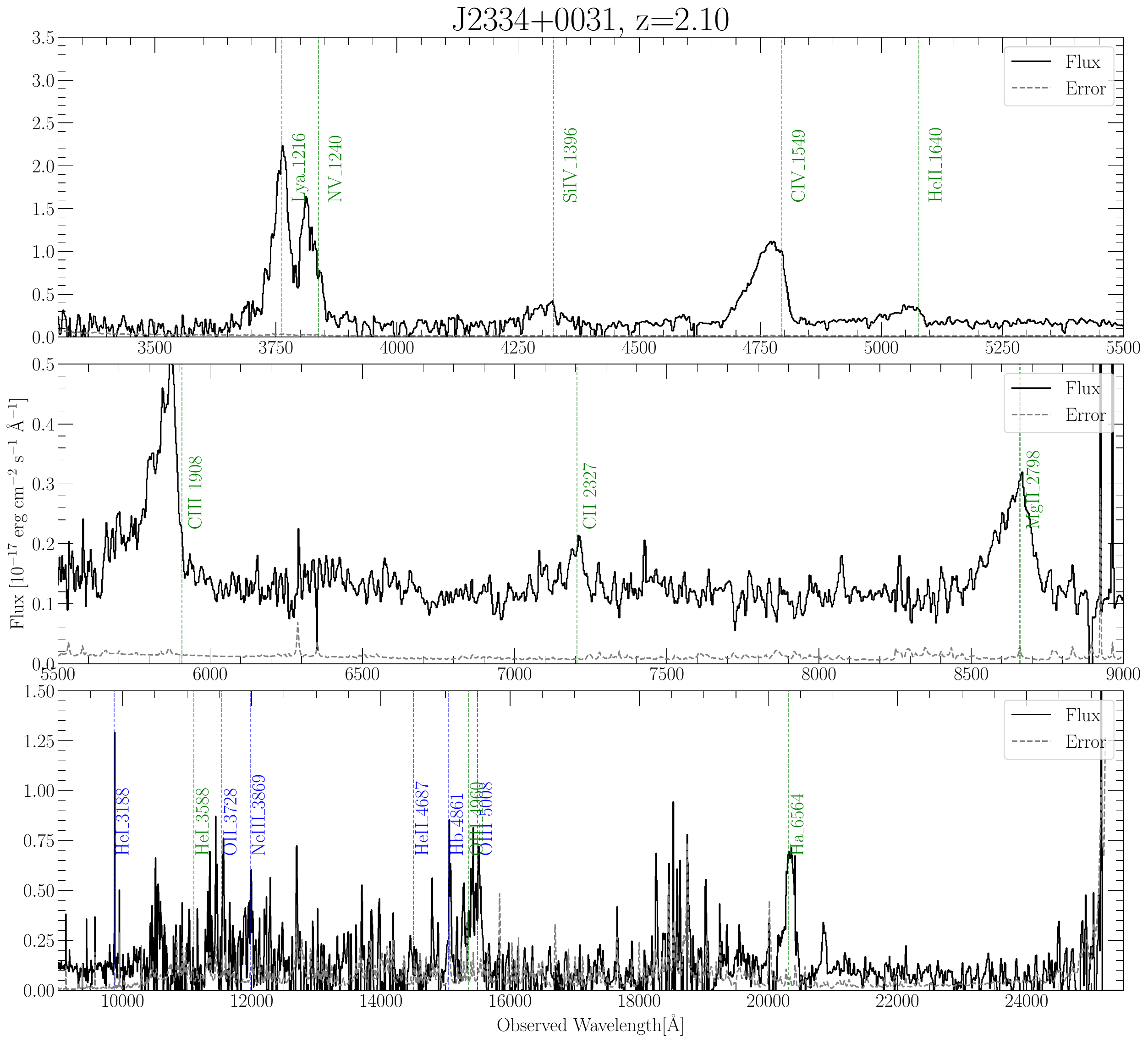}
    \end{subfigure}
     \begin{subfigure}[b]{0.34\linewidth}
         \includegraphics[width=\textwidth]{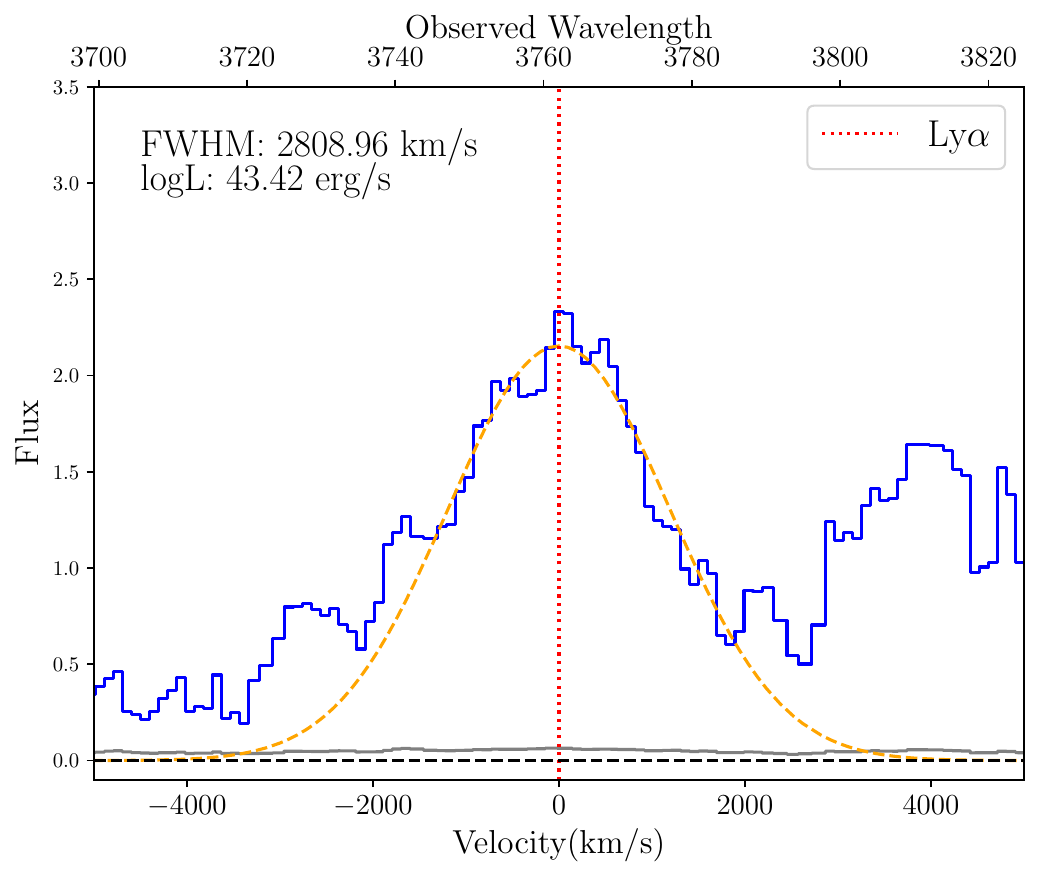}
         \hfill
         \includegraphics[width=\textwidth]{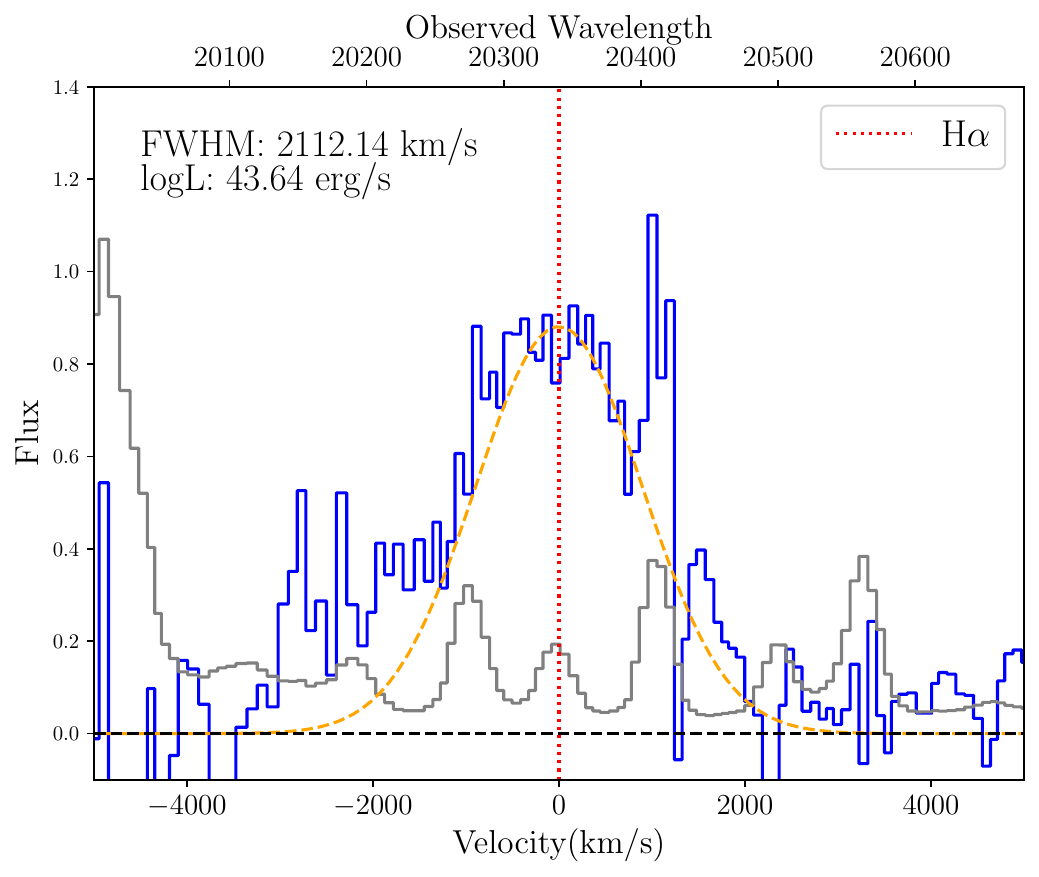}
    \end{subfigure}

\caption{Example spectra and line fitting for 1 target with both broad Ly$\alpha$ and H$\alpha$ emission lines. This target J2334+0031 is the only target in our sample with both broad Ly$\alpha$ and H$\alpha$ emission lines. This target also has broad CIV, CIII, and MgII emission lines. 
}.\label{fig:gnirs+lris_v4}     

\end{figure*}

Among our whole sample, 17 targets have GNIRS and LRIS spectra, two targets have GNIRS and KCWI spectra, and four targets only have GNIRS spectra. For the targets with multi-wavelength spectra from Keck and Gemini, we first rebin and coadd these spectra. We rebin all spectra to a pixel scale of  $\rm dv = 120 km/s$ in velocity space at the rest-frame.
For the overlapping region, we take the flux with higher SNR as the final flux. 
Figures~\ref{fig:gnirs+lris_v1} to \ref{fig:gnirs+lris_v4} present example spectra in three separate wavelength windows for better visibility of key lines.  Figures~\ref{fig:gnirs+lris_v1} shows the two highest-$z$ targets ($z>3$). Figures~\ref{fig:gnirs+lris_v2} shows the targets with narrow H$\alpha$ emission. Figures~\ref{fig:gnirs+lris_v3} and Figures~\ref{fig:gnirs+lris_v4} present some targets with broad H$\alpha$ or Ly$\alpha$ emissions. 


\section{Results} \label{sec:Results}
In this section, we present the emission line fitting results. We discuss the spectral diversity based on the line width of Ly$\alpha$ and H$\alpha$, and the relation between the bolometric luminosity and the [O III] line luminosity.

\subsection{Target Identification}
Among 24 targets, we see emission lines for 23 targets in the Gemini and Keck spectra. Based on the line detection, we identified 23 targets as real Type-2 or reddened Type-1 QSOs at z$\sim 2$, yielding a very high success rate ($\sim 96 \%$).  Of the 23 confirmed targets, 12 are at $z>2$ ($\sim 52\%$). 
One target, J0047+0003, remains inconclusive, since there are no obvious emission lines within the wavelength coverage of the Keck and Gemini spectra. It could be a very faint object or a spurious WISE source; identification of this target may require observations with JWST. 

\subsection{Line Fitting}
The redshift measurements are mainly based on the detection of Ly$\alpha$ and H$\alpha$ emission lines. For two targets, the H$\alpha$ emission is beyond the wavelength coverage of the Gemini/GNIRS, and we use Ly$\alpha$ and [O III] emission to measure the redshift. 
First, we manually check the spectra and identify the emission lines to determine the rough spectroscopic redshift. Then, we fit a Gaussian profile to every emission line to get the accurate redshift, line width, and line luminosity. For single emission lines like Ly$\alpha$ and H$\alpha$ we fit a single Gaussian profile. For other lines like [O III], two Gaussian profiles are used to fit the doublet. 


The fitted line profiles are shown in the right subplots from Figure~\ref{fig:gnirs+lris_v1} to Figure~\ref{fig:gnirs+lris_v4}. The observed flux is shown in blue, and the fitted Gaussian profiles are shown in orange. The red line marks the fitted line center. The FWHM and the line luminosity are labeled on the upper left of each plot.
 
\subsection{Ly$\alpha$ and H$\alpha$ Emission Lines}


Among the 23 identified targets, 13 targets have Ly$\alpha$ detection in the Keck spectra, and 21 targets have H$\alpha$ detection in the Gemini spectra. 
The fitted line properties of Ly$\alpha$ and H$\alpha$ are summarized in Table~\ref{tab:hl}. 

\begin{figure*}
  {\centering
  \includegraphics[width=0.9\textwidth]{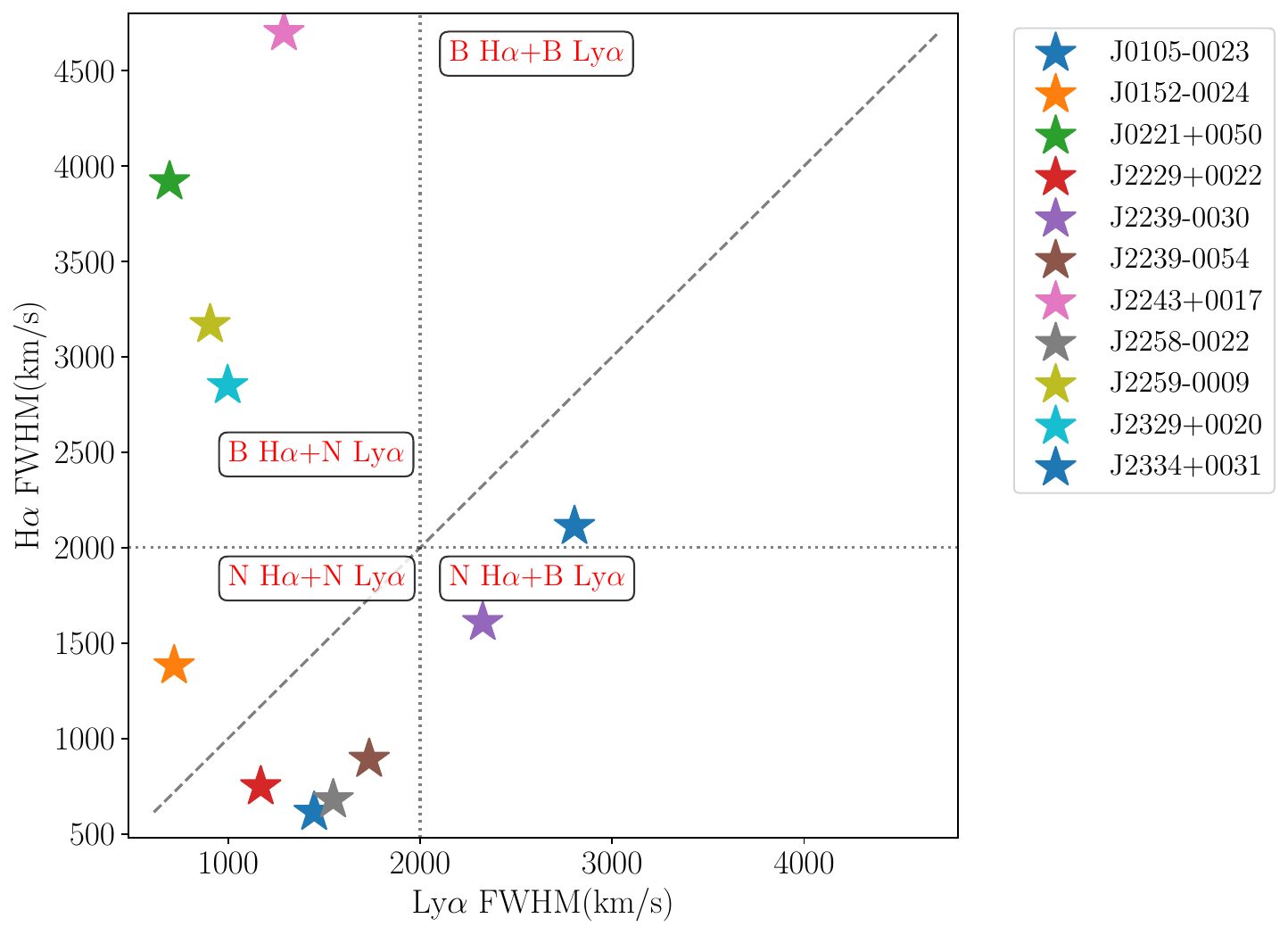}
  }
  \vspace{-0.2cm}
 \caption{The FWHM of H$\alpha$ and Ly$\alpha$ of 11 targets in our sample. The X-axis is the FWHM of the Ly$\alpha$ emission lines, and the y-axis is the H$\alpha$ FWHM. The 2000$\rm km/s$ is marked as black dashed lines. According to this definition, five targets have narrow H$\alpha$ and narrow Ly$\alpha$ emission lines, four targets have broad H$\alpha$ and narrow Ly$\alpha$ emission lines, one targets have narrow H$\alpha$ and broad Ly$\alpha$ emission lines, and one target has broad H$\alpha$ and broad Ly$\alpha$ emission lines. Our targets show spectral diversity in the line width. }.\label{fig:h-l}
\end{figure*}

For the 11 targets with both Ly$\alpha$ and H$\alpha$ detected, we can compare the line properties of these two lines. We define a broad line as $\rm FWHM > 2000 km/s$ and a narrow line as $\rm FWHM < 2000 km/s$ \citep{Zakamska2003,Hao2005}. According to this definition, six 
targets have narrow H$\alpha$ and narrow Ly$\alpha$ emission lines, three targets have broad H$\alpha$ and narrow Ly$\alpha$ emission lines, one target has narrow H$\alpha$ and broad Ly$\alpha$ emission lines, and one target has broad H$\alpha$ and broad Ly$\alpha$ emission lines. The targets in our sample show a diversity of this line width. The relation of Ly$\alpha$ and H$\alpha$ FWHM is shown in Figure~\ref{fig:h-l}. 


\begin{table*}
	\centering
	\caption{The redshift, emission lines used to get the redshift, line luminosity, and FWHM of Ly$\alpha$ and H$\alpha$ for all the targets in this sample.}
	\label{tab:hl}
	\begin{tabular}{lccccccr} 
		\hline
		Target & z & Ref Line & $log(\rm L_{Ly\alpha}(erg/s))$ & $\rm FWHM_{Ly\alpha}(km/s)$  & $log(\rm L_{H\alpha}(erg/s))$ & $\rm FWHM_{H\alpha}(km/s)$ & Spectra\\
		\hline
		J0024-0012 &  1.53  & [O III],H$\alpha$ &  -  &  -&  $10^{42.92}$  &  2144.99 & GNIRS+LRIS\\
		J0041-0029  &  2.09  &  H$\alpha$ &  -  &  - &  $42.92$  &  438.86& GNIRS+LRIS\\
		J0047+0003 &  -  & - & - & - &  -  &  -& GNIRS+LRIS\\
            J0054+0047 &  2.17  & H$\alpha$ & - & - &  $42.76$  &  879.38 & GNIRS+LRIS\\
            J0105-0023 &  1.87  & Ly$\alpha$,H$\alpha$ & $42.72$ & 1448.37 &  $42.81$  & 699.84 & GNIRS+LRIS\\
            J0112-0016 &  2.99  & Ly$\alpha$,[O III] & $43.27$ & 889.15&  -  &  -& GNIRS+LRIS+KCWI\\
            J0113+0029 &  2.33  & H$\alpha$ & - & -  &  $43.08$  &  1107.72 & GNIRS\\
            J0130+0009 &  2.51  & H$\alpha$ & - & -  &  $43.82$  &  3272.59 & GNIRS\\
            J0149+0052 &  1.85  & H$\alpha$ & - & -  &  $43.10$  &  924.22 & GNIRS\\
            J0150+0056 &  3.49  & Ly$\alpha$,[O III] & $43.38$ & 553.86  &  -  &  -& GNIRS+KCWI\\
            J0152-0024 &  2.78  & Ly$\alpha$,H$\alpha$ & $42.97$ & 725.51  &  $43.96$  &  1382.12 & GNIRS+LRIS\\
            J0213+0024 &  1.81  & H$\alpha$ & - & -  &  $43.74$  &  3849.89 & GNIRS+KCWI\\
            J0214-0000 &  1.63  & H$\alpha$ & - & - &  $43.21$  &  1145.28 & GNIRS+LRIS\\
            J0215+0042 &  0.88  & H$\alpha$ & - & -  &  $43.17$  &  2683.46 & GNIRS\\
            J0221+0050 &  2.48  & Ly$\alpha$,H$\alpha$ & $42.43$ & 695.90  &  $43.91$  &  3918.23 & GNIRS+LRIS\\
            J2229+0022 &  1.93  & Ly$\alpha$,H$\alpha$ & $42.95$ & 1173.02  &  $43.03$  &  747.91 & GNIRS+LRIS\\
            J2233-0004 &  1.60  & H$\alpha$ & - & - &  $43.21$  &  1237.80 & GNIRS+LRIS\\
            J2239-0030 &  1.97  & Ly$\alpha$,H$\alpha$ & $43.19$ & 2324.75  &  $42.52$  &  1511.77 & GNIRS+LRIS\\
            J2239-0054 &  2.09  & Ly$\alpha$,H$\alpha$ & $43.54$ & 1734.92  &  $42.78$  &  1295.67 & GNIRS+LRIS\\
            J2243+0017 &  1.91  & Ly$\alpha$,H$\alpha$ & $42.72$ & 1295.17  &  $43.57$  &  4776.12 & GNIRS+LRIS\\
            J2258-0022 &  2.42  & Ly$\alpha$,H$\alpha$ & $42.65$ & 1508.45  &  $43.07$  &  764.07 & GNIRS+LRIS\\
            J2259-0009 &  1.89  & Ly$\alpha$,H$\alpha$ & $42.39$ & 907.63 &  $43.68$  &  3009.32 & GNIRS+LRIS\\
            J2329+0020 &  2.67  & Ly$\alpha$,H$\alpha$ & $42.65$ & 1007.21  &  $44.15$  &  2850.25& GNIRS+LRIS\\
            J2334+0031 &  2.10  & Ly$\alpha$,H$\alpha$ & $43.42$ & 2808.96  &  $43.64$  &  2112.14 & GNIRS+LRIS\\
		\hline
	\end{tabular}
\end{table*}


\subsubsection{Narrow H$\alpha$ and Narrow Ly$\alpha$}
A narrow emission line ($\rm < 2000 km/s$) is the typical identifying spectral feature for Type-2 QSOs at low-$z$ \citep[e.g.][]{Zakamska2003,Alexandroff2013,Lacy2013}. We have five targets with both the Ly$\alpha$ and H$\alpha$ detections that are narrow, which is consistent with these previous Type-2 QSOs found at low redshift. 

\subsubsection{Broad H$\alpha$ and Broad Ly$\alpha$}
Presence of broad emission lines defines Type-1 QSOs. One target, J2334+0031,  has broad lines in both the Ly$\alpha$ and H$\alpha$ detections (see Fig~\ref{fig:gnirs+lris_v4}). This target also has broad CIV, CIII, and MgII emission lines. The SED and photometry of this target are obviously different from Type-1 QSOs, as it is redder than the extreme red QSOs in SDSS \citep{wb2025}. 

\subsubsection{Broad H$\alpha$ and Narrow Ly$\alpha$}
Four of our targets have broad H$\alpha$ emission lines but narrow Ly$\alpha$ emission lines. A broad H$\alpha$ emission line is usually not expected given the standard picture of Type-2 QSOs based on  low-$z$ ($z\lesssim 1)$ observations, but it is detected in many JWST AGNs (or LRDs) \citep[e.g.][]{Kokorev2023,Harikane2023}. In our sample, 7/21 targets have broad H$\alpha$ emission lines. It is puzzling that we still see the broad H$\alpha$ emission lines despite using red colors to select the most obscured QSOs. 
One possible scenario is that there is obscuration toward the H$\alpha$ emitting region,  
then the UV-emitting region would be fully obscured and Ly$\alpha$ would be produced on larger scales by a different mechanism and be narrower, or at least have different kinematics as they represent different regions.
In our SED fitting paper \citep{wb2025}, we show that the rest-UV/optical composite SEDs of our targets are very red and similar to LRDs. These broad-line targets may further reveal the potential connection of our Type-2 QSOs to JWST LRDs with broad emission lines. 

\subsubsection{Narrow H$\alpha$ and Broad Ly$\alpha$}
We have one target with narrow H$\alpha$ but broad Ly$\alpha$ emission lines. The mechanism to produce the inconsistency of these two lines is still unclear. One possible scenario is that the broad Ly$\alpha$ is emitted in the broad line region (BLR) while the narrow H$\alpha$ arises from the narrow line region \citep[NLR;][]{Zhou2006}.  It could also be due to so-called selective extinction, whereby internal reddening may suppress broad Balmer emission more than Ly$\alpha$ emission \citep{Netzer1982}.

\section{Discussion of Individual Targets} \label{sec:Discussion}
In this section, we discuss some special targets with interesting spectral features.

\subsection{J0150+0056}
This target is the highest redshift ($z = 3.49$) Type-2 QSO in our sample (lower panel in Figure~\ref{fig:gnirs+lris_v1}). Initially, this target only had a Gemini/GNIRS spectrum, and we considered the single emission line to be H$\alpha$ \citep{Ishikawa2023}. 
However, after the Keck/KCWI observations, we identified a strong Ly$\alpha$ emission line in KCWI, and the emission line in GNIRS spectra turns out to be [O III]. The redshift of this target is $z=3.49$ by fitting the line center of Ly$\alpha$ and [O III]. This target is also the brightest target in our sample ($\rm L_{bol} = 10^{48.08}erg/s$, see \citet{wb2025} for the $\rm L_{bol}$ estimate). Assuming the Eddington limit, the black hole mass should be above $\rm 10^{9.98}M_{\odot}$, placing it at the bright and massive end of SMBHs at this redshift. 

\subsection{J0152-0024}
This target is a $z=2.77$ Type-2 QSOs confirmed by the Keck/LRIS and Gemini/GNIRS spectra (upper panel in Figure~\ref{fig:gnirs+lris_v2}). The H$\alpha$ line of this target is located at the edge of the GNIRS coverage, so it is difficult to measure the full line FWHM. 
There is a strong absorption line feature at Ly$\alpha$ and the NV$\lambda1240$ emission line. We measure the Ly$\alpha$ emission line has FWHM about $\rm 720.62 km/s$, but this could be underestimated, 
due to the strong absorption feature. We also notice that a similar absorption feature is seen at the SiIV$\lambda1396$ emission line. There is also an $\rm 2480 km/s$ 
velocity
 offset between the Ly$\beta$ emission line and the expected line wavelength. 

\subsection{J0221+0050}
J0221+0050 object is a $z=2.48$ Type-2 QSO confirmed by both the Keck and Gemini spectra (upper panel in Figure~\ref{fig:gnirs+lris_v3}). Only H$\alpha$ and Ly$\alpha$ emission lines are detected in the spectra. The H$\alpha$ has a broad $\rm \sim 3918 km/s$
component, but the Ly$\alpha$ is narrow ($\rm  \sim 695 km/s$). Such a broad H$\alpha$ emission line is not expected in typical Type-2 QSOs, but is commonly detected in many JWST LRDs. Considering the similar UV SED shape of these targets and the JWST LRDs \citep{wb2025}, such broad-line obscured QSOs could be the lower-$z$ analogs of JWST LRDs. 


\subsection{J2334+0031}
Target J2334+0031 is the only object with both broad H$\alpha$ and Ly$\alpha$ in our sample (see Figure~\ref{fig:gnirs+lris_v4}). This target is confirmed to be a Type-2 QSO at $z=2.09$ by Gemini and then further observed by Keck/LRIS. The FWHM of Ly$\alpha$ for this target is $\rm 2808 km/s$ and the H$\alpha$ is $\rm 2112 \space km/s$. Such a line width is similar to the typical Type-1 QSOs. However, the photometry and SED of this target are different from the typical UV-bright Type-1 QSO. 
Besides, there is a blueshift of the NV emission line. Due to the low resolution of Keck/LRIS, we don't know if there is an strong absorption between the Ly$\alpha$ and NV emission line. We are also not sure if this causes the blueshift of the NV emission line. The CIV, CIII] and MgII emission line show an asymmetric line profile, and the line peak of  CIII] shows a $\sim \rm 1776 \space km s^{-1}$ offset from the expected position. 
The broad and asymmetric line profiles are reported in some ERQs \citep{Perrotta2019,Villar2020,Gillette2024}. The broad-line regions in ERQs are typically outflow-dominated, and these outflows can produce strong blueshifted or broad emission lines \citep[e.g.][]{Gillette2024}. 
ERQs could also be very luminous Type-1 quasars viewed at an intermediate orientation, and this orientation allows a direct view of the outer part of the large broad-line region \citep[e.g.][]{Villar2020}. These two scenarios can explain the broad and asymmetric line profiles for this target J2334+0031. 



\section{The $L_{\rm [O III]}-L_{\rm bol}$ Scaling Relation}
The [O III] 
line is forbidden and is expected to arise from the narrow-line region, because its critical density, $n_{\rm crit}\simeq 7\times10^{5} \rm cm^{-3}$ 
is exceeded by the much higher density gas, $n_{\rm H|} \sim 10^{9-11} \rm cm^{-3}$ broad-line region \citep{Osterbrock2006}. In both Type-1 and Type-2 QSOs, [O III] is powered by the same UV emission from the AGN \citep{Stern2012}. However, the UV continuum is obscured and reprocessed by the mid-IR emission in Type-2 QSOs. Thus, a relation between the [O III] line luminosity and the mid-IR luminosity (or bolometric luminosity determined from the mid-IR emission) is expected \citep[e.g.][]{Zakamska2003,Lamastra2009,Lacy2013}.

\begin{figure*}
  {\centering
  \includegraphics[width=0.9\textwidth]{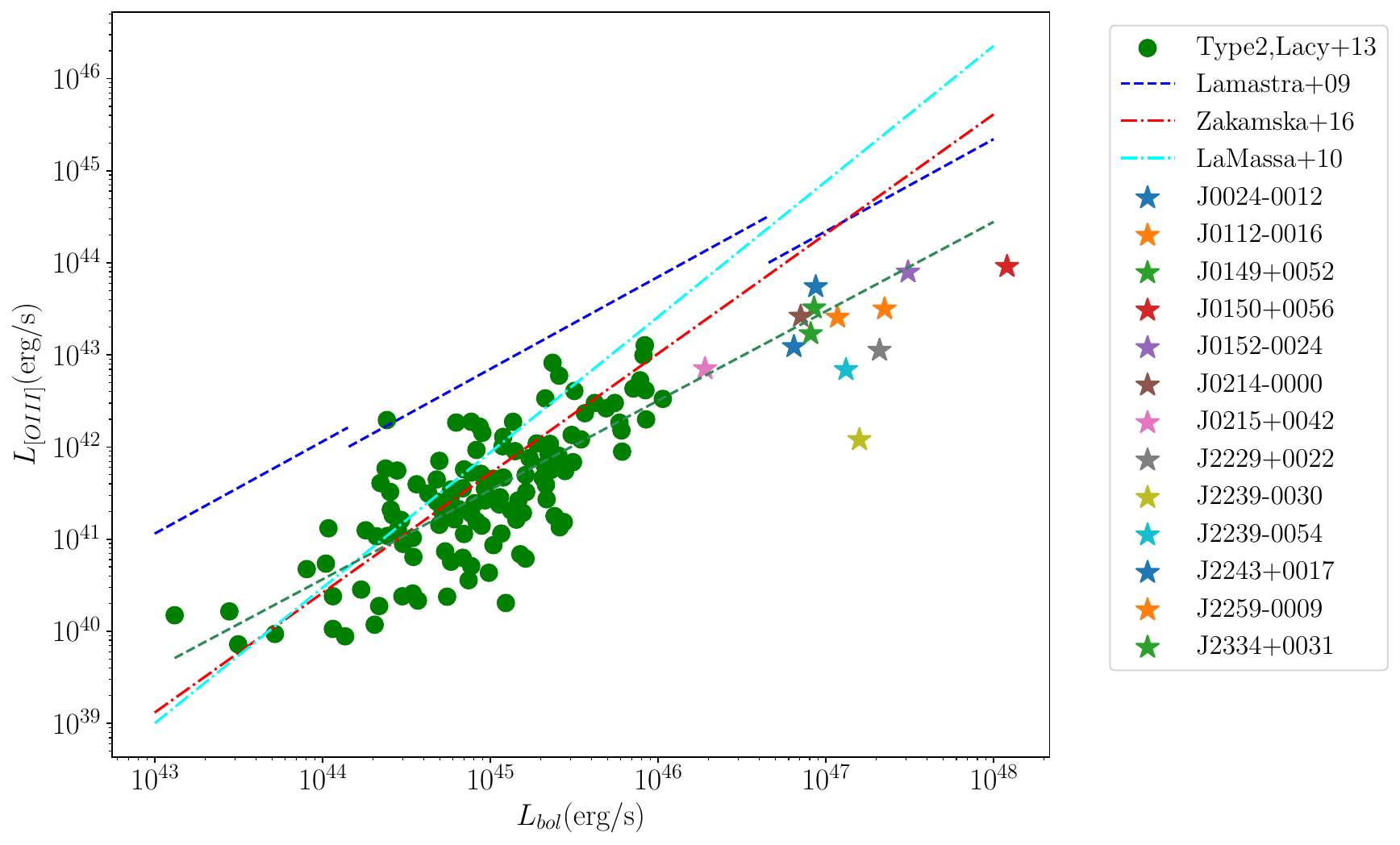}
  }
  \vspace{-0.2cm}
 \caption{The relation between the luminosity of the [O III] emission line and the bolometric luminosity. The relations from \citet{Lamastra2009}, \citet{LaMassa2010}, and \citet{Zakamska2016} are shown as blue, cyan, and red lines. 
 The green dots are the Type-2 AGN measurements from \citet{Lacy2013}. The luminosities at 15$\rm \mu m$ of these targets are scaled to estimate the bolometric luminosity. The colored stars are the measurements of 13 targets with [O III] detection in our sample. The green dashed line presents a power-law fit (linear fit on log scale) to the green data points. The [O III] luminosities in relations from \citet{Lamastra2009}, \citet{LaMassa2010}, and \citet{Zakamska2016} are all extinction corrected. The $\rm L_{[O III]}$ in \citet{Lacy2013} and our sample is not dust extinction corrected due to the non-detection of H$\beta$ or H$\alpha$ emissions. 
Compared to the three corrected relations, our targets show a lower [O III] luminosity because of the dust extinction. 
However, even compared to the uncorrected sample in \citet{Lacy2013}, the [O III] luminosity is still below the fitting (green dotted line). 
Our targets are about 100$\times$ brighter than the low-$z$ objects in \citet{Lacy2013}, but the [O III] luminosity doesn't increase accordingly.   }.\label{fig:O-B}
\end{figure*}

\citet{Lacy2013} studied this relation with a sample of $\sim 100$ Type-2 QSOs at $z<2$. 
Specifically, they observed that the luminosity of the [O III]$\rm \lambda5007$ line, 
without any extinction correction, strongly correlates with 
the mid-IR luminosity, $\nu L_{\nu}$,  at 15$ \rm \mu m$. We use the torus template from \citet{Stalevski2016} and our composite SED in \citet{wb2025} 
to determine the relation between the luminosity at 15$\rm \mu m$ (observed frame at $z\sim2$) and the bolometric luminosity\footnote{We have a composite SED for Type-2 QSOs in \citet{wb2025}. From that composite SED, we can get the luminosity at 15$\rm \mu m$. The composite SED contains a torus model using the template from \citet{Stalevski2016}, and then we calculate the torus luminosity by integrating the torus model.
The bolometric luminosity is estimated using Equation 4 in \citet{wb2025} assuming R=0.5. After that, we apply the relation between the 15$\rm \mu m$ luminosity and bolometric luminosity.  }:
\begin{equation}
    \text{L}_{\text{bol}} = 6.03 \times \text{L}_{15 \rm \mu m}. 
\end{equation}
Using this relation, the luminosity at 15$\rm \mu m$ in \citet{Lacy2013} can be used to compute the bolometric luminosity, which assumes their targets have the same mid-IR SED shape as our targets. The results are shown as the green dots in Figure~\ref{fig:O-B}. A green dotted line shows a simple power law fit (linear fit in the log) 
to these data.

In our Type-2 sample, 13/23
targets have an [O III] detection. We use a double-Gaussian profile with wavelength separation and relative line strength set by atomic physics to fit the emission line.
The FWHM and the luminosity of this double-line can be estimated from the fitting. 
The [O III] luminosity is here defined to be the luminosity of [O III]$\rm \lambda5007$ alone. 
The fitted parameters are shown in Table~\ref{tab:OIII}, and the relation between the [O III] luminosity and the bolometric luminosity for our targets is shown as the colored stars in Figure~\ref{fig:O-B}. Since our targets don't have $\rm H\beta$ line detection, we couldn't conduct the  extinction correction using .

We also show the $\rm L_{[O III]} - L_{bol}$ relation from \citet{Lamastra2009}, \citet{Zakamska2016} and \citet{LaMassa2010}. 
\citet{Lamastra2009} estimate the bolometric correction 
$\rm C_{[O III]} = L_{bol}/L_{[O III]}^{c}$
in different [O III] luminosity ranges. 
The bolometric luminosity is estimated from the X-ray luminosity. The $\rm L_{[O III]}^{c}$ here is the extinction corrected [O III] luminosity using:
\begin{equation}
    {\text{L}_{\text{[O III]}}^\text{c}} = {\text{L}_{\text{[O III]}}}\left( \frac{\left( \text{H$\alpha$}/\text{H$\beta$} \right)_{\text{obs}}}{3.0} \right)^{2.94}
\end{equation}
They found a mean value of $\rm C_{[O III]}$ in the luminosity
ranges log $\rm L_{[O III]}$ = 38-40, 40-42, and 42-44 of 87, 142, and 454, respectively. 
 However, we still plot this relation in Figure~\ref{fig:O-B} as for comparison. 

\citet{Zakamska2016} and \citet{LaMassa2010} summarized the relation between the luminosity at 13.5 $\rm \mu m$ and the [O III] luminosity for Type-2 QSOs to be: $\rm log \nu L_{\nu}[13.5\mu m] = 12.1 + 0.77logL[O III]$ and $\rm log \nu L_{\nu}[13.5\mu m] = 15.7 + 0.68logL[O III]$. We assume their targets have a similar SED shape to ours, then we can use our SED to adopt the relation between the  $\rm L_{13.5 \mu m}$ and the $\rm L_{bol}$:
\begin{equation}
    \text{L}_{\text{bol}} = 6.01 \times \text{L}_{13.5 \rm \mu m}
\end{equation}
We plot the adopted bolometric luminosity and [O III] line luminosity as red and cyan dotted lines in Figure~\ref{fig:O-B}. 

The [O III] luminosities in relations from \citet{Lamastra2009}, \citet{LaMassa2010}, and \citet{Zakamska2016} are all extinction corrected. The $\rm L_{[O III]}$ in \citet{Lacy2013} and our sample is not dust extinction corrected due to the fact that we do not always detect both 
the H$\beta$ or H$\alpha$ lines. 
Compared to the three extinction corrected relations, our targets show a lower [O III] luminosity, 
which is likely due to extinction. 
However, even comparing to the uncorrected sample in \citet{Lacy2013}, the [O III] luminosity is still below our fit to their data (green dotted line). 
Our targets are about 100$\times$ brighter than the low-$z$ objects in \citet{Lacy2013}, but the [O III] luminosity doesn't increase accordingly.


\begin{table*}
	\centering
	\caption{Theredshift, [O III] line property, and bolometric luminosity for 13 targets.}
	\label{tab:OIII}
	\begin{tabular}{lccccr} 
		\hline
		Target & z  & $log(\rm L_{[O III]}(erg/s))$ & $\rm FWHM_{4960}(km/s)$  & $\rm FWHM_{5007}(km/s)$  & $log(\rm L_{bol})$\\
		\hline
		J0024-0012 &  1.53 & $43.04$ & - & 1045.61 & $46.81$\\
            J0112-0016 &  2.99  & $43.55$ & 336.24 & 324.74 &$47.35$\\
            J0149+0052 &  1.85  & $43.57$ & 1289.20 & 1417.45 &$46.93$\\
            J0150+0056 & 3.49 & $43.91$ & - & 2712.18 &$48.08$\\
            J0152-0024 &  2.78  & $43.93$ & - & 1047.80 &$47.49$\\
            
            J0214-0000 &  1.63  & $43.43$ & 616.39 & 450.66 &$46.85$\\
            J0215+0042 &  0.88  & $42.81$ & - & 1125.59 &$46.28$\\
            J2229+0022 &  1.93  & $43.03$ & - & 690.97 &$47.32$\\
            J2239-0030 &  1.97  & $42.08$ & - & 369.25 &$47.20$\\
            J2239-0054 &  2.09  & $42.79$ & - & 924.68 &$47.12$\\
            J2243+0017 & 1.91 & $43.74$ & 1216.76 & 1022.27 & $46.94$\\
            J2259-0009 &  1.89  & $43.29$ & - & 1711.87 &$47.07$\\
            J2334+0031 &  2.10  & $43.23$ & 572.67 & 653.70 &$46.91$\\
		\hline
	\end{tabular}
\end{table*}

\section{Composite Spectra}\label{sec:composite}
\begin{figure*}

  {\centering
  \includegraphics[width=\textwidth]{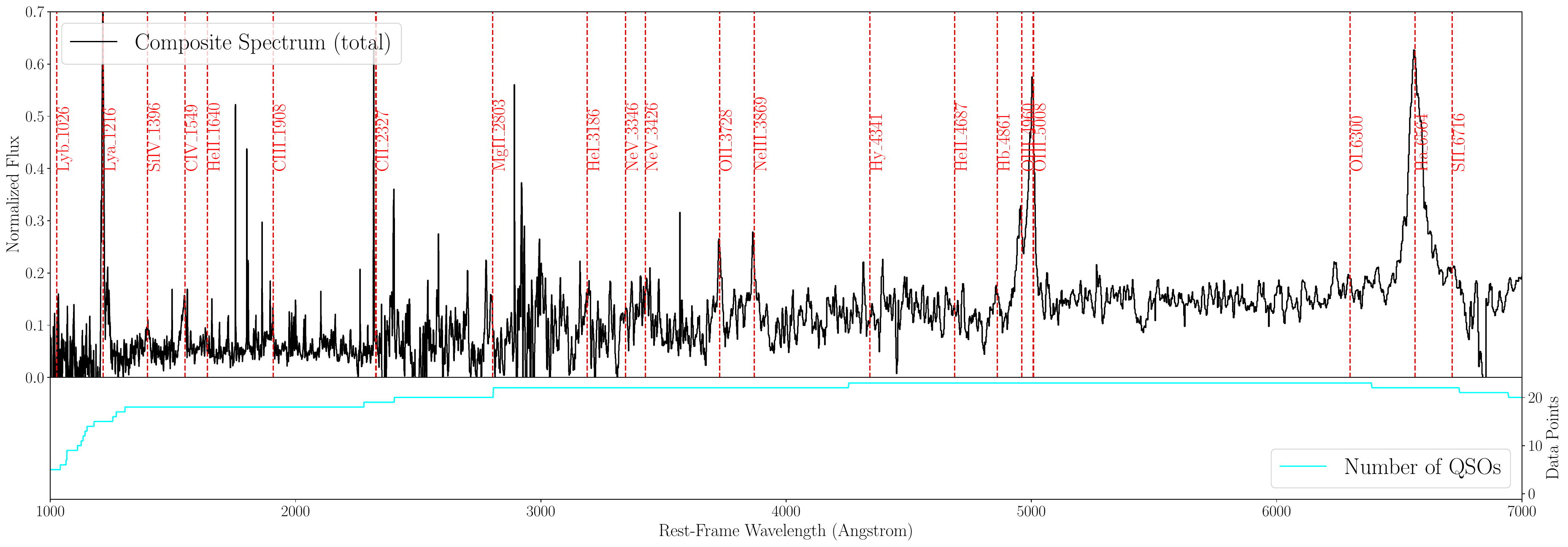}
  \includegraphics[width=\textwidth]{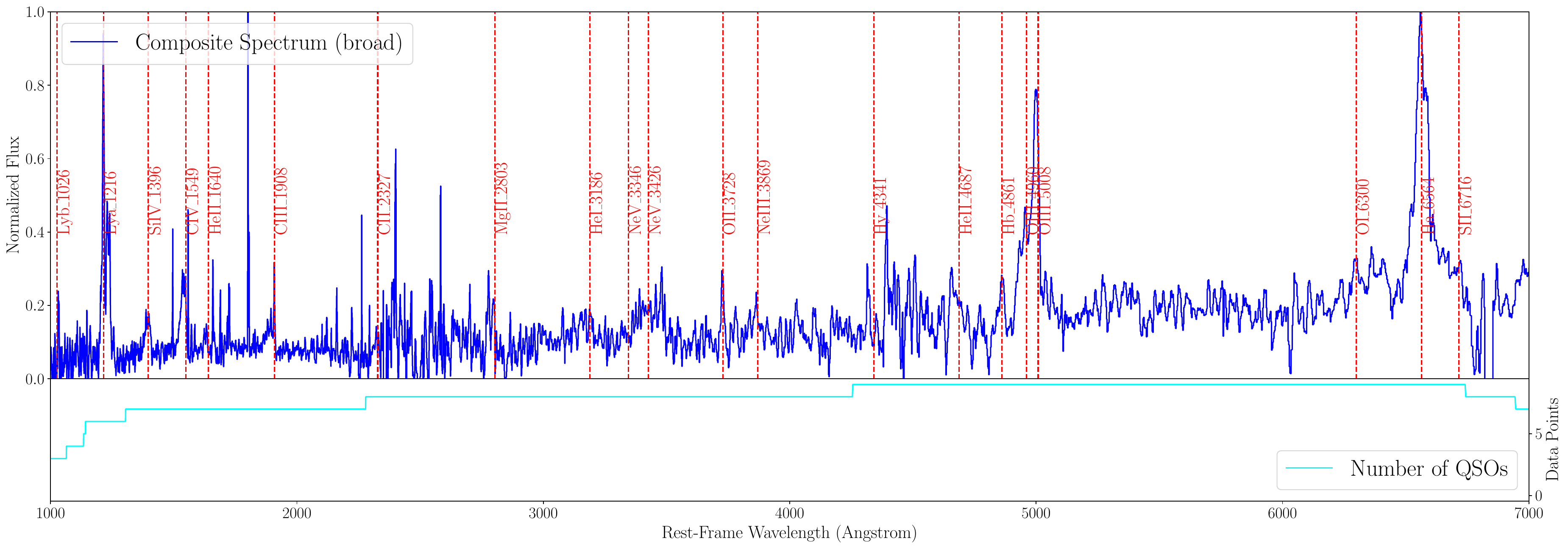}
  \includegraphics[width=\textwidth]{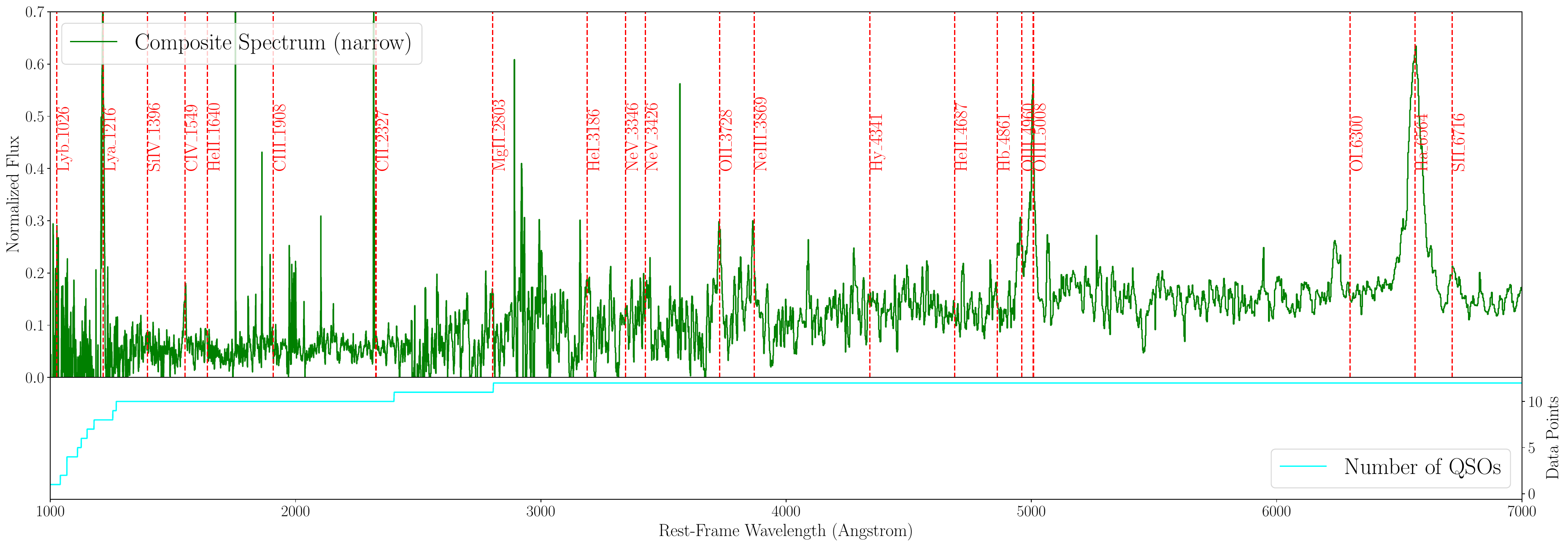}
  }
  \vspace{-0.7cm}
 \caption{The composite spectra for 23 targets in our sample. The top panel shows the composite spectra for all 23 targets, and the median panel shows the composite spectra for nine targets with broad H$\alpha$ emissions. The bottom panel shows the composite spectrum only for 12 targets with narrow H$\alpha$ emission lines. 
The red dashed lines mark the typical emission lines for QSOs. The bottom subplot shows the number of QSOs contributing to each wavelength }.\label{fig:composite}
\end{figure*}

In this section, we generate composite spectra for all 23 identified targets to better understand the spectral properties of Type-2 QSOs. 
To generate the composite spectra, we firstly define the scaling parameter using the WISE W4 flux for each target. All 23 targets are scaled to the median value of the WISE W4 flux ($\rm 6.53 mJy$), and then the same scaling parameter is applied to the spectra. Then we take the median flux value at each wavelength to get the composite spectra. The final composite spectra is shown in Figure~\ref{fig:composite}. 
The top panel shows the composite spectra for all 23 targets (black), and the middle panel shows the composite spectra for nine targets with broad H$\alpha$ emission (blue). Here we further include target J0152-0024 into the broad H$\alpha$ group. Since the H$\alpha$ emission for this target fall  at the edge of the wavelength coverage, making the fitted FWHM with a single Gaussian profile could be an underestimate of the intrinsic value. 
The bottom panel shows the composite spectrum only for 12 targets with narrow H$\alpha$ emission lines (green). 

The red dashed lines mark the typical emission lines for QSOs. The bottom sub-plots show the number of QSOs contributing to each wavelength. 

From our composite spectra, we see the tentative detection of some relative weak rest-optical lines like:[OII]$\lambda3782$, [NeIII]$\lambda3869$,  H$\beta$, [OI]$\lambda6300$, and [SII]$\lambda6716$. The Type-2 composite spectra in \citet{Lacy2004} show both strong high-ionization narrow lines and low-ionization lines such as [O I], but it lacks coronal lines. \citet{Hainline2011} generated the rest-UV composite spectra for Type-2 AGNs and find a median velocity difference between Ly$\alpha$ and He II to be $\Delta v = 208 \space \rm km s^{-1}$. They also find a red UV continuum slope produced by dust extinction and some rare emission lines like N IV]$\lambda1486$. Compare to these two composite spectra from previous surveys, our composite spectra don't show a strong detection of HeII emission line. Our composite cover the rest-optical emissions which \citet{Hainline2011} doesn't cover, and a comparison between our composite and the composite in \citet{Lacy2013} could give us a idea if the obscured population is evolving with redshift. 
A detailed discussion between different Type-2 composite spectra will be presented in a future paper.


The weak lines ([OII]$\lambda3782$, [NeIII]$\lambda3869$,  H$\beta$, [OI]$\lambda6300$, and [SII]$\lambda6716$) are usually used to build the BPT line ratio diagram to identify AGNs or galaxies \citep[e.g.][]{Mazzolari2024}. However, even the most powerful ground-based telescopes like Keck and Gemini can't detect these weak lines individually for the most luminous Type-2 QSOs at $z\sim2$. Only  JWST observations can further reveal these weak lines and help us better understand the relation between these obscured populations and other galaxies or AGNs. 


\section{Conclusion} \label{conclusion}

In this paper, we present the spectroscopic observation results for 23 IR-luminous Type-2 QSOs using Keck and Gemini. 
These targets are selected using an $\rm r - W4$ color cut, and the Gemini and Keck spectroscopic observations confirm that these targets are at $z = 0.88 - 3.49$ (12 are at $z>2$).
The SED fitting results are published in a previous paper \citep{wb2025}.  
Our main conclusions based on these spectra are:
\begin{enumerate}
\item Our targets show significant spectral diversity in the relative widths of their emission lines. Five of our targets have narrow Ly$\alpha$ and H$\alpha$ emission lines, which is consistent with some typical Type-2 QSOs. One of our targets J2334+0041 has both broad Ly$\alpha$ and H$\alpha$ emission, which is more similar to typical Type-1 QSOs. However, the UV-faintness 
and the SED shape make it different from the UV-bright Type-1 QSOs; we consider this target 
to be like some ERQs. Four of our targets have broad H$\alpha$ and narrow Ly$\alpha$ emission. This feature is commonly seen in some JWST LRDs or broad-line AGNs. One of our targets has narrow H$\alpha$ and broad Ly$\alpha$ emission, which could be caused by the viewing angle or the lines are raised  
from different regions (the broad Ly$\alpha$ is emitted in the broad line region (BLR) while the narrow H$\alpha$ arises from the narrow line region). 

\item Three out of our targets have broad H$\alpha$ and narrow Ly$\alpha$ emission. Such a broad line is not usually seen in the low-z Type-2 QSOs, but it is a common feature for some JWST broad-line AGNs. Considering the similar UV SED shape (see \citet{wb2025}), these targets could be the analog of JWST LRDs at lower-z.  Our selection can uncover the diverse obscured population and could bridge the gap between low-$z$ obscured QSOs and the JWST AGNs.

\item Compared to the previous Type-2 QSO samples at lower-$z$ ($z<2$) in \citet{Lacy13b}, our targets have bolometric luminosity over a hundred times larger. Since most of our targets don't have H$\beta$ detections, it is not possible to perform extinction corrections of the [OIII] line luminosity via the  
the Balmer decrement. Compared to the uncorrected [O III] luminosity and bolometric luminosity relation in \citet{Lacy13b}, our targets show  relatively lower [O III] luminosity. The physics behind this low [O III] luminosity is not yet clear.  


\item We constructed composite spectra for all 23 targets, only broad H$\alpha$ targets, and only narrow H$\alpha$ targets. The composite spectra allow us to detect many weak lines, which is useful for us to better understand the spectral properties of this obscured population. 
However, the detection of such weak lines for individual objects, key for understanding the spectral diversity, 
would require higher signal-to-noise ratio observations with JWST. 
\end{enumerate}

Prior to the launch of JWST, multi-wavelength surveys have build a large sample of obscured QSOs at $z<2$. The recent JWST observations suggest the presence of large obscured population at $4<z<9$. Our Type-2 QSO survey can bridge the gap between these two obscured populations at cosmic noon ($2<z<4$) and uncover the diverse obscured population. 
Building a census of Type-2 QSOs across cosmic time is key to understanding two questions: 1)if the obscured population is evolving; 2) what is the true nature of obscured QSOs. 
Gemini and Keck are the most powerful spectroscopic telescopes on the ground. They are sufficiently sensitive to detect the strongest emission lines and measure the redshifts of these heavily obscured QSOs with exposure times of 0.5-1 hour. Gemini spectra cover the rest-frame optical that is usually covered by low-$z$ surveys, but our S/N is quite poor. So it is hard to make a comparison and reveal any differences. At low-$z$ one does not have the rest-UV spectra that we cover in this paper, making the low-$z$ to high-$z$ comparison challenging. We see significant spectral diversity in the relative widths of the emission lines. Some targets show very broad emissions despite the obscuration. They don't seem to follow from a simple unification explanation. The higher S/N JWST spectra may allow us to do a comparison across the cosmic time to study if the obscured population is evolving. Besides, the detection of the weak emission lines will allow us to build a new BPT diagram, to better understand the relation of the obscured populations and other galaxies and AGNs.

\section{Acknowledgements}

BW and ZC are supported by the National Key R\&D Program of China (grant no.~2023YFA1605600), the National Science Foundation of China (grant no.~12073014), the science research grants from the China Manned Space Project with No.~CMS-CSST-2021-A05, and Tsinghua University Initiative Scientific Research Program (No.~20223080023). 

JTS is supported by the Deutsche Forschungsgemeinschaft (DFG, German Research Foundation) - Project number 518006966.



\section{Data availability}
The Gemini/GNIRS spectra are public in the Gemini Observatory Archive under the proposal ID GN-2017B-Q-51. 
The Keck spectra are all public on the Keck Data Archival (KOA). 
The reduced version of all the spectra can be found at \url{https://github.com/samwang141224/Type-2QSOs/tree/main/spec-data}
.



\bibliographystyle{mnras}
\bibliography{example} 



\clearpage
\appendix

\section{Gemini and Keck Spectra for all the targets}

\begin{figure*}
\begin{subfigure}[b]{0.65\linewidth}
         \includegraphics[width=\textwidth]{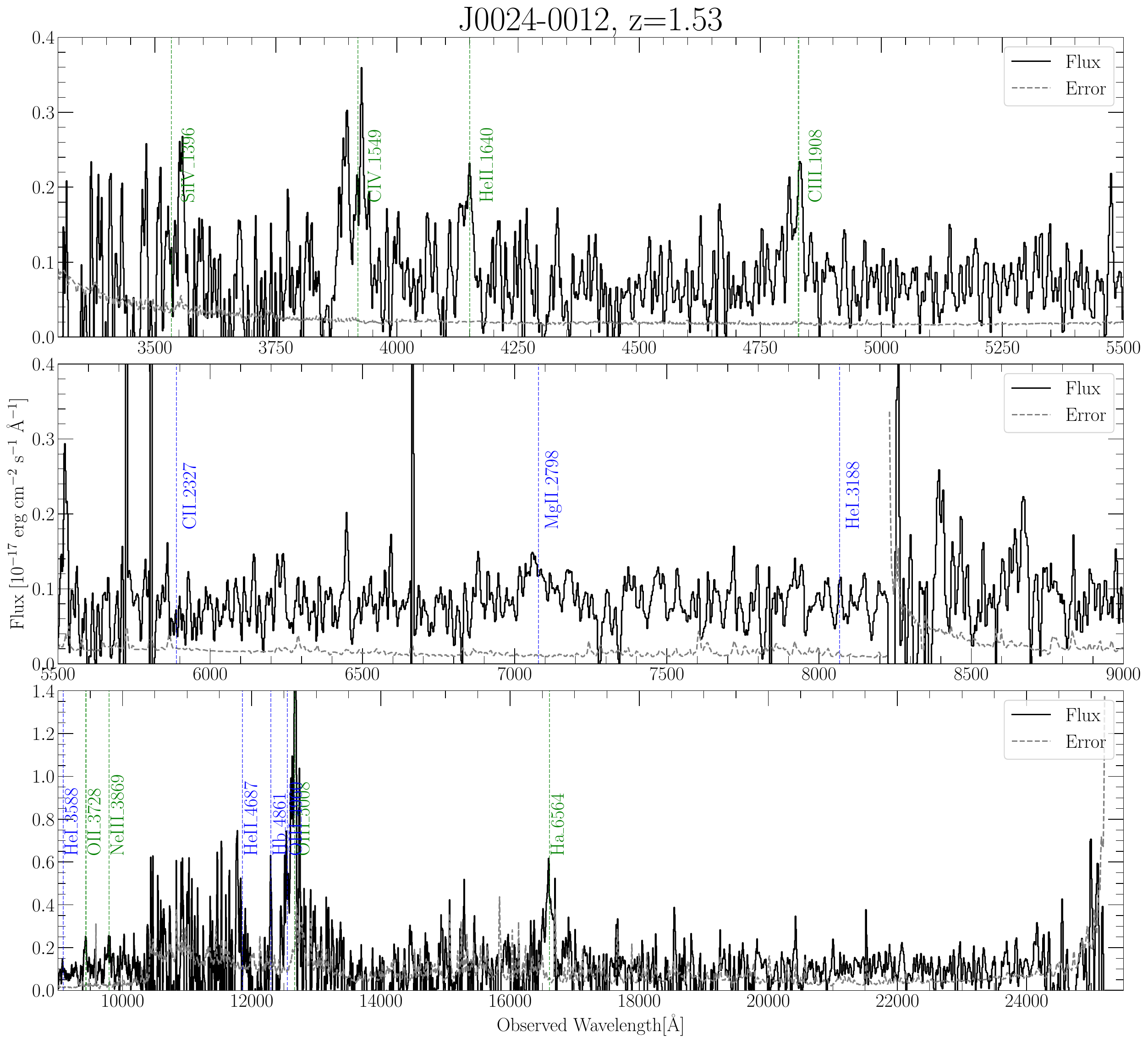}
    \end{subfigure}
     \begin{subfigure}[b]{0.34\linewidth}
         \includegraphics[width=\textwidth]{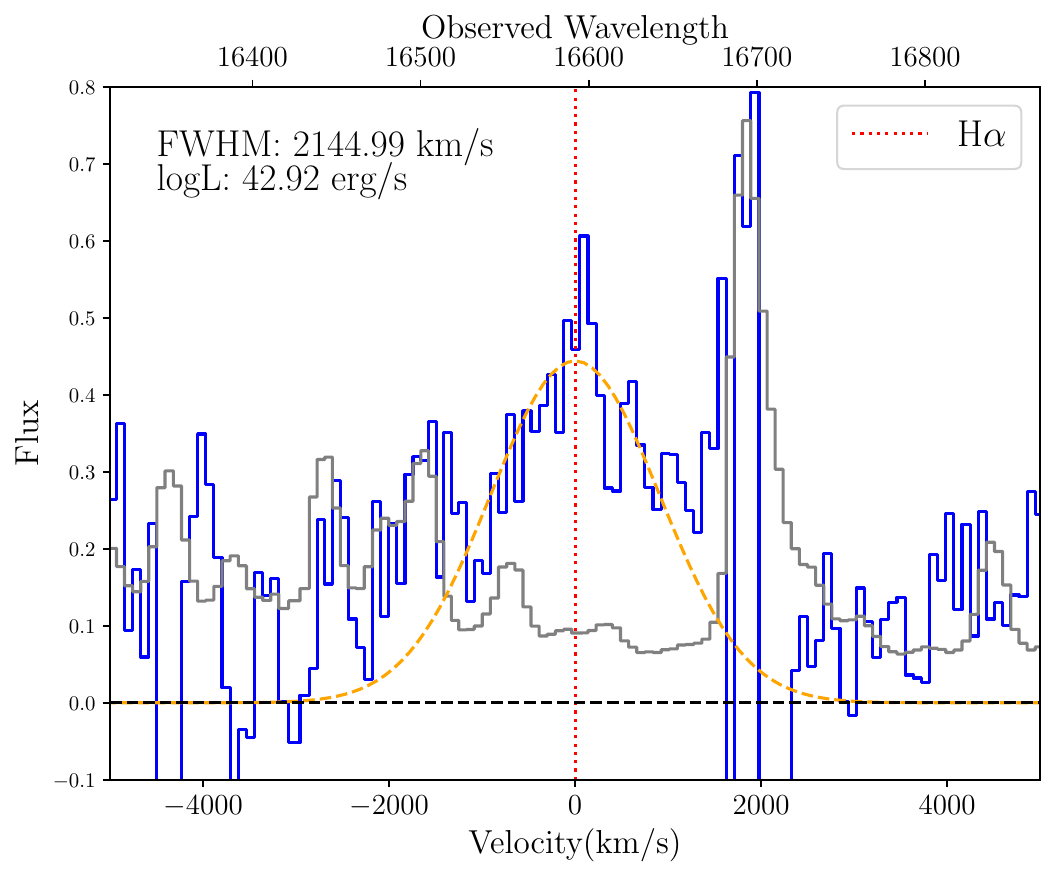}
         \hfill
         \includegraphics[width=\textwidth]{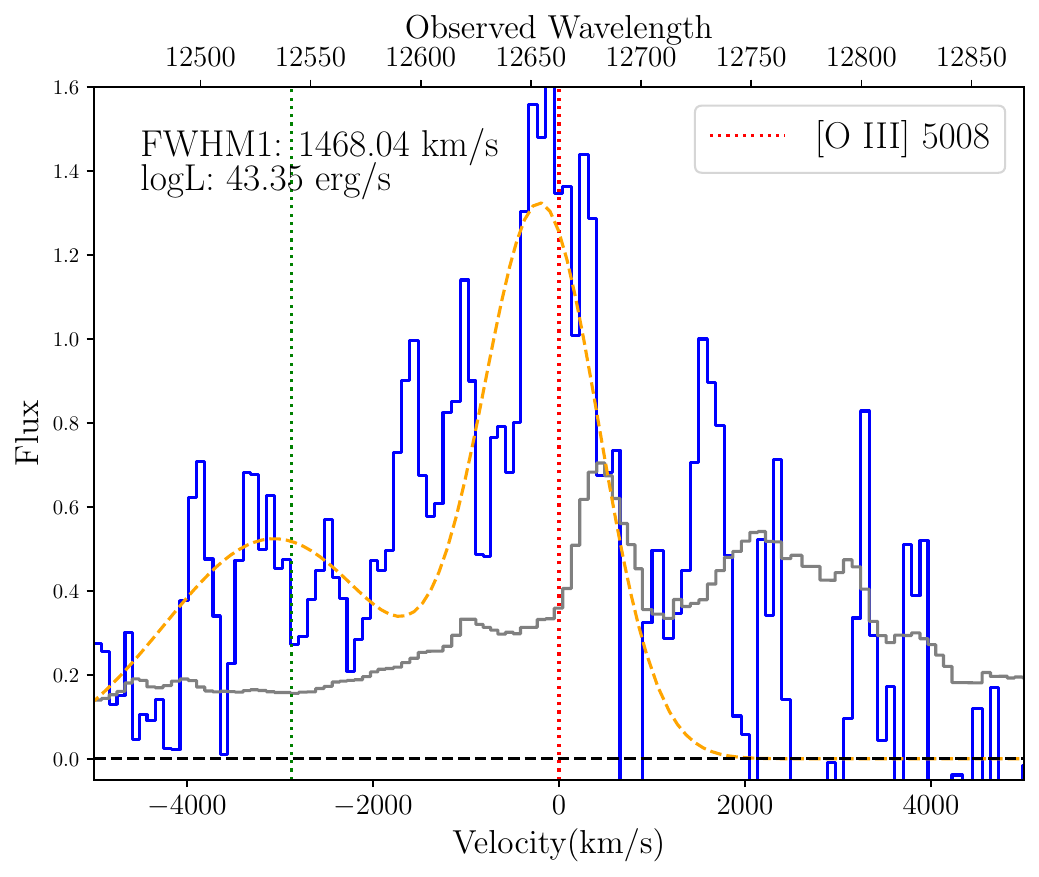}
    \end{subfigure}    
    \hfill
    \begin{subfigure}[b]{0.65\linewidth}
         \includegraphics[width=\textwidth]{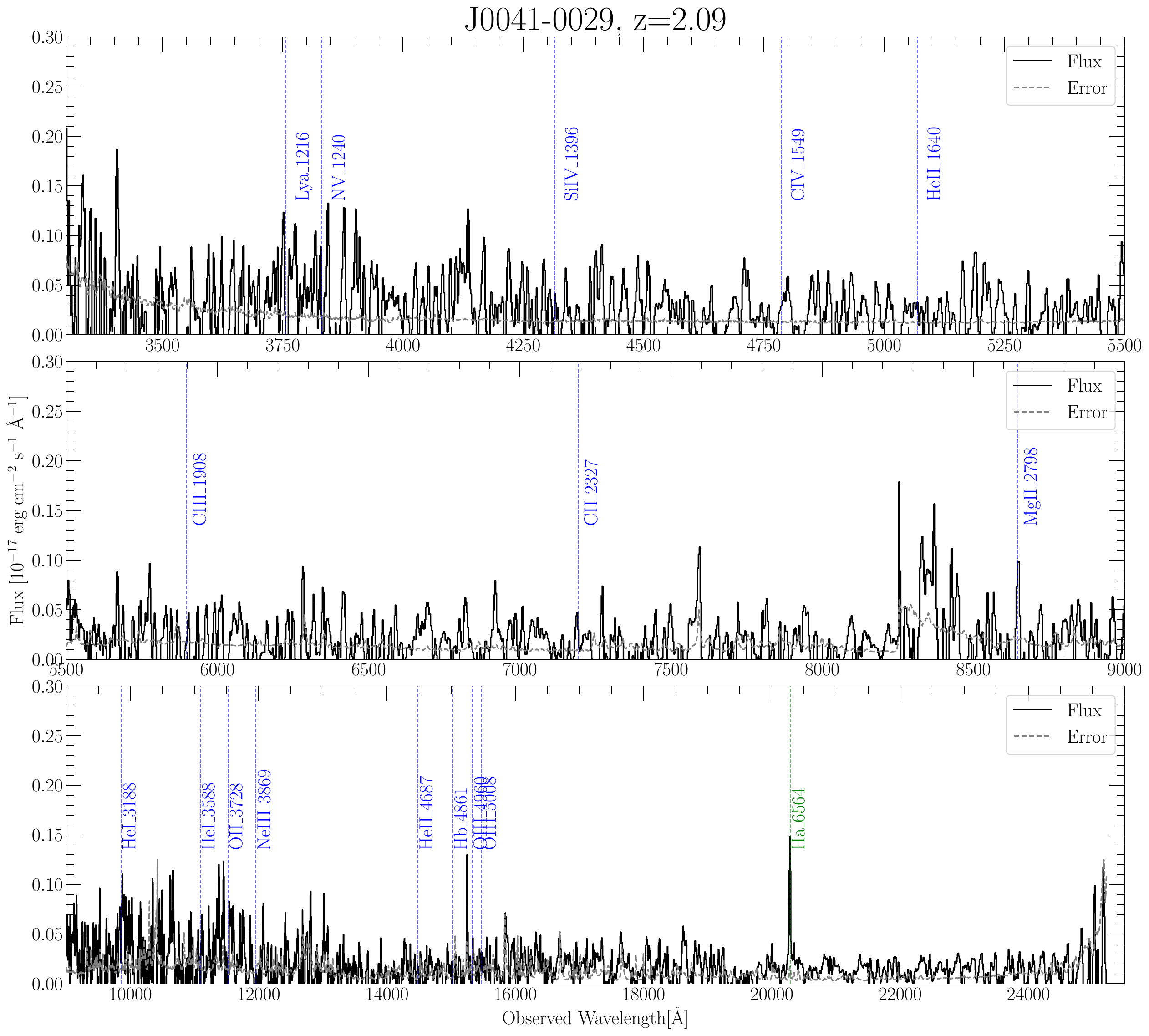}
    \end{subfigure}
     \begin{subfigure}[b]{0.34\linewidth}
         \includegraphics[width=\textwidth]{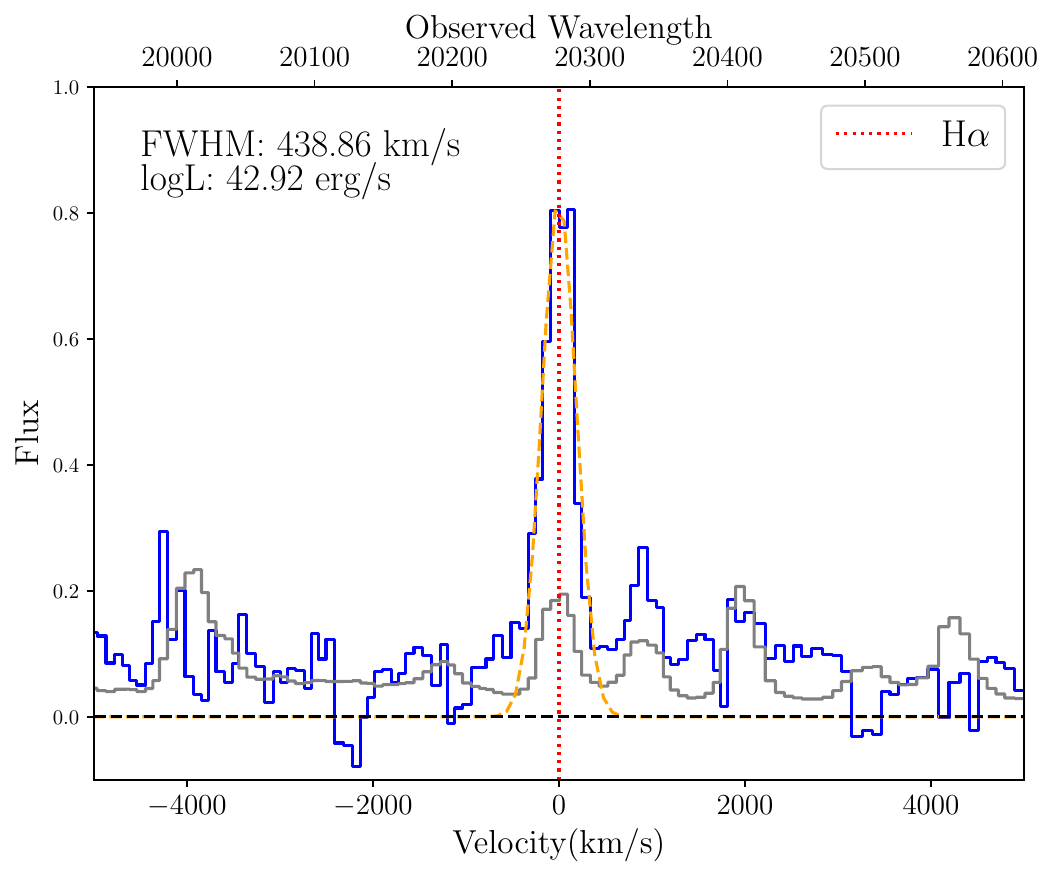}
         \hfill
         
    \end{subfigure}     

\caption{Reduced spectra and the fitted emission line for all our targets. Each left panel shows the coadded spectra from Keck and Gemini into three windows for better visibility. The dashed blue lines show the expected location of typical emission lines, and the dashed green lines show the detected lines. The right panel shows the fitting for Ly$\alpha$, H$\alpha$ or [O III] emissions. The red dashed line shows the center of the emission line and the orange shows the Gaussian fitting. 
}.\label{fig:A1}     

\end{figure*}

\begin{figure*}
    \begin{subfigure}[b]{0.79\linewidth}
         \includegraphics[width=\textwidth]{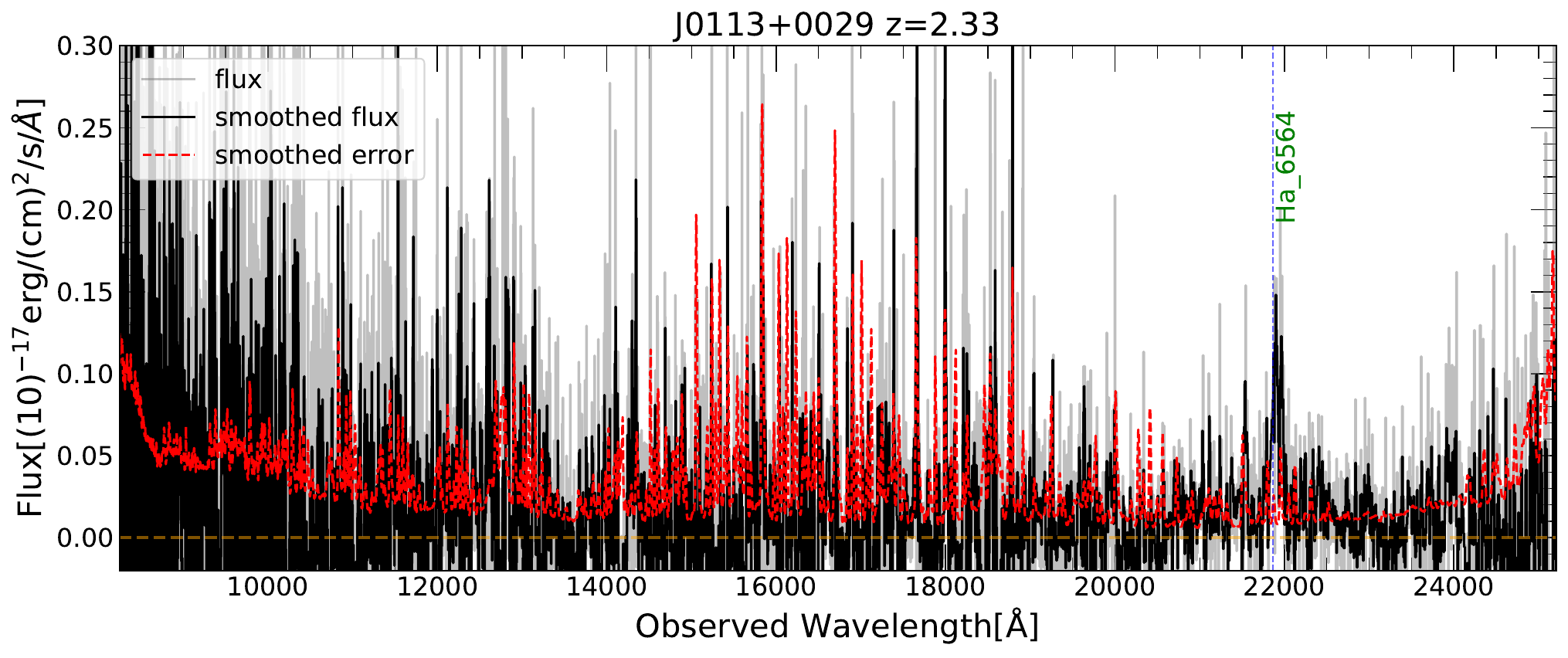}
    \end{subfigure}
     \begin{subfigure}[b]{0.79\linewidth}
         \includegraphics[width=\textwidth]{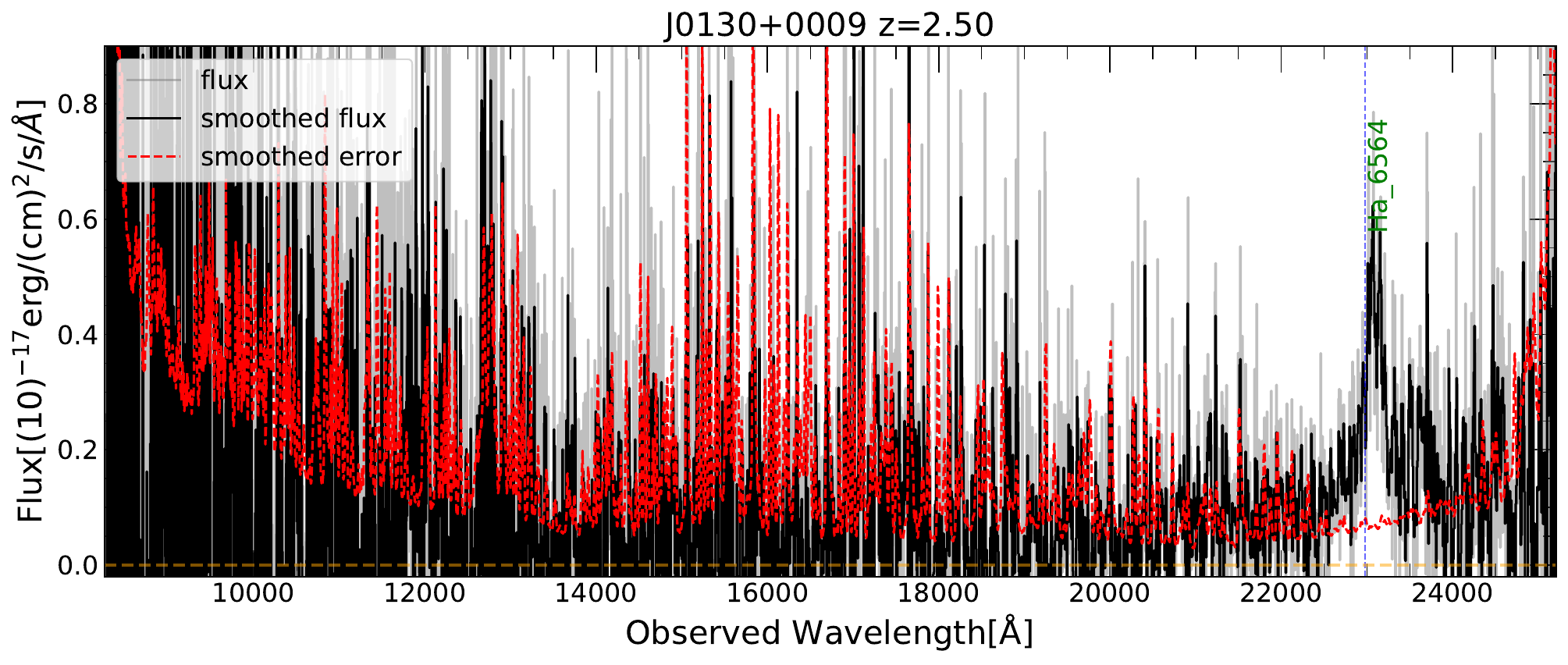}
    \end{subfigure}     
     \hfill   
     \begin{subfigure}[b]{0.79\linewidth}
         \includegraphics[width=\textwidth]{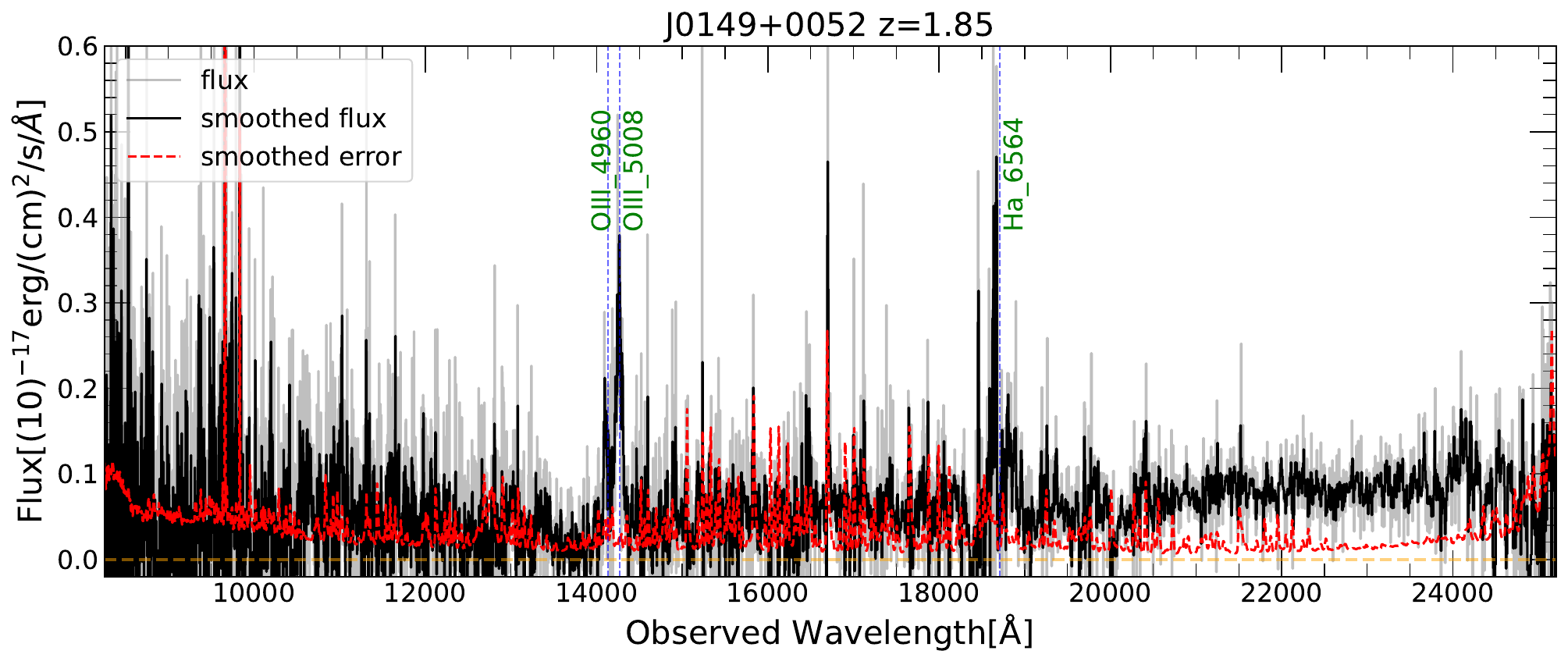}
    \end{subfigure}
     \begin{subfigure}[b]{0.79\linewidth}
         \includegraphics[width=\textwidth]{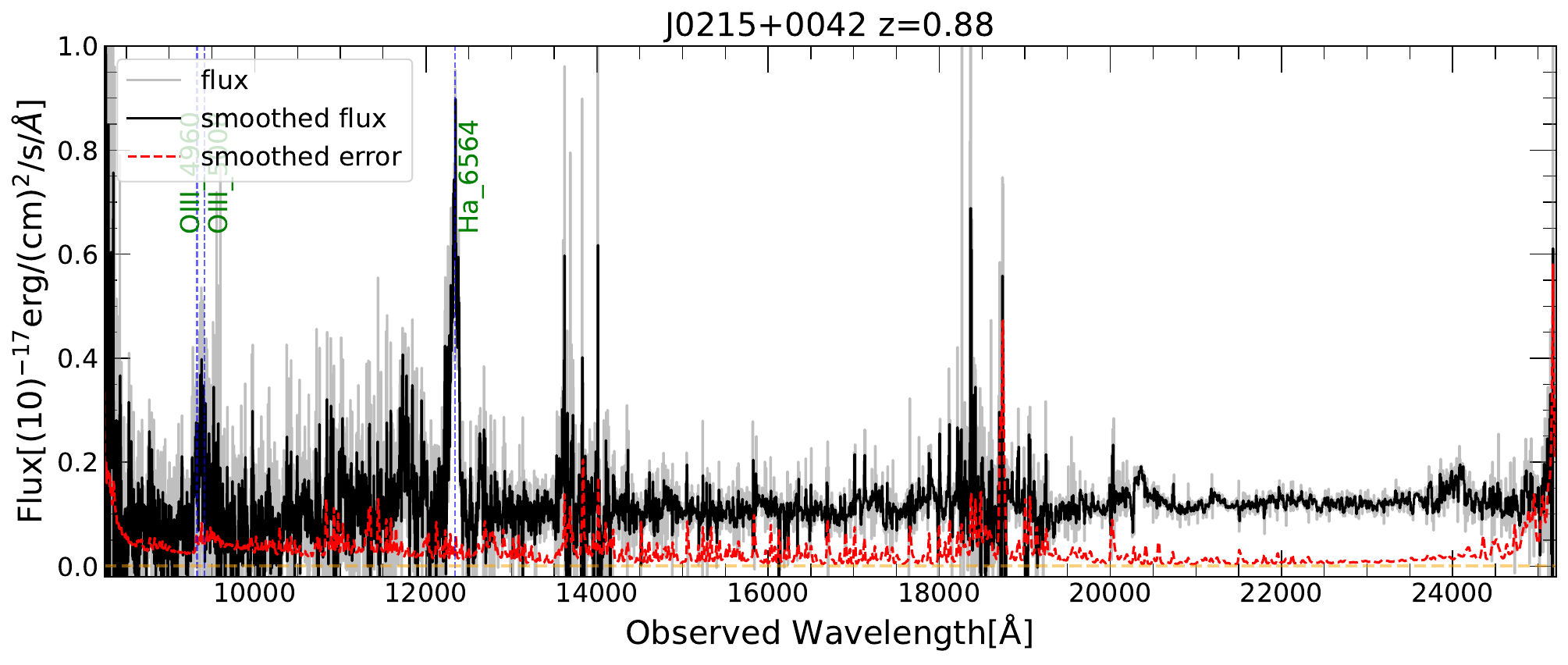}
    \end{subfigure}

\caption{Four targets in our sample with Gemini/GNIRS spectra only. 
}.\label{fig:A2}
\end{figure*}

Here we show the Gemini and Keck spectra for the rest of the targets. Targets with both Keck and Gemini spectra are shown in Figure~\ref{fig:A1}. The four targets with Gemini spectra only are shown in Figure~\ref{fig:A2}. 





\begin{figure*}
\begin{subfigure}[b]{0.65\linewidth}
         \includegraphics[width=\textwidth]{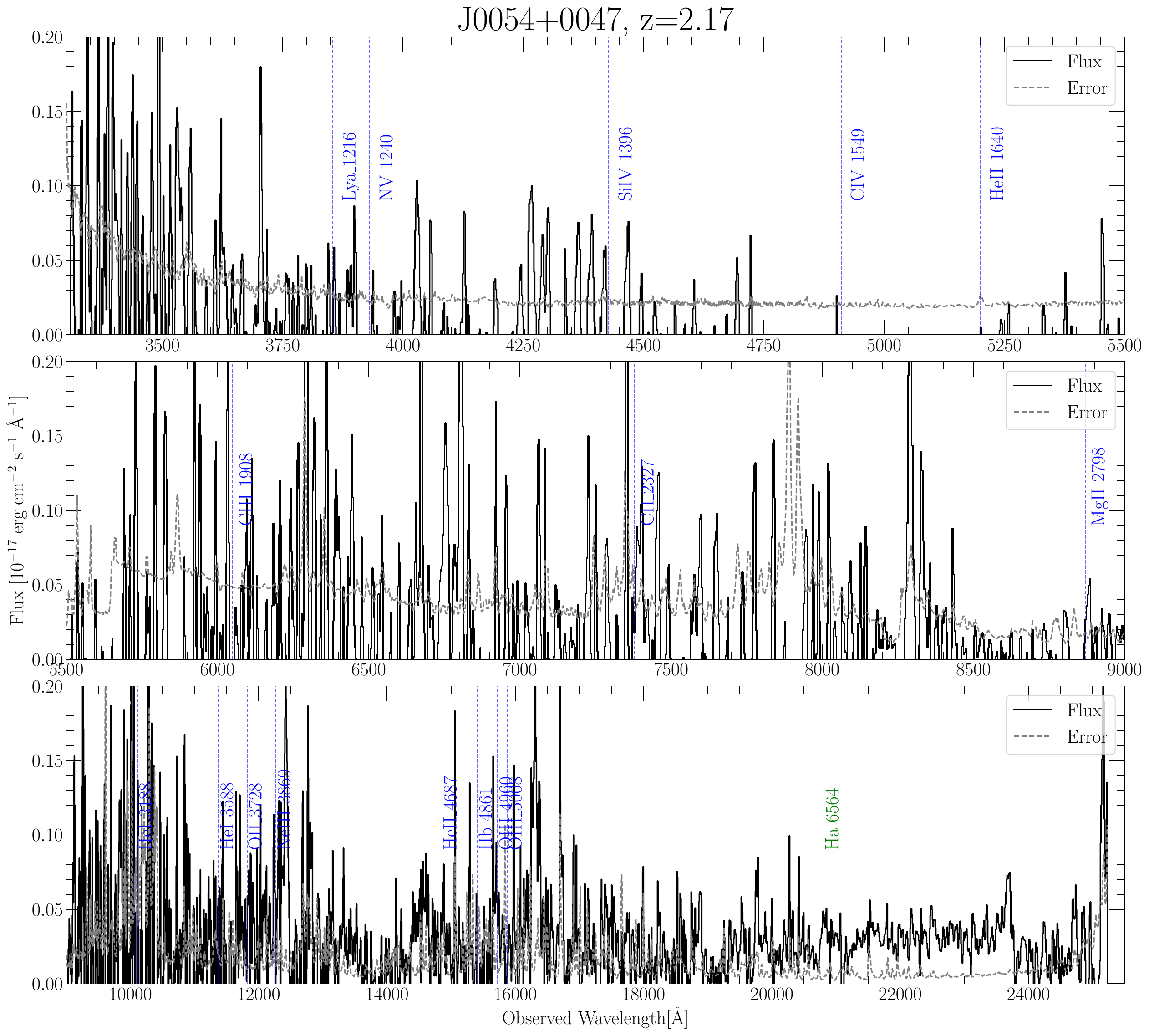}
    \end{subfigure}
     \begin{subfigure}[b]{0.34\linewidth}
         \includegraphics[width=\textwidth]{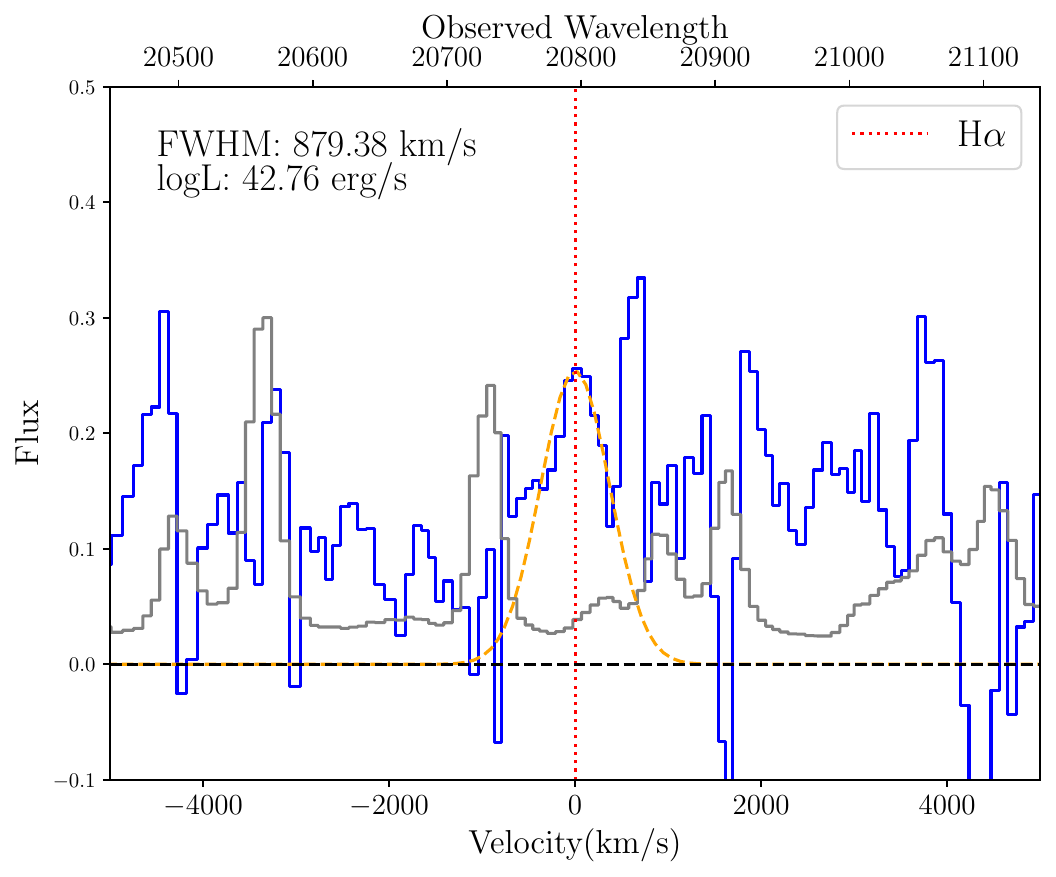}

    \end{subfigure}    
    \hfill
    \begin{subfigure}[b]{0.65\linewidth}
         \includegraphics[width=\textwidth]{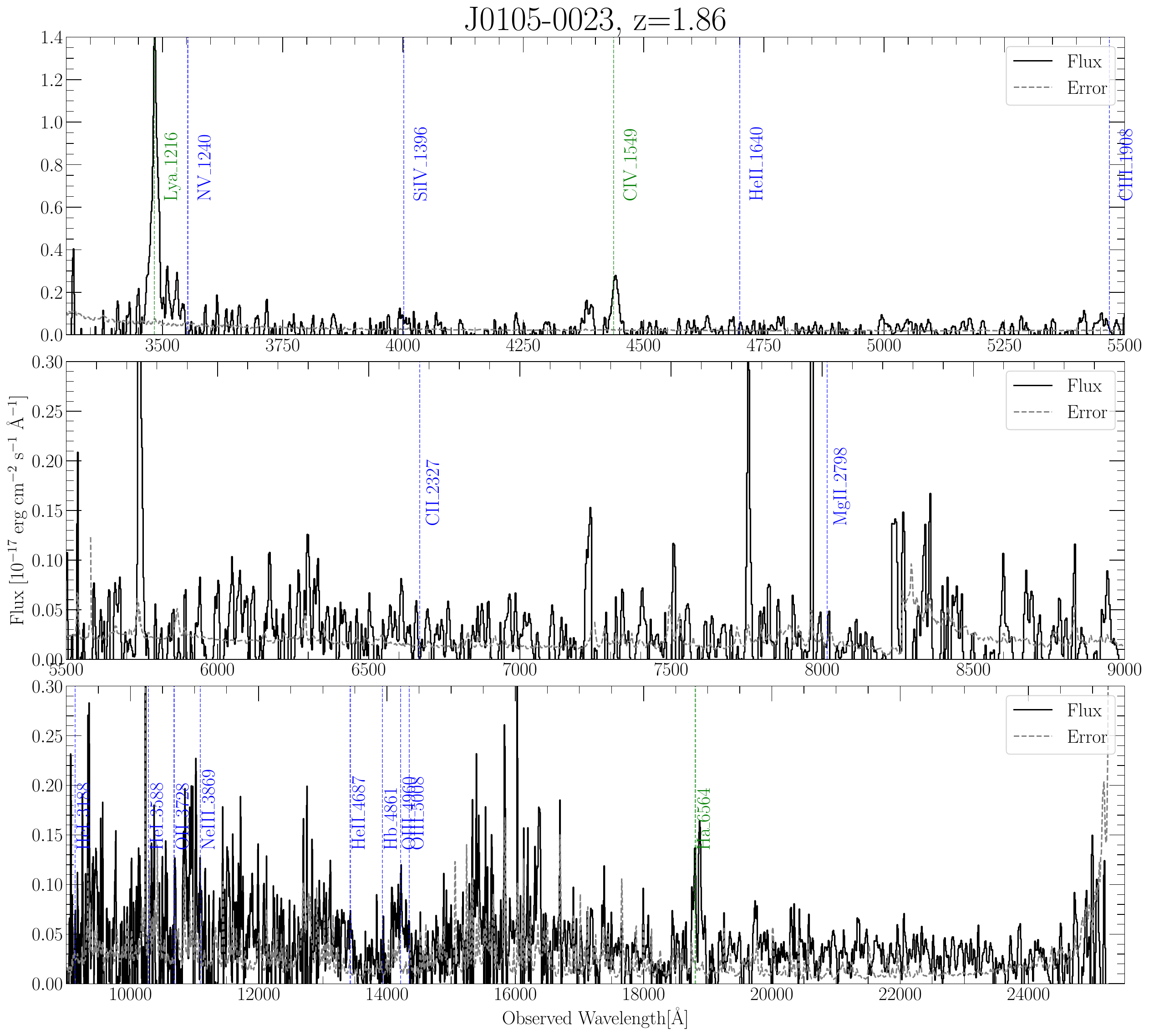}
    \end{subfigure}
     \begin{subfigure}[b]{0.34\linewidth}
         \includegraphics[width=\textwidth]{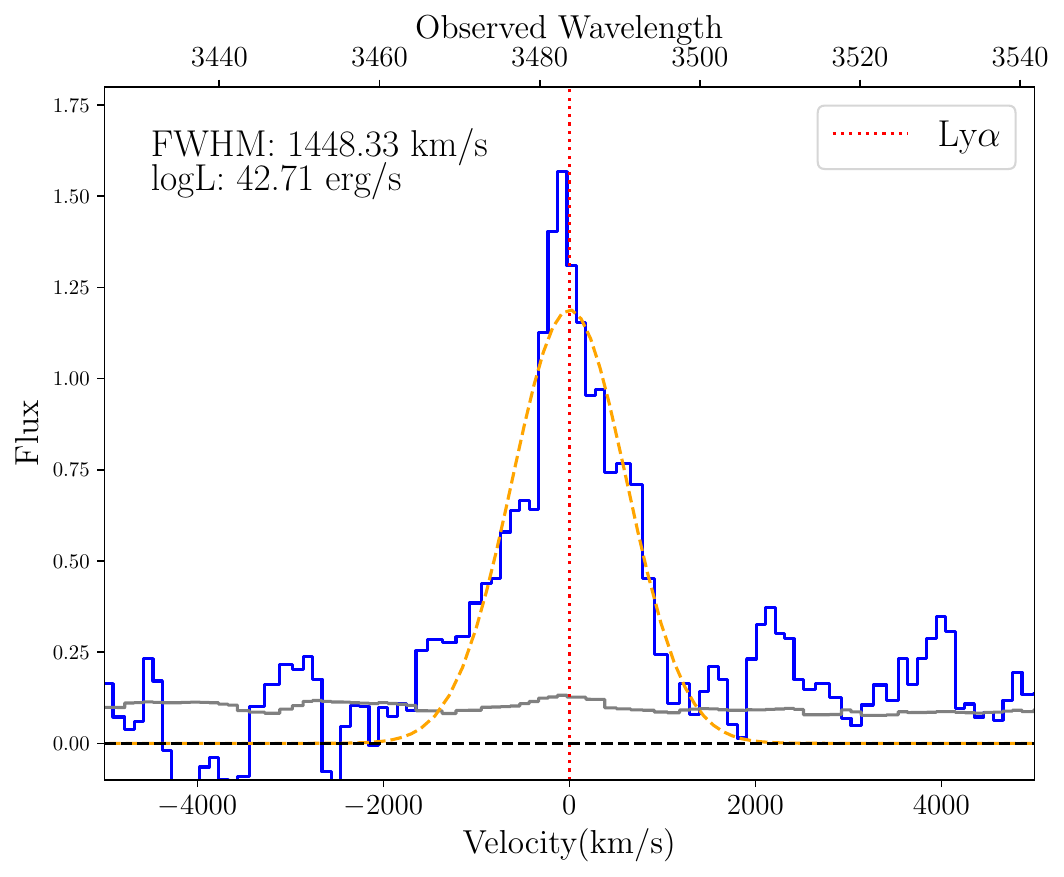}
         \hfill
         \includegraphics[width=\textwidth]{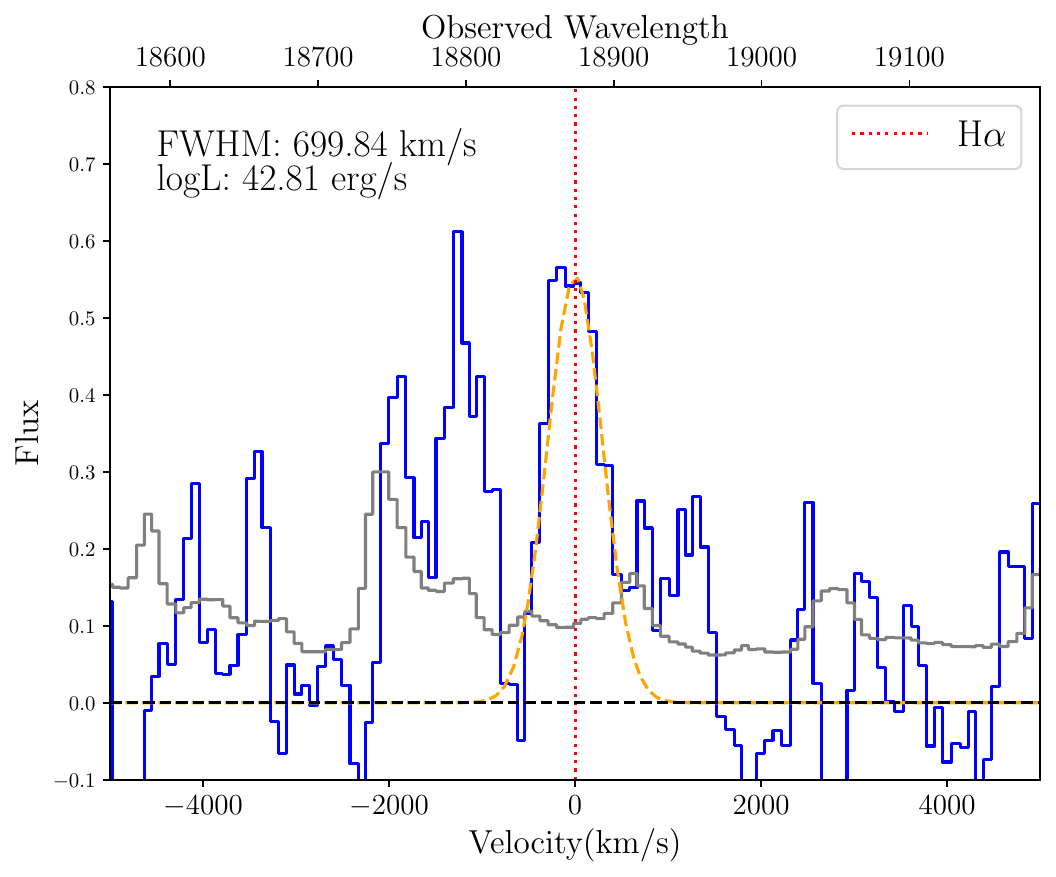}
    \end{subfigure}     

\end{figure*}

\begin{figure*}
\begin{subfigure}[b]{0.65\linewidth}
         \includegraphics[width=\textwidth]{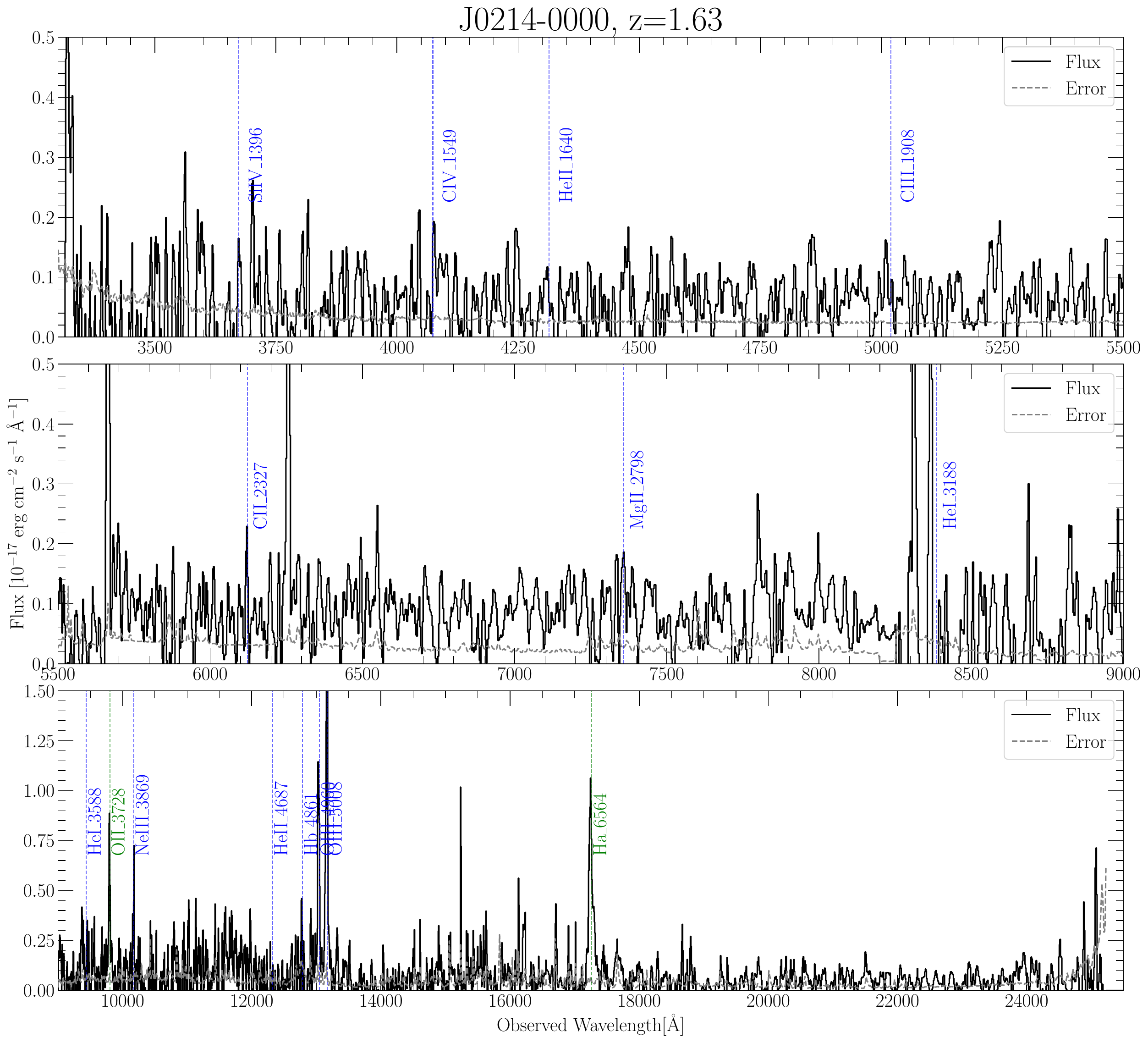}
    \end{subfigure}
     \begin{subfigure}[b]{0.34\linewidth}
         \includegraphics[width=\textwidth]{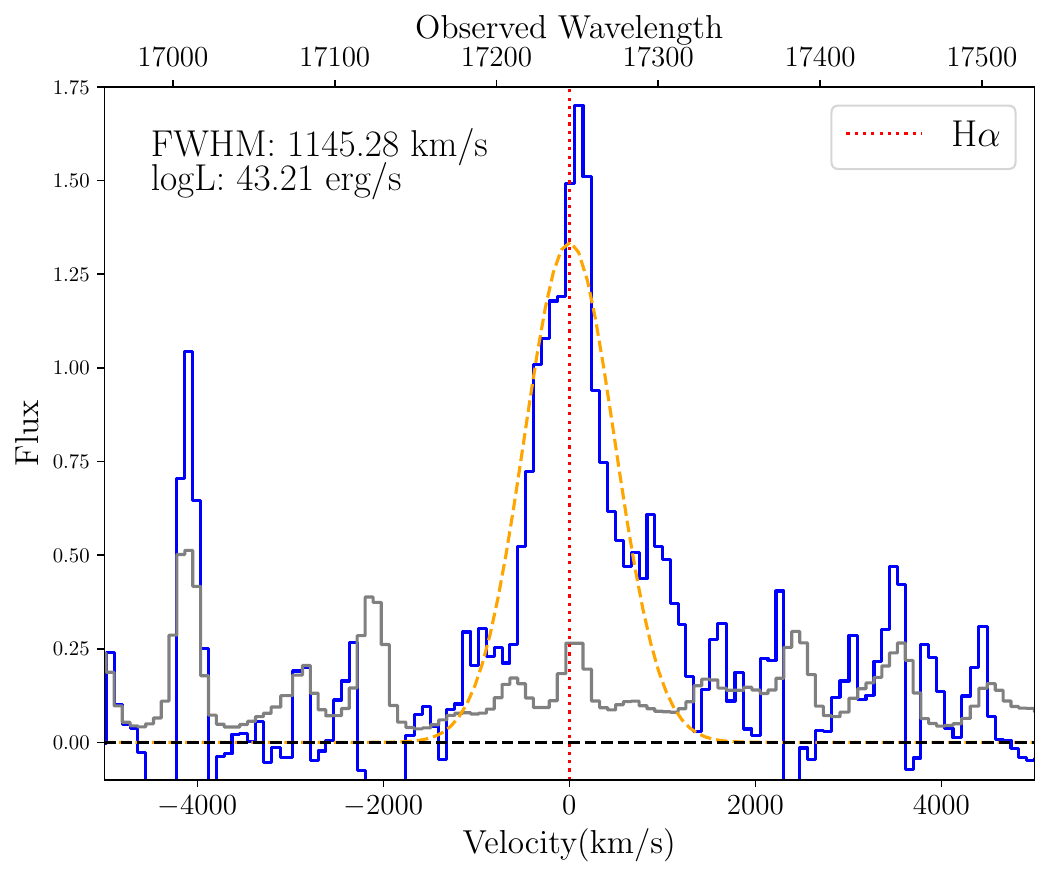}
         \hfill
         \includegraphics[width=\textwidth]{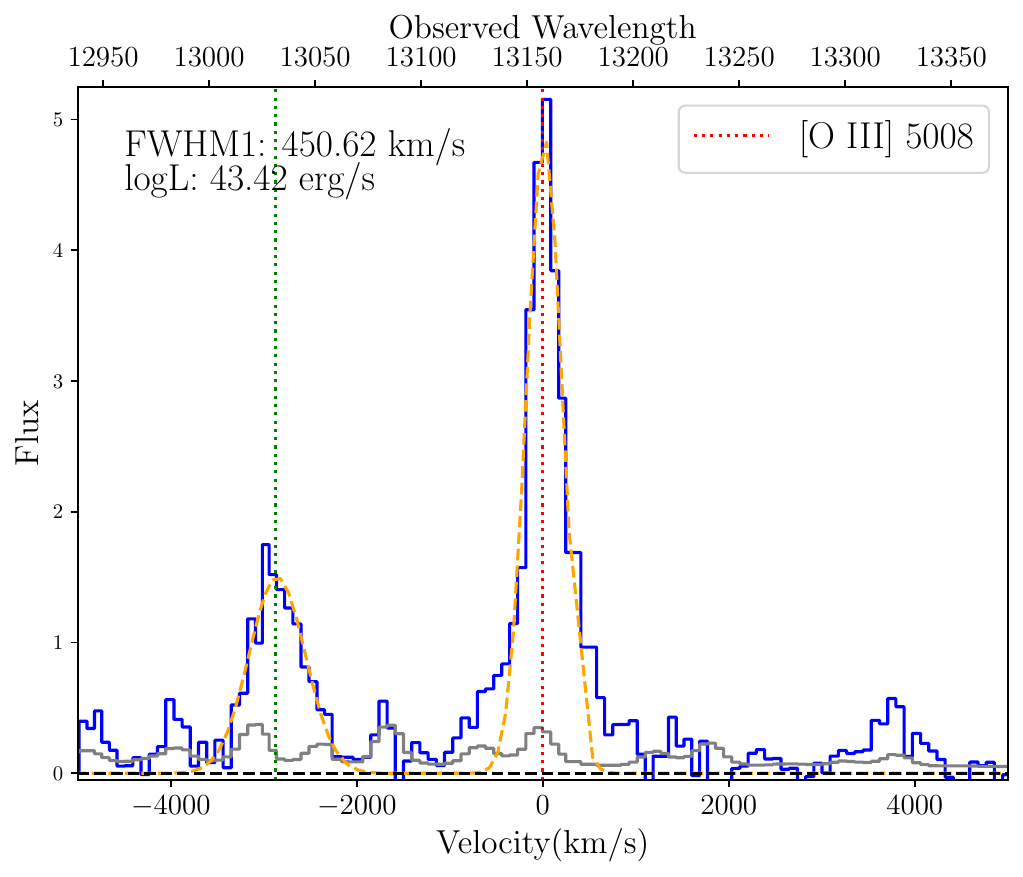}
    \end{subfigure}    
    \hfill
    \begin{subfigure}[b]{0.65\linewidth}
         \includegraphics[width=\textwidth]{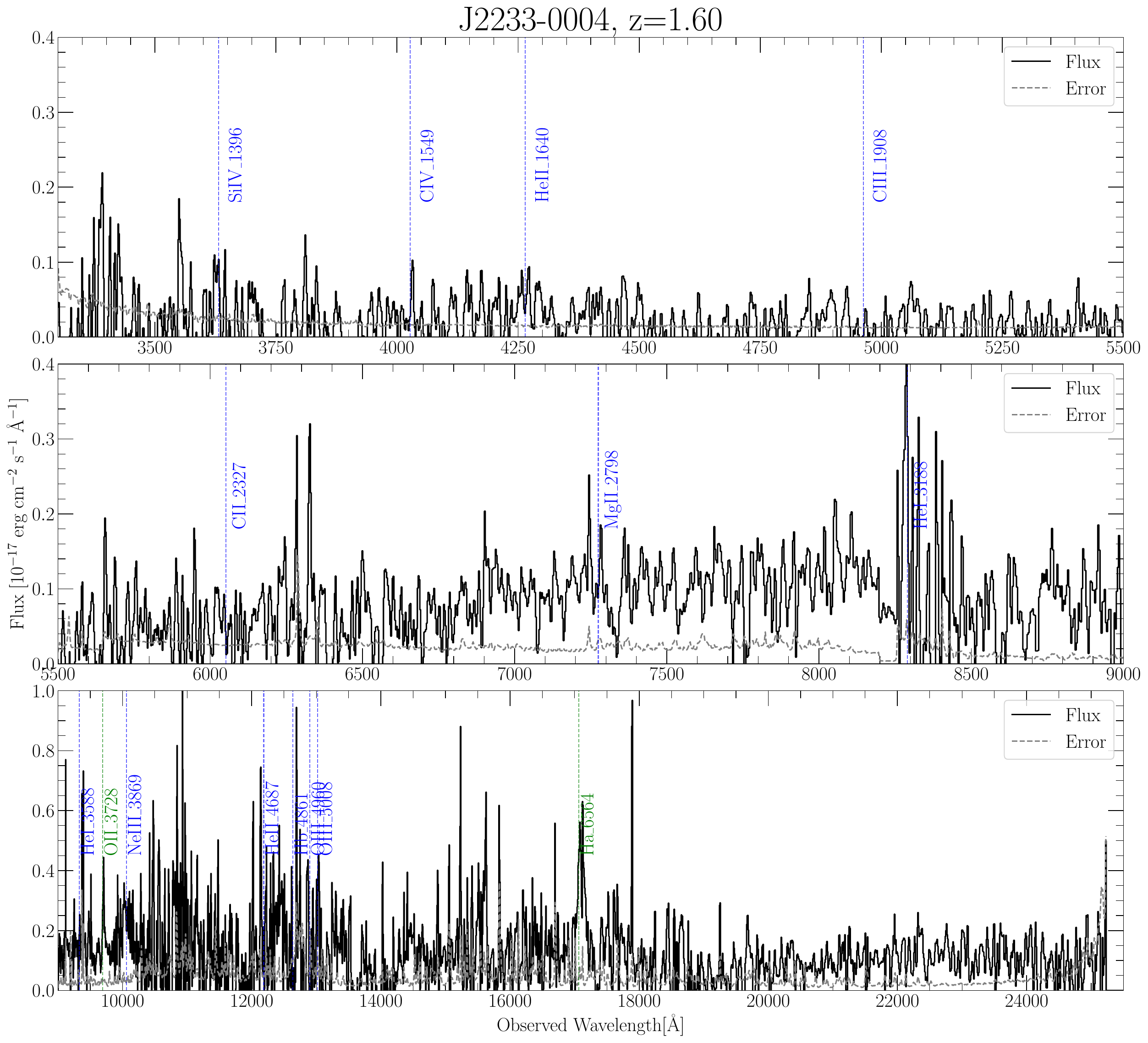}
    \end{subfigure}
     \begin{subfigure}[b]{0.34\linewidth}
         \includegraphics[width=\textwidth]{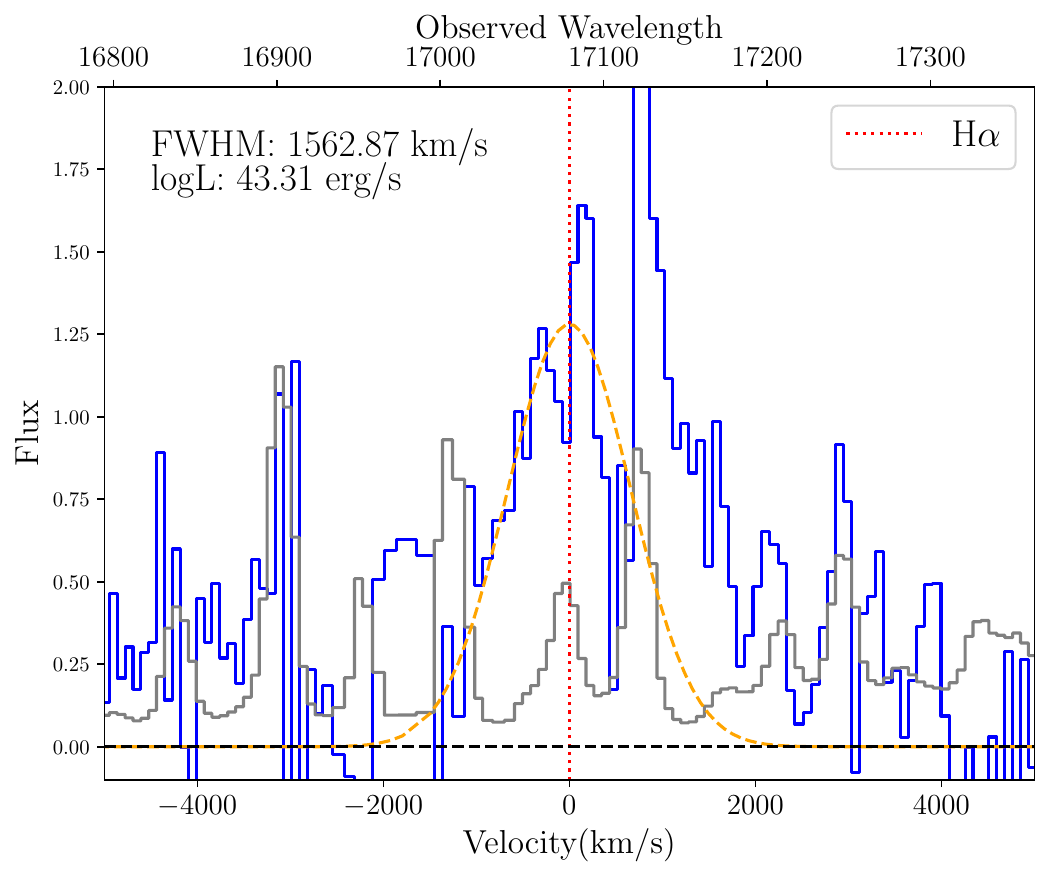}
         
    \end{subfigure}     

\end{figure*}

\begin{figure*}
\begin{subfigure}[b]{0.65\linewidth}
         \includegraphics[width=\textwidth]{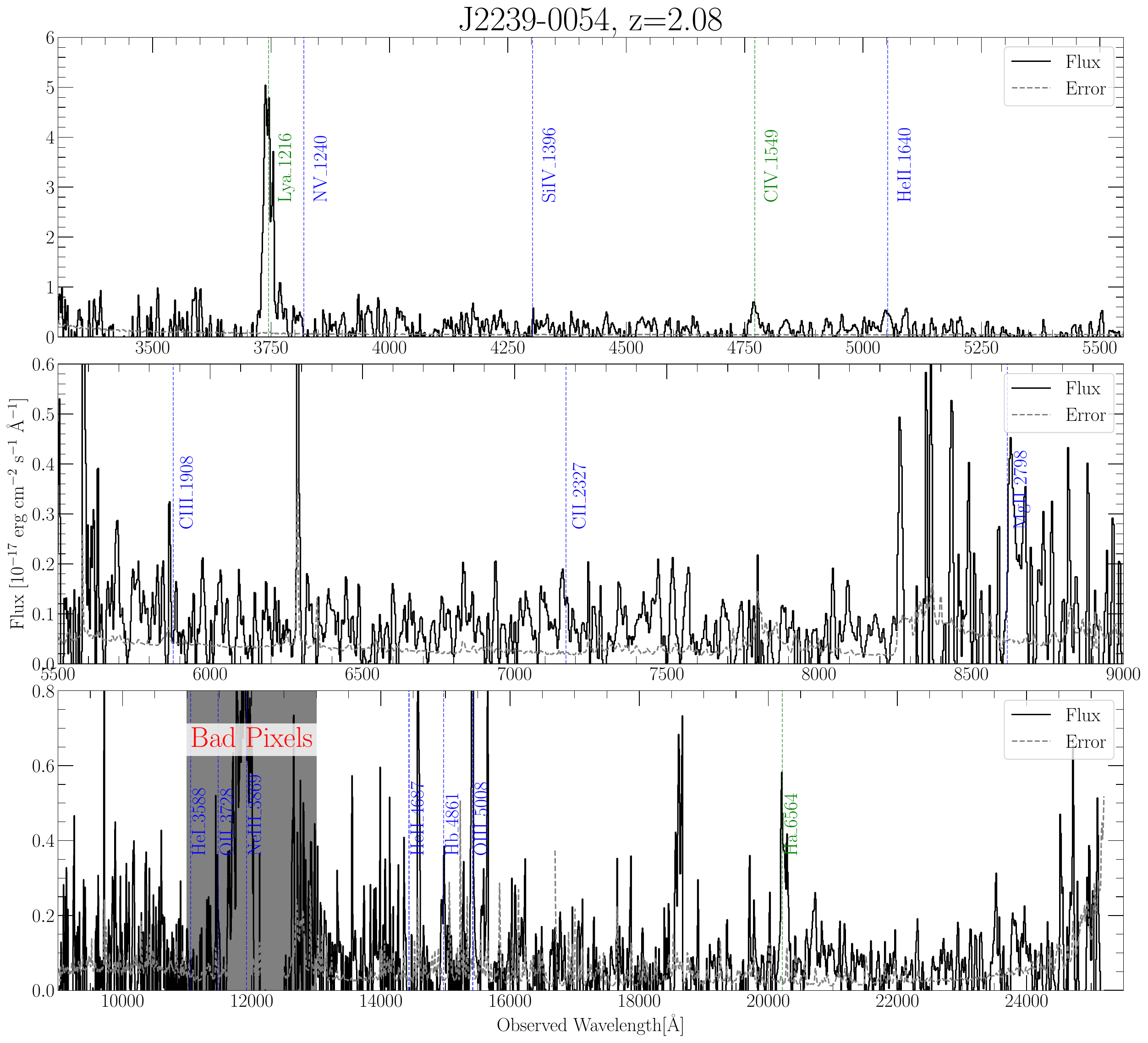}
    \end{subfigure}
     \begin{subfigure}[b]{0.34\linewidth}
         \includegraphics[width=\textwidth]{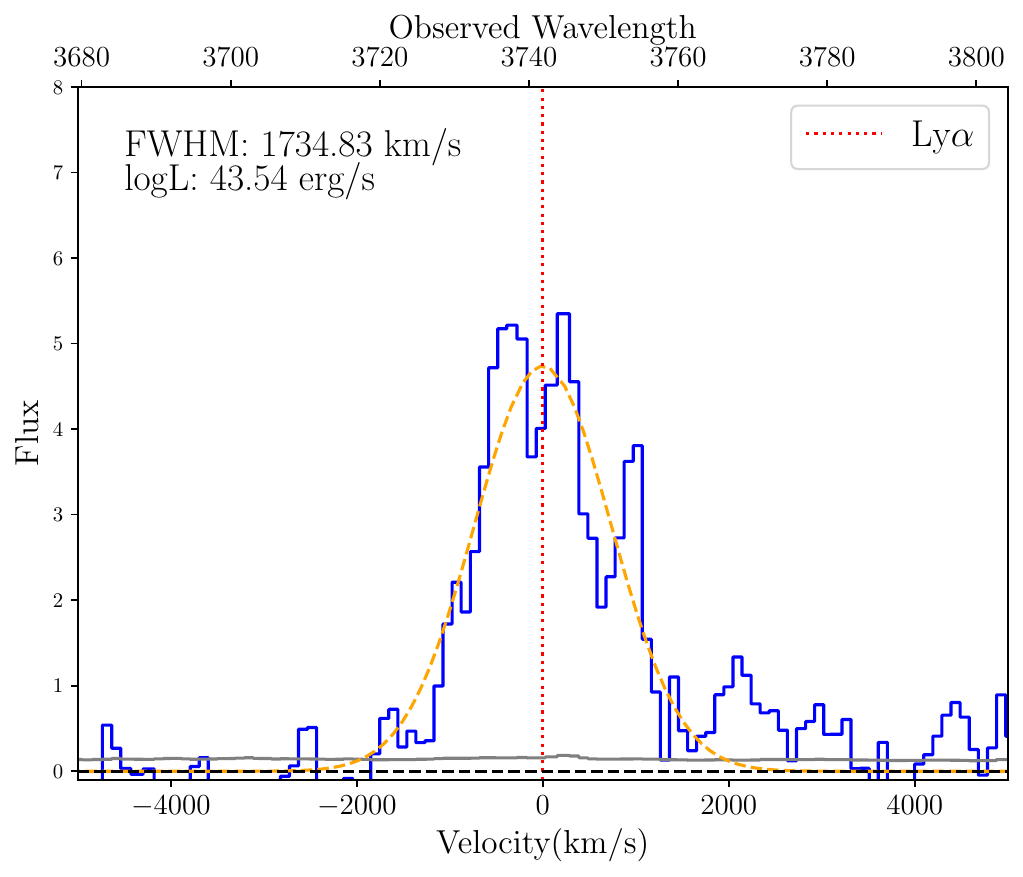}
         \hfill
         \includegraphics[width=\textwidth]{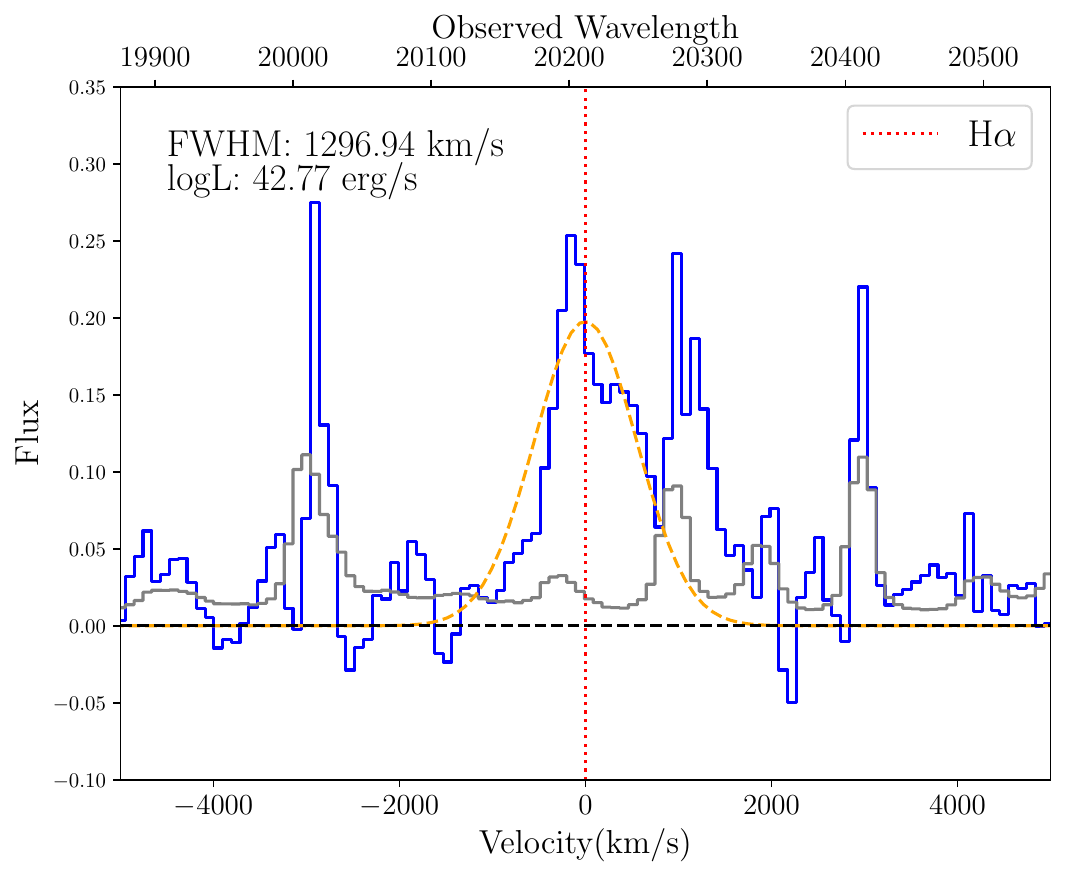}
    \end{subfigure}    
    \hfill
    \begin{subfigure}[b]{0.65\linewidth}
         \includegraphics[width=\textwidth]{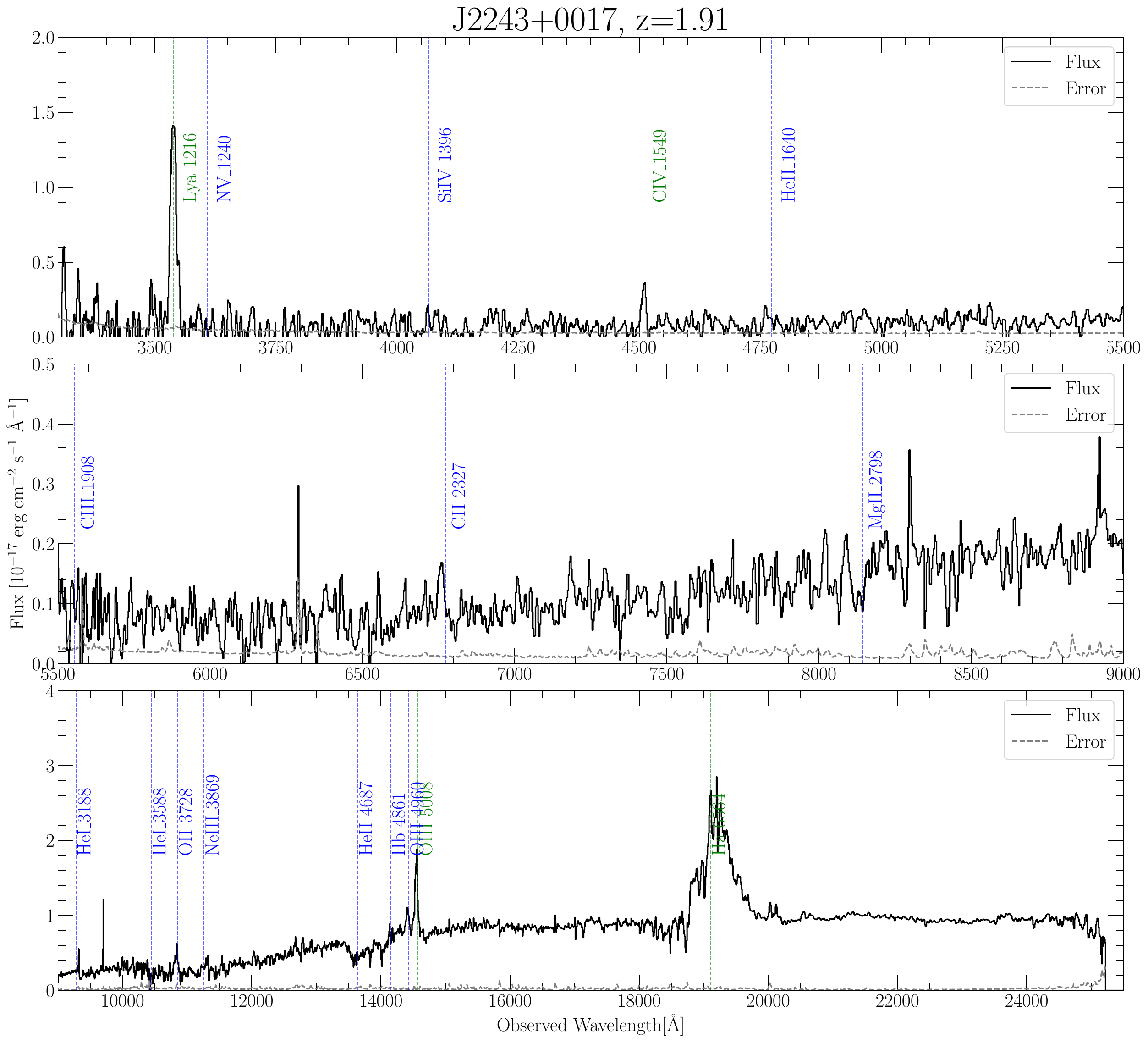}
    \end{subfigure}
     \begin{subfigure}[b]{0.34\linewidth}
         \includegraphics[width=\textwidth]{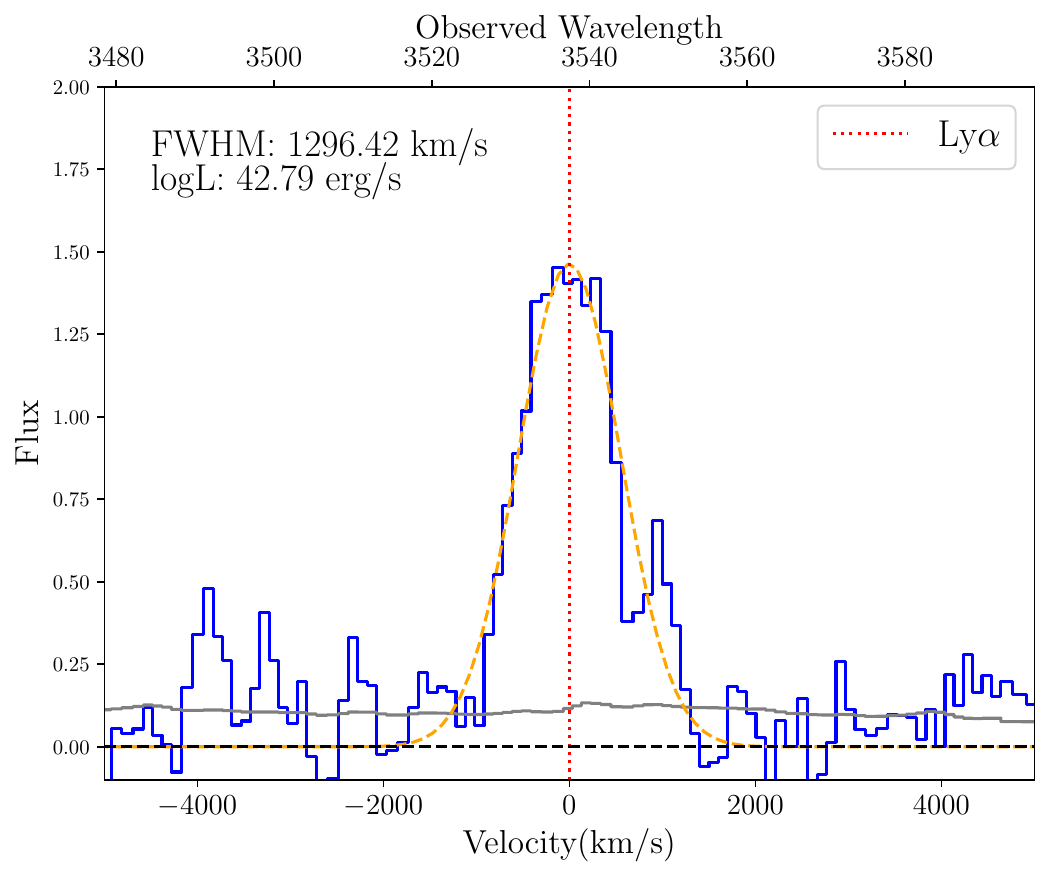}
         \hfill
         \includegraphics[width=\textwidth]{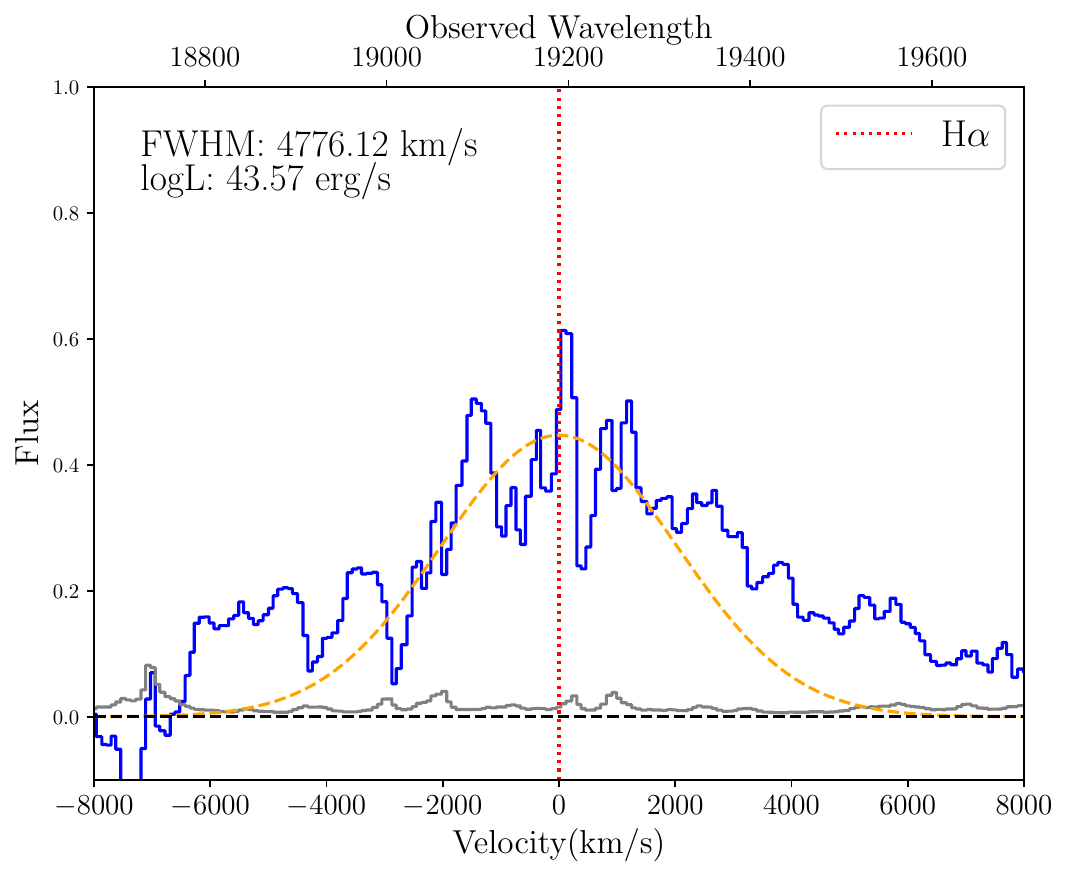}
    \end{subfigure}     

\end{figure*}

\begin{figure*}
\begin{subfigure}[b]{0.65\linewidth}
         \includegraphics[width=\textwidth]{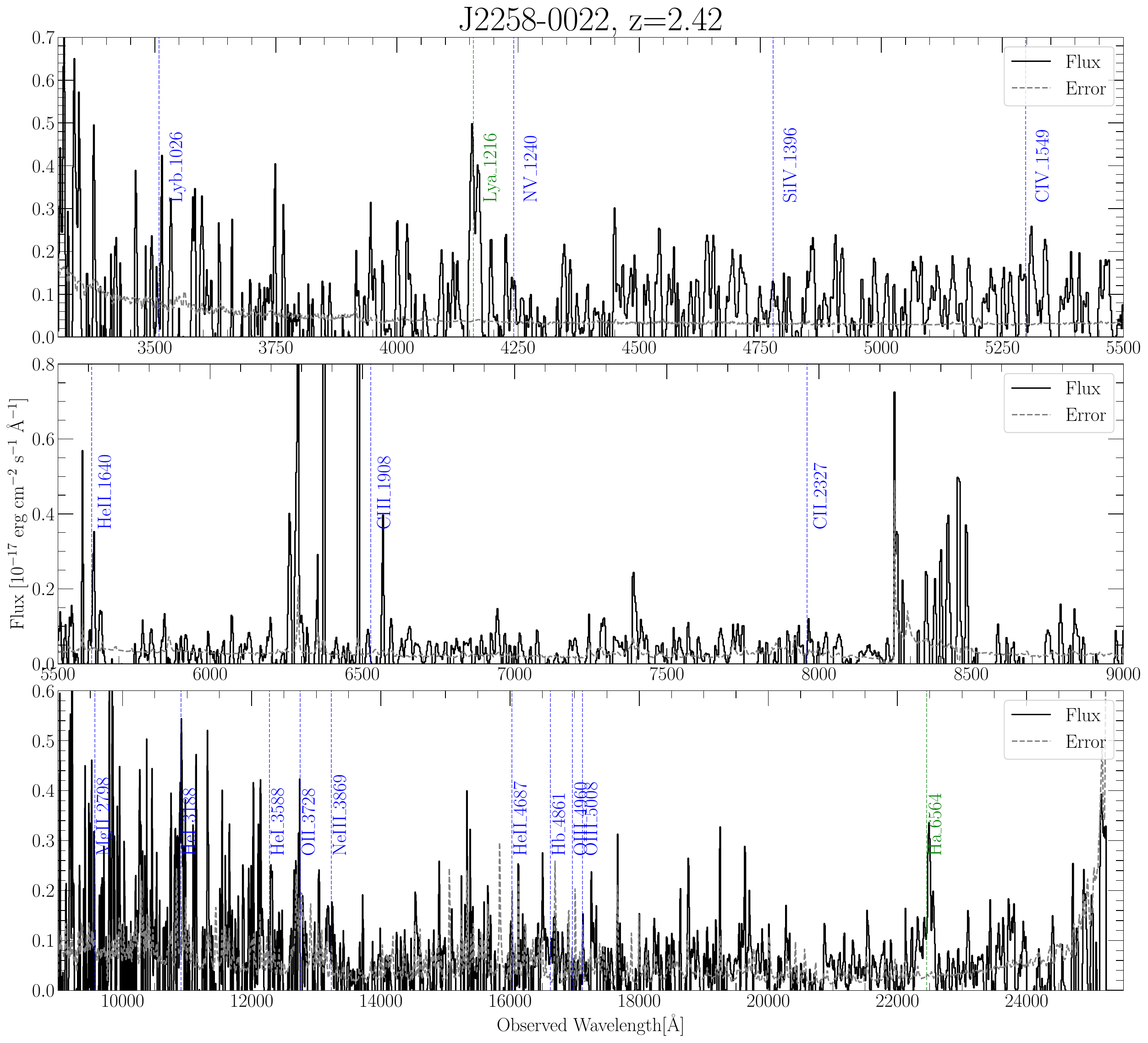}
    \end{subfigure}
     \begin{subfigure}[b]{0.34\linewidth}
         \includegraphics[width=\textwidth]{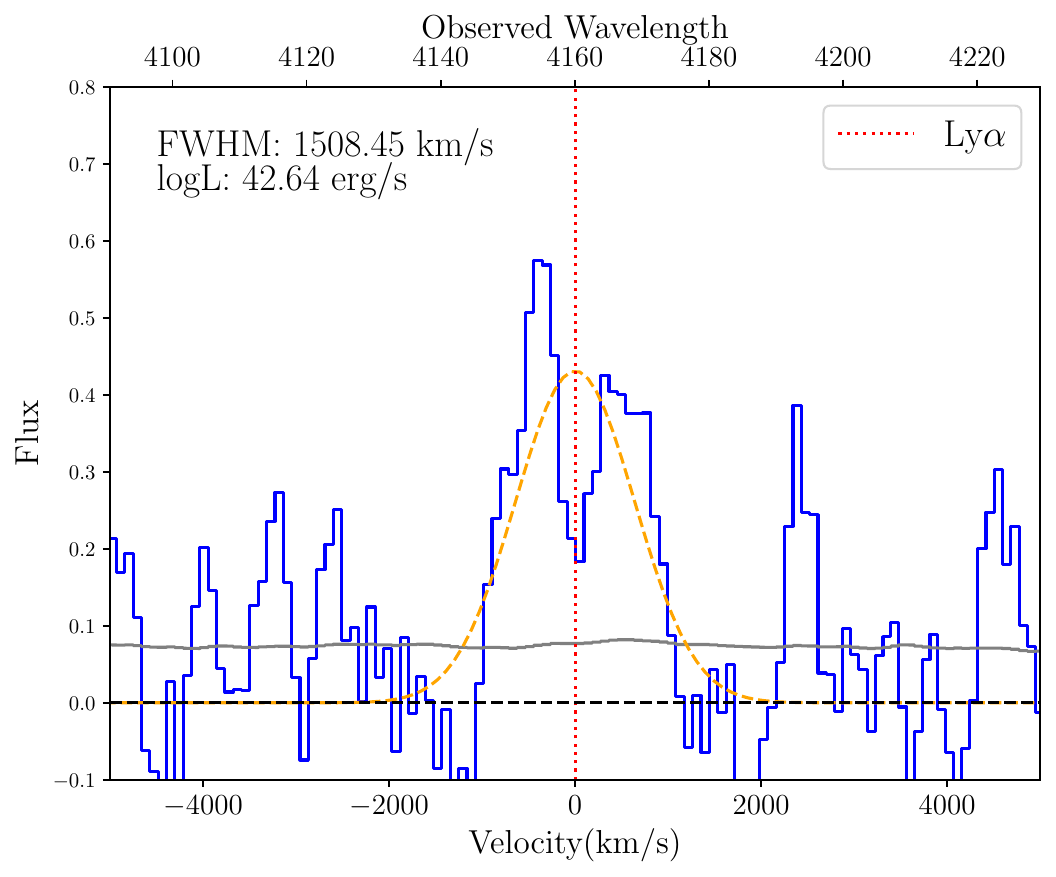}
         \hfill
         \includegraphics[width=\textwidth]{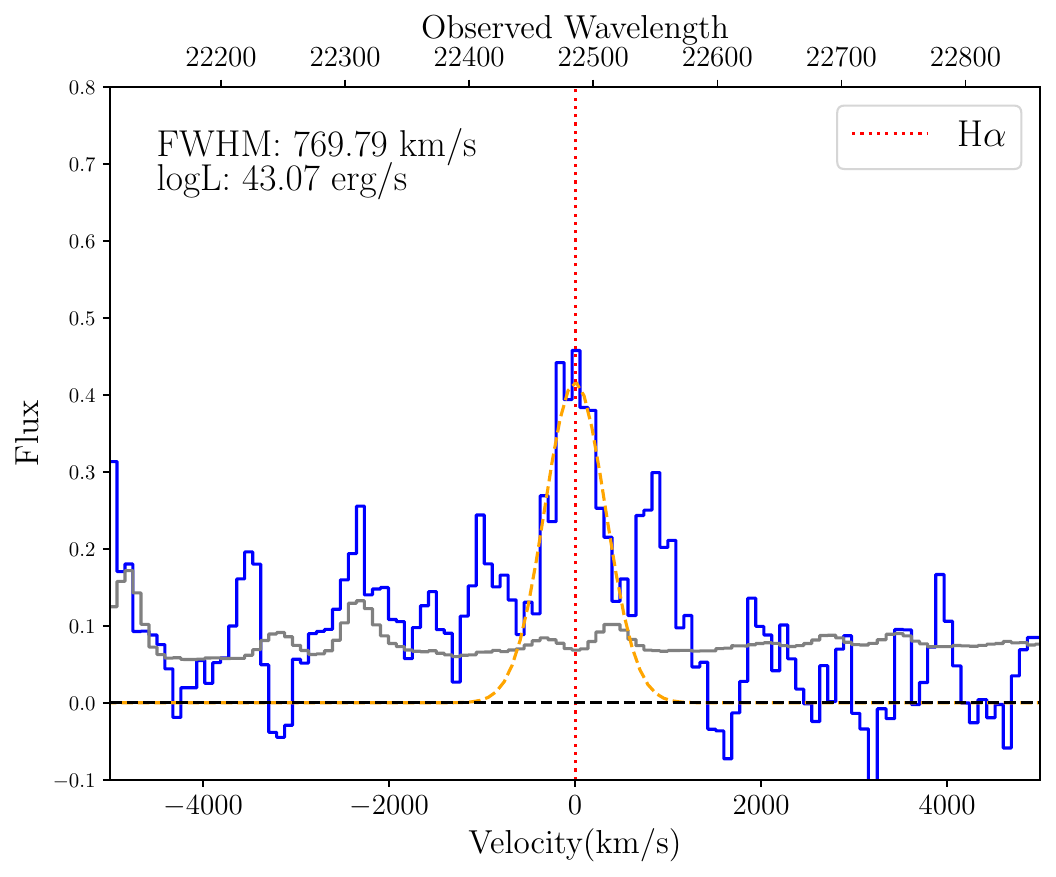}
    \end{subfigure}    
    \hfill
    \begin{subfigure}[b]{0.65\linewidth}
         \includegraphics[width=\textwidth]{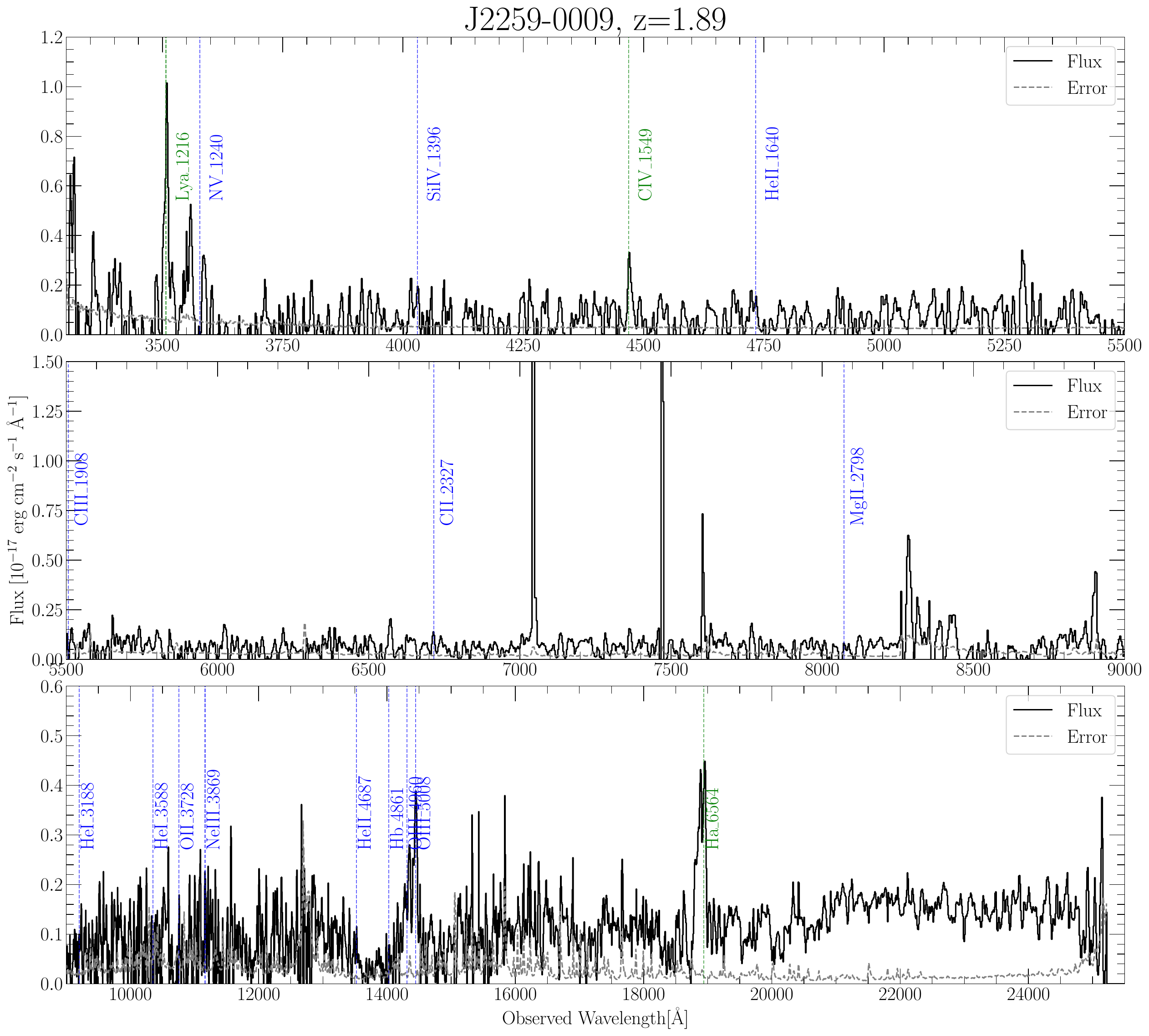}
    \end{subfigure}
     \begin{subfigure}[b]{0.34\linewidth}
         \includegraphics[width=\textwidth]{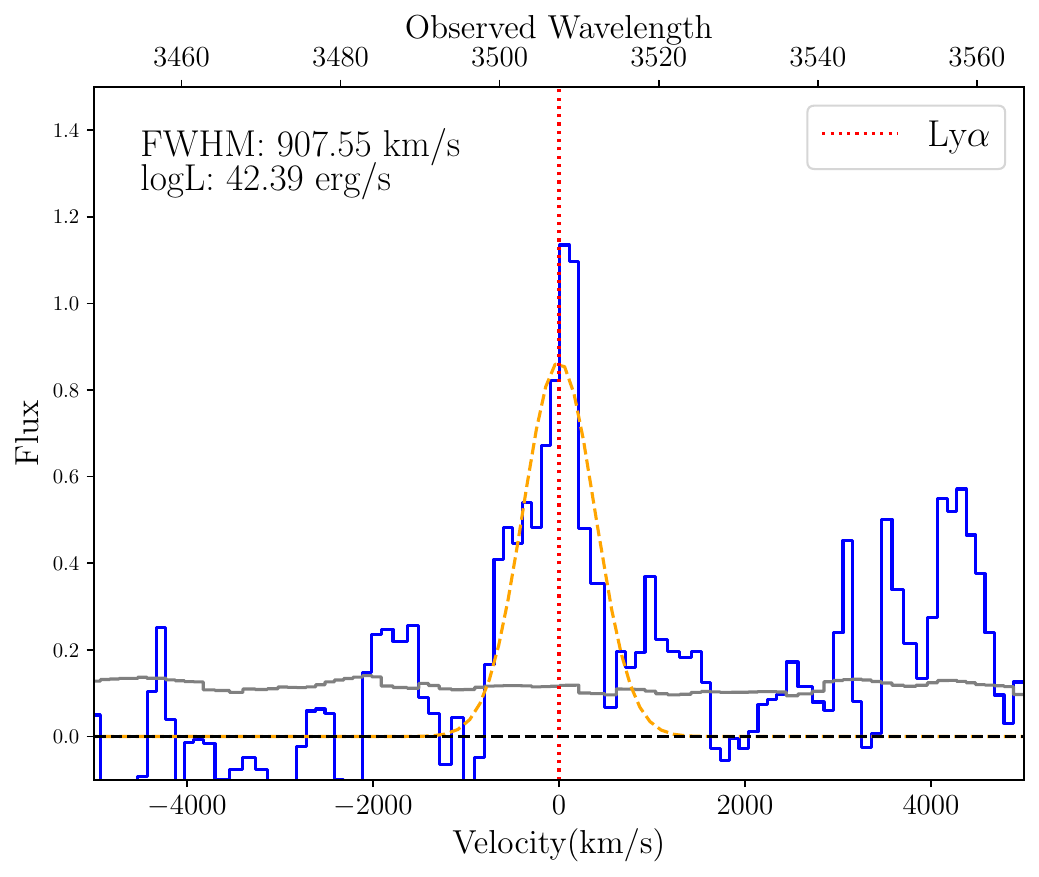}
         \hfill
         \includegraphics[width=\textwidth]{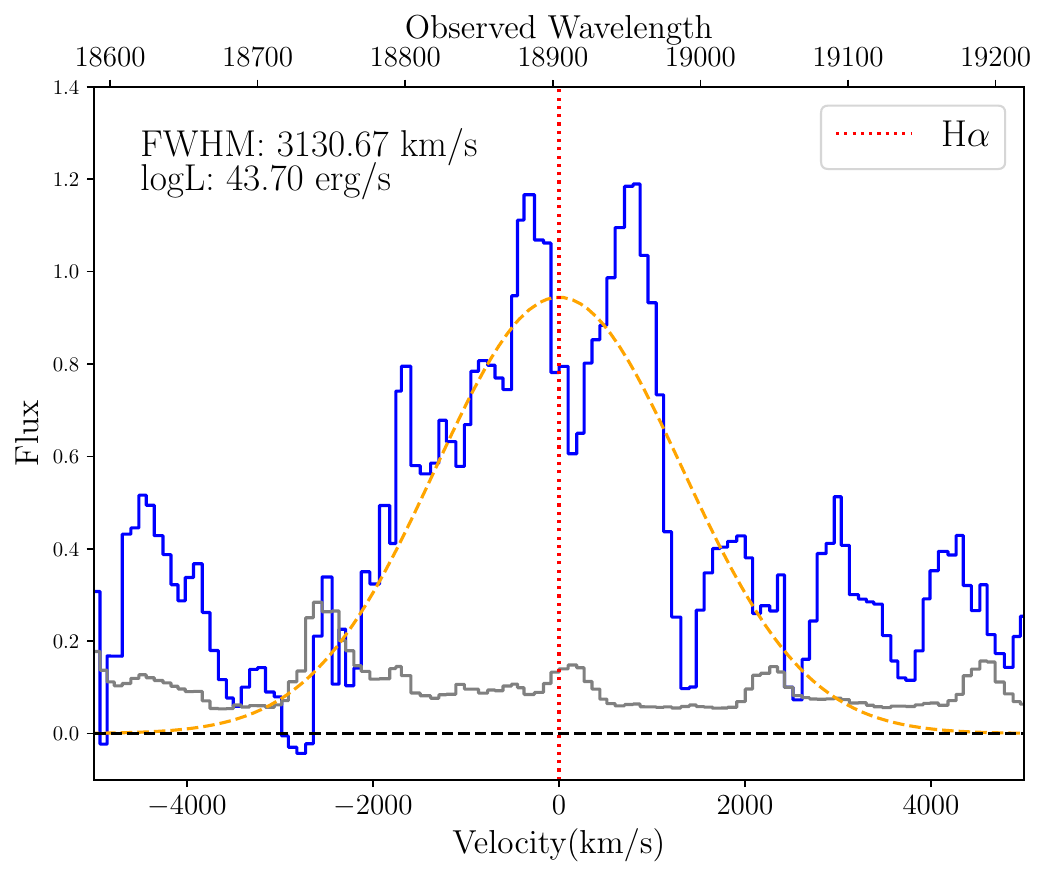}
    \end{subfigure}     

\end{figure*}

\begin{figure*}
\begin{subfigure}[b]{0.65\linewidth}
         \includegraphics[width=\textwidth]{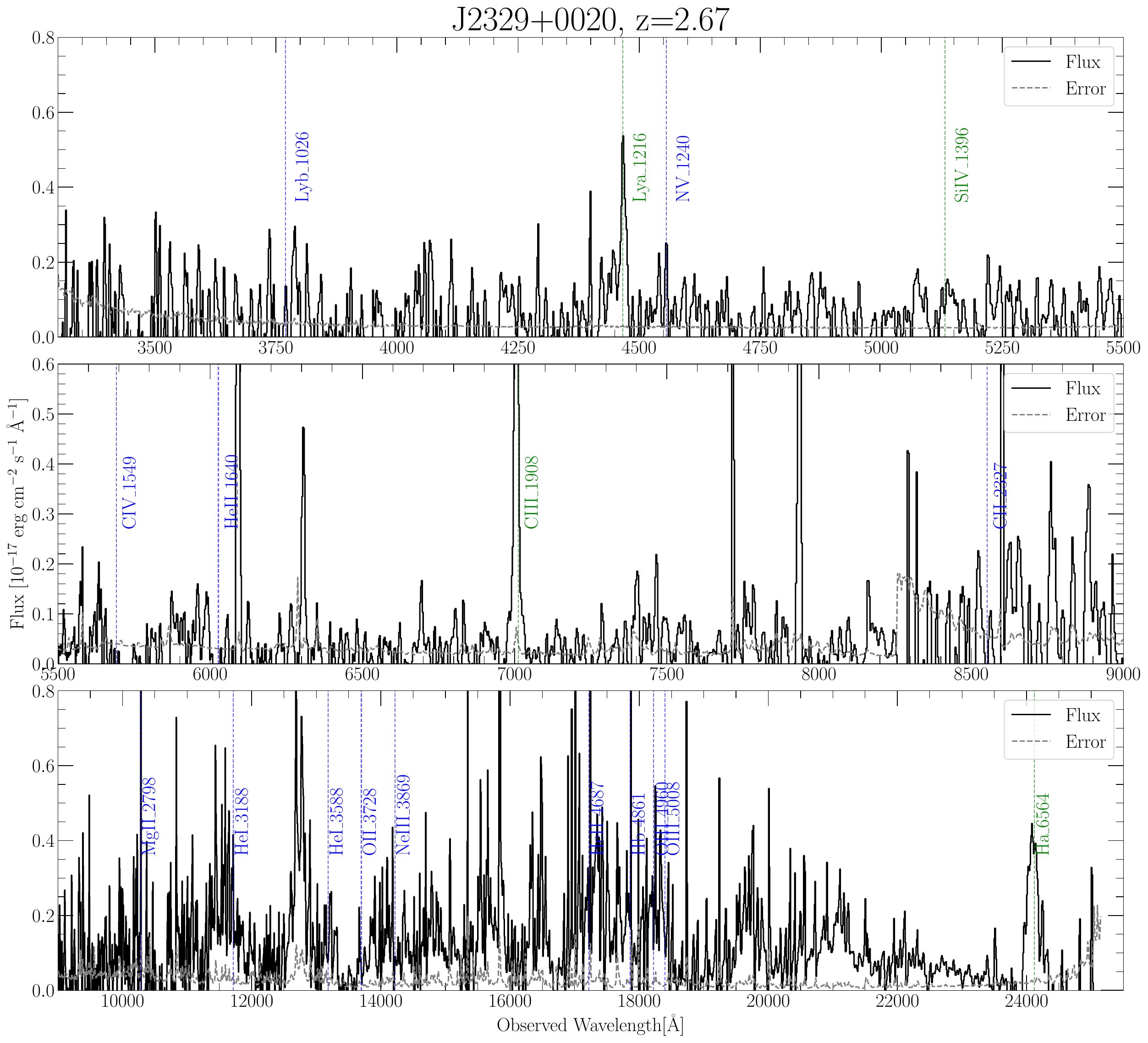}
    \end{subfigure}
     \begin{subfigure}[b]{0.34\linewidth}
         \includegraphics[width=\textwidth]{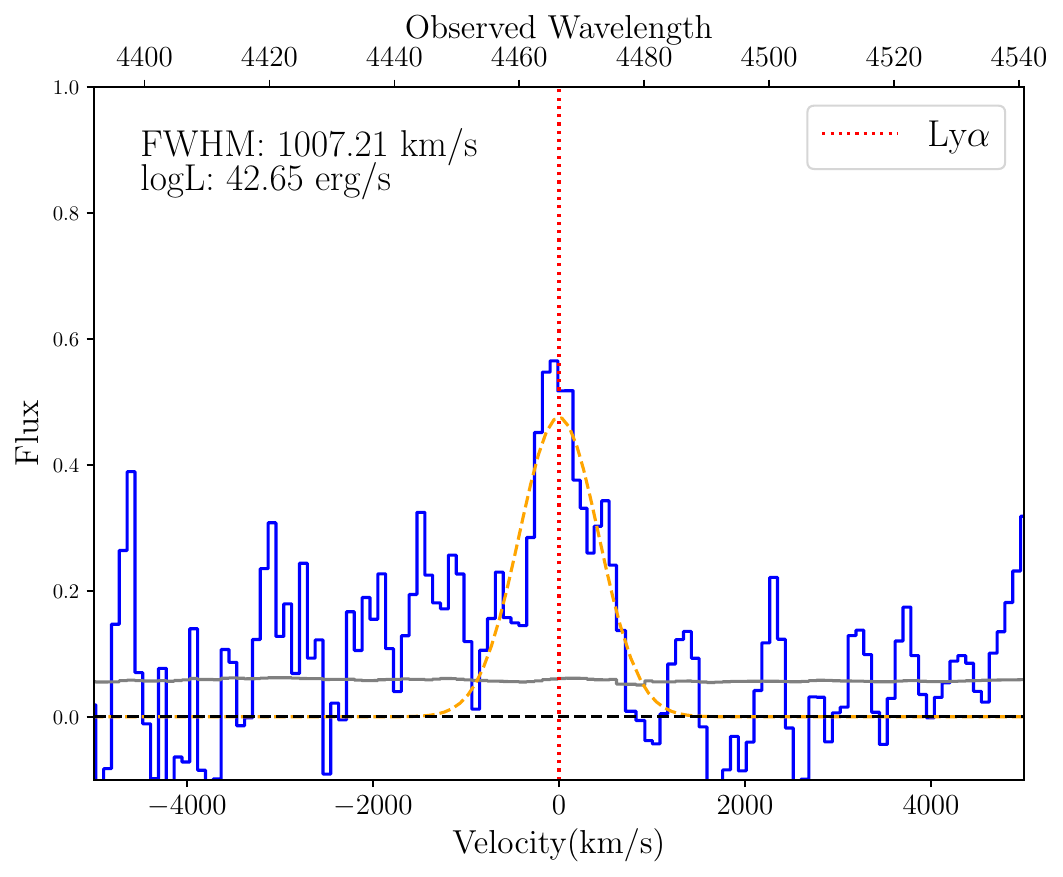}
         \hfill
         \includegraphics[width=\textwidth]{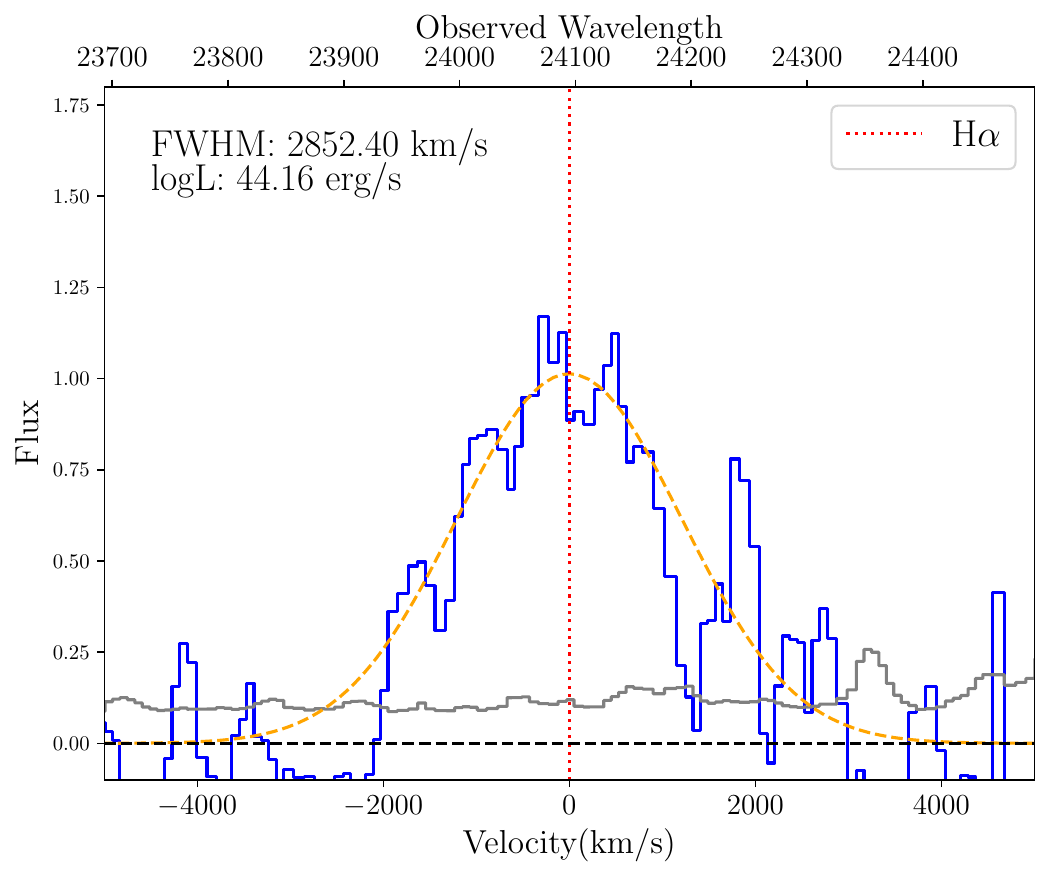}
    \end{subfigure}    
    \hfill
    \begin{subfigure}[b]{0.65\linewidth}
         \includegraphics[width=\textwidth]{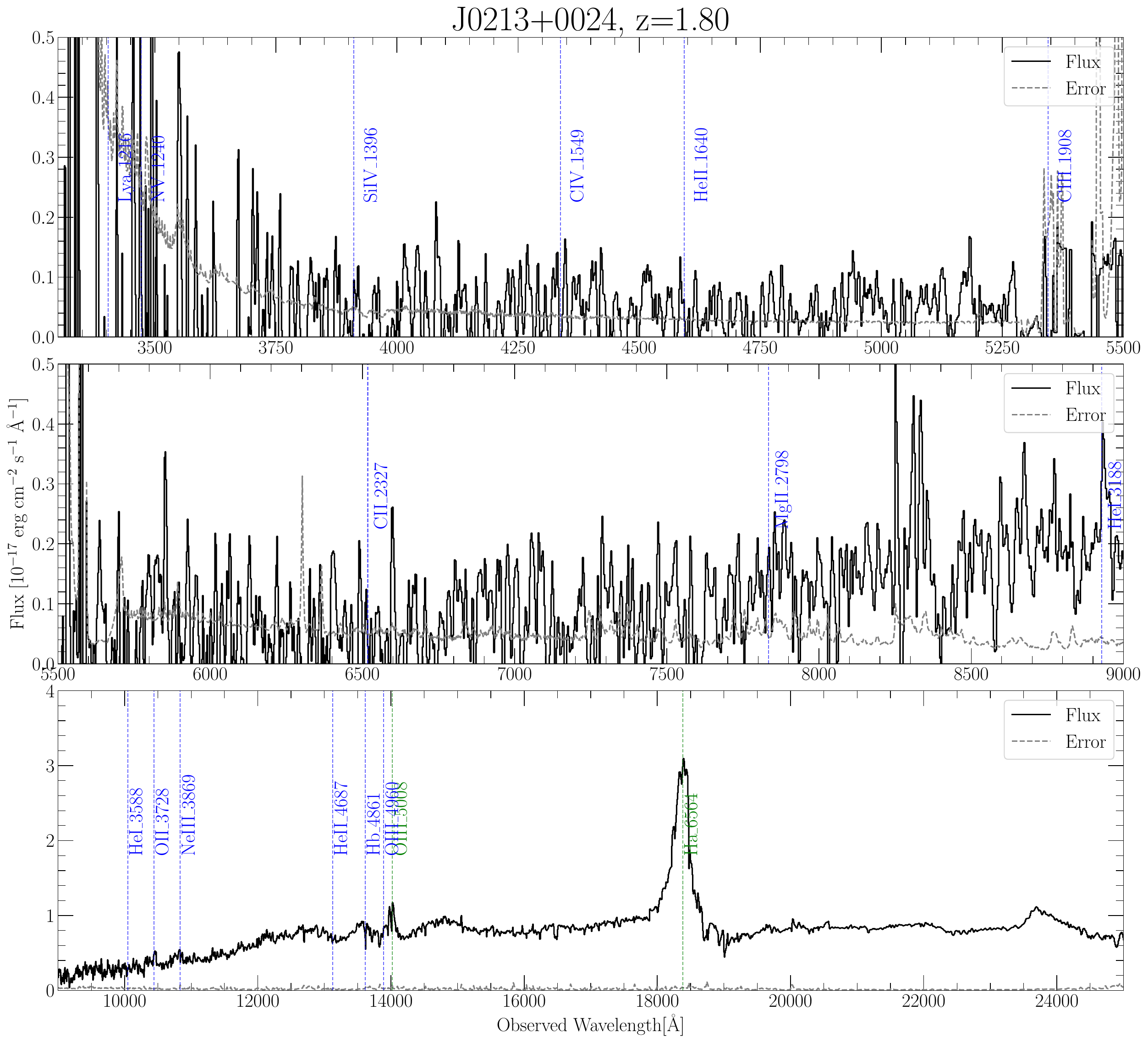}
    \end{subfigure}
     \begin{subfigure}[b]{0.34\linewidth}
         \includegraphics[width=\textwidth]{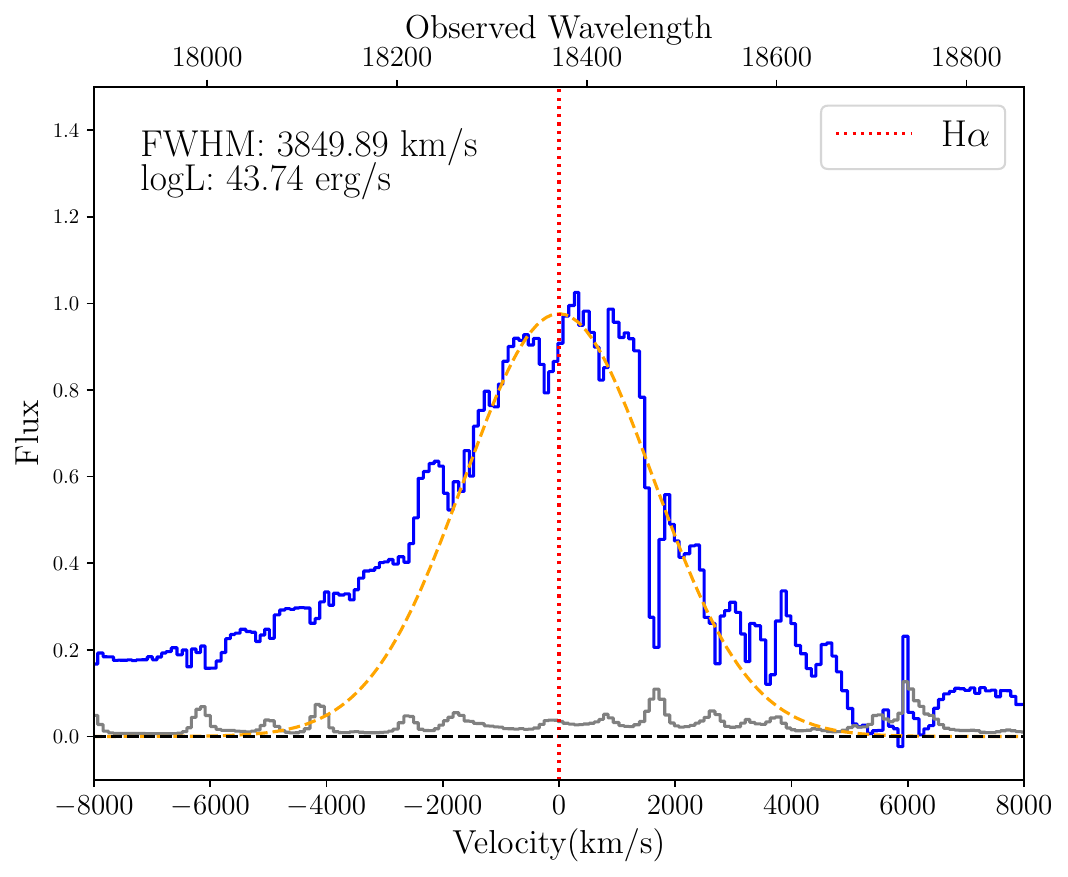}
         \hfill
         
    \end{subfigure}     
\caption{Spectra and line fittings for all the targets in our sample. 
}
\end{figure*}

\label{lastpage}
\end{document}